\documentclass[reprint,amsmath,amssymb,aps,nofootinbib]{revtex4-1}
\usepackage{graphicx}
\usepackage{amsmath}
\usepackage{dcolumn}
\usepackage{bm}

\usepackage{mathtools}
\usepackage{blkarray}
\usepackage{footnote}
\usepackage{hyperref}
\hypersetup{
    colorlinks,
    citecolor=blue,
    filecolor=blue,
    linkcolor=blue,
    urlcolor=blue,
    pagebackref=true,
    pdfdisplaydoctitle=true}
\usepackage[toc,page]{appendix}
\usepackage{adjustbox}
\usepackage{float}
\usepackage{color}

\usepackage{scalerel}
\usepackage{tikz}
\usetikzlibrary{svg.path}

\definecolor{orcidlogocol}{HTML}{A6CE39}
\tikzset{
  orcidlogo/.pic={
    \fill[orcidlogocol] svg{M256,128c0,70.7-57.3,128-128,128C57.3,256,0,198.7,0,128C0,57.3,57.3,0,128,0C198.7,0,256,57.3,256,128z};
    \fill[white] svg{M86.3,186.2H70.9V79.1h15.4v48.4V186.2z}
                 svg{M108.9,79.1h41.6c39.6,0,57,28.3,57,53.6c0,27.5-21.5,53.6-56.8,53.6h-41.8V79.1z M124.3,172.4h24.5c34.9,0,42.9-26.5,42.9-39.7c0-21.5-13.7-39.7-43.7-39.7h-23.7V172.4z}
                 svg{M88.7,56.8c0,5.5-4.5,10.1-10.1,10.1c-5.6,0-10.1-4.6-10.1-10.1c0-5.6,4.5-10.1,10.1-10.1C84.2,46.7,88.7,51.3,88.7,56.8z};
  }
}

\newcommand\orcidicon[1]{\href{https://orcid.org/#1}{\mbox{\scalerel*{
\begin{tikzpicture}[yscale=-1,transform shape]
\pic{orcidlogo};
\end{tikzpicture}
}{|}}}}

\definecolor{myred}{rgb}{1,0,0}
\colorlet{lightred}{black}
\colorlet{darkred}{myred!60!black}

\definecolor{mypurple}{rgb}{1,0,1}
\colorlet{lavender}{mypurple!50!white}

\newcommand{\jbmsec}[1]{\textcolor{lightred}{#1}}
\newcommand{\jbmsectwo}[1]{\textcolor{lightred}{#1}}
\newcommand{\jbmt}[2]{\textcolor{lightred}{#2}}

\newcommand{\misec}[1]{\textcolor{lightred}{#1}}

\newcommand{\jbm}[1]{#1}
\newcommand{\jbmb}[1]{#1}
\newcommand{\oldmichael}[1]{#1} 
\definecolor{orange}{rgb}{1,0.5,0.25}
\newcommand{\oldyy}[1]{#1} 
\newcommand{\oldmh}[1]{#1} 

\begin{document}

\title{\textit{Ab initio} low-energy effective Hamiltonians for high-temperature superconducting cuprates Bi$_2$Sr$_2$CuO$_6$, Bi$_2$Sr$_2$CaCu$_2$O$_8$\jbmsec{, HgBa$_2$CuO$_4$} and CaCuO$_2$
}
\author{
Jean-Baptiste Mor\'ee$^{1}$ \orcidicon{0000-0002-0710-9880}, 
Motoaki Hirayama$^{2,3,4}$ \orcidicon{0000-0002-1817-3810}, 
Michael Thobias Schmid$^{1}$ \orcidicon{0000-0003-2724-0621}, 
Youhei Yamaji$^{5}$ \orcidicon{0000-0002-4055-8792}, 
and Masatoshi Imada$^{1,6}$ \orcidicon{0000-0002-5511-2056}
}

\affiliation{
$^{1}$ Waseda Research Institute for Science and Engineering, Waseda University, 3-4-1, Okubo, Shinjuku, Tokyo 169-8555, Japan\\
\oldmh{$^{2}$ Department of Applied Physics, University of Tokyo, 7-3-1 Hongo, Bunkyo-ku, Tokyo 113-8656, Japan}\\
\oldmh{$^{3}$ RIKEN Center for Emergent Matter Science, Wako, Saitama 351-0198, Japan}\\
\oldmh{$^{4}$ JST, PRESTO, Hongo, Bunkyo-ku, Tokyo 113-8656, Japan}\\
$^{5}$ Center for Green Research on Energy and Environmental Materials, National Institute for Materials Science, Namiki, Tsukuba-shi, Ibaraki, 305-0044, Japan\\
$^{6}$ Toyota Physical and Chemical Research Institute, 41-1, Yokomichi, Nagakute, Aichi 480-1192, Japan
}


\begin{abstract}
We derive \textit{ab initio} low-energy effective Hamiltonians (LEH) for high-temperature superconducting (SC) copper oxides Bi$_2$Sr$_2$CuO$_6$ (Bi2201, $N_{\ell}=1$, $T_c^{\rm exp} \sim 10$ K), Bi$_2$Sr$_2$CaCu$_2$O$_8$ (Bi2212, $N_{\ell}=2$, $T_c^{\rm exp} \sim 84$ K), HgBa$_2$CuO$_4$ (Hg1201, $N_{\ell}=1$, $T_c^{\rm exp} \sim 90$ K) and CaCuO$_2$ (Ca11, $N_{\ell}=\infty$, $T_c^{\rm exp} \sim 110$ K), with different experimental optimal SC transition temperature $T_c^{\rm exp}$ and number $N_{\ell}$ of laminated CuO$_2$ planes between the two neighboring block layers. We apply the latest methodology of the multiscale \textit{ab initio} scheme for correlated electron systems (MACE), and focus on the LEH consisting of one antibonding (AB) Cu$3d_{x^2-y^2}$/O$2p_{\sigma}$ orbital centered on each Cu atom. We discuss prominent features of this LEH: (1) The ratio $U/|t_1|$ between the onsite effective Coulomb repulsion (ECR) $U$ and amplitude of nearest neighbour hopping $t_1$ increases with $T^{\rm exp}_c$ and $N_{\ell}$, consistently with the expected increase in $d$-wave SC correlation function $P_{dd}$ with $U/|t_1|$. One possible cause of the increase of $U/|t_1|$ is the replacement of apical O atoms by Cu atoms from neighbouring CuO$_2$ planes when $N_{\ell}$ increases. Furthermore, we show that the increase in distance between Cu and apical O atoms decreases the effective screening (ES) by electrons outside of the LEH and increases $U/|t_1|$. (2) For Hg1201 and Ca11, we show that $U/|t_1|$ decreases when hole doping per AB orbital $\delta$ increases, which may partly account for the disappearance of SC when $\delta$ exceeds the optimal value in experiment. (3) For $N_{\ell} \geq 2$, off-site inter-CuO$_2$ plane ECR is comparable to off-site intra-CuO$_2$ plane ECR. We discuss contributions of inter-CuO$_2$ plane ECR to both $P_{dd}$ and the stability of the SC state.
\end{abstract}

\maketitle

\section{Introduction}

Unconventional superconductivity (SC) occurs in cuprates \cite{Bednorz1986} with the transition temperature $T_c$ reaching the maximal \jbmb{value} 
$T_c^{\rm exp}\sim 138$ K at ambient pressure for HgBa$_2$Ca$_2$Cu$_3$O$_8$ \cite{Dai1995}, and even higher values ($\gtrsim 150$ K) for Hg-based compounds under pressure \cite{Nunez1993,Gao1994}.
It is empirically observed that $T_c^{\rm exp}$ has a correlation with $N_{\ell}$, the number of CuO$_2$ layers sandwiched by the neighboring block layers; in the Bi-based cuprates Bi$_2$Sr$_2$Ca$_{N_{\ell}-1}$Cu$_{N_{\ell}}$O$_{2N_{\ell}+4}$; 
$T_c^{\rm exp} < 10$ K for $N_{\ell}=1$ (Bi2201) \cite{Maeda1988,Chu1988}, up to $\sim 40$ K under optimal substitution \cite{ARAO2005351}; $T_c^{\rm exp} \sim 84$ K for $N_{\ell}=2$ (Bi2212) \cite{Hazen1988,Subramanian1988,Tarascon1988,Tallon1988}; $T_c^{\rm exp} \sim 110$ K for $N_{\ell}=3$ (Bi2223) \cite{Michel1987} and HgBa$_2$Ca$_2$Cu$_3$O$_8$ also belongs to $N_{\ell}=3$.
Namely, $T_c^{\rm exp}$ increases progressively with $N_{\ell}$ and this trend is also satisfied for carrier doped CaCuO$_2$, which is interpreted as $N_{\ell}=\infty$, and reaches
$T_c^{\rm exp}\sim 110$ K 
\cite{Azuma1992}. However, the microscopic mechanism that causes this trend is not well understood.

If appropriate LEHs that correctly describe low-energy physics could be derived systematically for compounds that have different $N_{\ell}$ and \oldmichael{the relevant material dependent parameters are revealed, it would provide hints to the origin of the difference in $T_c^{\rm exp}$ and eventually the SC mechanism in cuprates.}

Historically, single-orbital Hubbard model\oldmichael{s} with adjustable parameters of onsite \misec{ECR} $U$ and the hopping $t$ have been extensively studied to understand the cuprate superconductors. This single orbital is expected to \oldmichael{be an}
antibonding \oldmichael{orbital originating}
from strongly hybridized atomic Cu 3$d_{x^2-y^2}$ and O $2p_{\sigma}$ orbitals\jbmb{, which is} centered on each Cu atom \jbmb{and} denoted as $\rm AB$ hereafter.
However, it is not trivial whether the degrees of freedom beyond the antibonding orbital $\rm AB$ play only minor roles in understanding physics of SC.
The band structure derived from
density functional theory (DFT) suggests atomic Cu 3$d_{x^2-y^2}$, 3$d_{3z^2-r^2}$ and O $2p_{\sigma}$ orbitals\oldmichael{,} abbreviated as $x$, $z$ and $p$ orbitals hereafter, are located relatively close to the Fermi level and potentially contribute to low-energy physics. 
For instance, charge transfer energy $\Delta E_{xp}$ between atomic $x$ and $p$ orbitals smaller than the onsite \misec{ECR} 
for $x$ was claimed to lead to essential insufficiency of the single-band Hubbard-type model~\cite{PhysRevLett.55.418} including the negative correlation of $\Delta E_{xp}$ to $T_c^{\rm exp}$~\cite{Weber2012}. The positive correlation between $T_c^{\rm exp}$ and $\Delta E_{xz}$ defined between $x$ and $z$ orbitals \jbmb{was} also addressed~\cite{Sakakibara2010}.
An inapplicability of the $\rm AB$ Hubbard model to overdoped cuprates was also claimed from earlier X-ray absorption spectra~\cite{Peets2009,Peets2009erratum}.
Furthermore, theoretical studies of the Hubbard model~\jbmb{\cite{White2000,Yokoyama2004,Capone2006,Yokoyama2013,Misawa2014,Corboz2014,Otsuki2014,Zhao2017,Zheng2017,Ido2017,Ido2018,Darmawan2018}} 
showed the increasing superconducting order for larger $|U/t_1|$ in the superconducting solution of the simple Hubbard model \oldyy{at finite hole doping}, where $U$ is the onsite interaction and $t_1$ denotes the nearest neighbour hopping within a CuO$_2$ layer\oldmichael{.}
However, this  solution becomes an excited state for large $|U/t_1|$ and the true ground state is
dominated by \oldmichael{severely competing charge-ordered (CO) states}
contrary to the widely spread SC phase found in the doping concentration dependence of the experiments\cite{White2000,Misawa2014,Corboz2014,Otsuki2014,Zhao2017,Zheng2017,Ido2017,Ido2018,Darmawan2018}. 

Nonetheless, 
\oldyy{a} recent study~\cite{Ohgoe2020} of the \textit{ab initio} single-orbital $\rm AB$ Hamiltonian~\cite{Hirayama2019} for 
Hg1201 ($T_c^{\rm exp} \sim 90$ K \cite{Putilin1993}) \jbmb{derived by MACE, without adjustable parameters and beyond the simple Hubbard model,} was able to reproduce the experimental phase diagram at zero temperature, including the dominant SC phase under hole doping. 
Off-site interaction parameters beyond the Hubbard model have turned out to be crucially important: Although they reduce the long-range $d$-wave superconducting order parameter \misec{$\Delta_{\rm SC}=\sqrt{\lim_{r\rightarrow \infty}P_{dd}(r)}$ monitored by the $d$-wave Cooper pair correlation function $P_{dd}(r)$ at distance $r$}, they allow \oldmichael{the} stabilization of the SC state over the competing CO states, because
the CO is more severely damaged, so that the SC ground state is successfully predicted.
This supports that the {\it ab initio} single-orbital $\rm AB$ Hamiltonian offers \oldmichael{a} promising framework for the in-depth understanding of the SC mechanisms in the cuprates, provided that the \textit{ab initio} LEH is carefully derived beyond the Hubbard picture.
In this paper, we extend the work along this line and derive the {\it ab initio} Hamiltonians systematically for several compounds that have different $N_{\ell}$ to gain insights into the substantial dependence of $T_c^{\rm exp}$ on $N_{\ell}$.

In the early stage of the derivation of the \textit{ab initio} LEH for the cuprates within the MACE~\cite{Imada2010}, the hopping parameters (one-particle part) were at the level of the local density approximation (LDA) or generalized gradient approximation (GGA), and the effective interactions (two-particle part) were at the level of the constrained version (cRPA) of the random phase approximation (RPA)~\cite{Aryasetiawan2004,Aryasetiawan2006}\jbmt{.}{, where the screening by the electrons contained in the effective Hamiltonian is excluded and is called cRPA screening.}
At this level, the $T_c^{\rm exp}$ dependence of the LEH parameters \jbmb{was}
studied~\cite{Werner2015,Jang2016,Teranishi2018,Nilsson2019,Teranishi2021,Teranishi2021effect}, without the recent improvement of the MACE by the constrained $GW$ (c$GW$) method \cite{Hirayama2013,Hirayama2015,Hirayama2017,Hirayama2018}, self-interaction correction (SIC) \cite{Hirayama2015} and level renormalization feedback (LRFB) \cite{Hirayama2019}.
\jbmt{}{There, the partial screening nature of the cRPA screening is retained even for the screening taken into account in the c$GW$ level, 
and we call it c$GW$ screening, 
or c$GW$+LRFB screening if we start from the $GW$ electronic structure improved by the LRFB.}
Recent LDA/GGA+cRPA studies \cite{Teranishi2018,Nilsson2019,Teranishi2021} at this level reported that the high-$T_c$ SC is favored by a higher value of $|U/t_1|$. 
However, they studied the Hamiltonian without considering the off-site interaction 
and did not consider the competition with the CO. 
For instance, in Ref. \onlinecite{Teranishi2018}, the cRPA value of $U$ for TlBa$_2$CuO$_5$ (\oldyy{Tl1201,} $N_{\ell}=1$, $T_c^{\rm exp} \sim 50$ K) is $\sim 40\%$ lower than their estimate
for Hg1201 ($N_{\ell}=1$, $T_c^{\rm exp} \sim 90$ K) and Ca11 ($N_{\ell}=\infty$, $T_c^{\rm exp} \sim 110$ K). In addition, their value of $U$ for Hg1201 is $\sim 2.9$ eV, which is substantially underestimated with respect to $\sim 3.8$ eV in Ref. \onlinecite{Hirayama2018}.
It is known that an insufficient treatment of the disentanglement procedure~\cite{Miyake2009} \oldmichael{can cause the underestimate.}
\jbmb{Still, t}he enhanced SC for larger $|U/t_1|$ is consistent with the Hubbard model study mentioned above~\cite{White2000,Capone2006,Misawa2014,Corboz2014,Otsuki2014,Zhao2017,Zheng2017,Ido2017,Ido2018,Darmawan2018}.

In this paper, we apply the state-of-the-art methodology of the MACE \cite{Hirayama2019}, by using the \texttt{RESPACK} code \cite{Nakamura2020}. We use the pseudopotential (PP) and plane wave formalisms, to reduce the computational cost compared to the all-electron (AE) implementation. This allows us
to treat compounds with more atoms in the unit cell such as Bi2201 and Bi2212 even with the improved MACE scheme mentioned above~\cite{Hirayama2019}.
It enables to derive LEHs for different $N_{\ell}$ and to study the systematic dependence of the LEH parameters on 
\jbmsec{Bi2201 ($T_c^{\rm exp} \sim 10$ K, $N_{\ell}=1$), Bi2212 ($T_c^{\rm exp} \sim 84$ K, $N_{\ell}=2$), Hg1201 ($T_c^{\rm exp} \sim 90$ K, $N_{\ell}=1$) and Ca11 ($T_c^{\rm exp} \sim 110$ K, $N_{\ell}=\infty$)},
which helps reaching our present goal to understand the microscopic origin of strongly increasing $T_c$. 
Our comparison of the LEH for Hg1201 
\jbmsec{with} the literature using the all-electron implementation in Refs. \onlinecite{Hirayama2018,Hirayama2019} is useful to establish the accuracy and reliability of our PP framework.

\jbmsectwo{We separate 
the comparison of Hg1201 and Ca11 (Hg/Ca compounds) on the one hand, 
and the comparison of Bi2201 and Bi2212 (Bi compounds) on the other hand.
As explained below, it is possible to compare Hg/Ca compounds on equal footing, then Bi compounds on equal footing ;
however, the comparison of Hg/Ca with Bi compounds altogether is not reliable, due to two main complications:
}
\paragraph{\jbmsectwo{Difference in optimal hole concentration $\delta_{\rm opt}$ between Hg/Ca and Bi compounds.}}
\jbmsectwo{
A difficulty in estimating the optimal hole concentration is due to the uncertainty in the oxygen deficiency. 
By keeping in mind the uncertainty, still, the optimal hole concentration for Hg1201 was estimated 
to be between $\delta_{\rm opt}=0.10$ and 0.15 from the Seebeck coefficient \cite{Yamamoto2000}.
In the case of Ca11, we consider the experimental structure from Ref.\cite{KARPINSKI1994}, 
while the optimum concentration may have a similar value to Hg1201 by considering the oxygen deficiency \cite{Azuma1992}.
In order to 
reproduce $\delta_{\rm opt}$ in the experimental SC phase
while keeping the comparison of Hg/Ca compounds on equal footing,
we employ the hole concentration per AB orbital $\delta=0.1$ in the derivation of the LEH for Hg/Ca compounds.
For completeness, we also consider $\delta=0.0$\footnote{\jbmsec{In addition, we derive and present LEHs at $\delta=0.2$ for Hg/Ca compounds in the Supplemental Material. 
However, we stress that the latter LEHs may correspond to the overdoped regime, 
and may not be able to reproduce the SC ground state when the LEHs are solved by the low-energy solver.}}.
However, in the case of Bi compounds, the optimal value of $\delta$ in the experimental SC phase is estimated as 
$\delta_{\rm opt} \sim 0.19$ for Bi2201 and $\delta_{\rm opt} \sim 0.27$ for Bi2212 \cite{ARAO2005351,fukase1990ultrasonic,Fang1992}.
Here, $\delta_{\rm opt}$ appears to be closer to $0.2$,
so that we derive the LEH for $\delta=0.2$ for both Bi compounds}\footnote{\jbmt{}{For Bi2212, we also give the LEH for $\delta=0.1$ in the Supplemental Material.}}.
\paragraph{\jbmsectwo{\misec{Uncertainty} on atomic coordinates in Bi compounds.}}
\jbmsectwo{
In the SC phase of Bi compounds, experimental estimates of the distance $d^{z}_{\rm Oap}$ between Cu and apical O atoms along $z$ direction
vary between $\sim 2.26 - 2.60$~\AA~for Bi2201 \cite{Torrance1988,Torardi1988,Ito1998,Schlogl1993}, 
and between \misec{$\sim 2.30 - 2.50$~\AA~for} Bi2212 \cite{Torrance1988,Beskrovnyi1990,Cicco1993,Levin1994}.
Similarly, in the case of Bi2212, there is an \misec{uncertainty} on the amplitude $d^{z}_{\rm buck}$ of the buckling of Cu-O-Cu bonds,
which varies between $d^{z}_{\rm buck} \lesssim 0.07$~\AA~\cite{Beskrovnyi1990,Levin1994} and $0.27$~\AA~\cite{Torrance1988}.
This further complicates the comparison of Bi compounds with Hg/Ca compounds.
Nonetheless, our treatment of Bi compounds on equal footing
allows to investigate the effect of variations in $d^{z}_{\rm Oap}$ and $d^{z}_{\rm buck}$,
at least in the comparison within the two Bi compounds.
We show that the \misec{uncertainty} on $d^{z}_{\rm Oap}$ and $d^{z}_{\rm buck}$ 
causes significant variations in $U/|t_1|$ and the ES.
However, this does not change the trend,
that is, $U/|t_1|$ is at least $\sim 10\%$ larger for Bi2212 compared to Bi2201.
}
\\

We \jbmsectwo{first} analyze a three-orbital LEH, called the ABB LEH below, consisting of the antibonding $\rm AB$ orbital as well as two counterpart bonding orbitals denoted by $\rm B$.
By examining the ABB LEH, we \jbmsec{suggest} 
that the single-orbital $\rm AB$ Hamiltonian looks enough, by showing that the lower Hubbard band (LHB) from the $\rm AB$ manifold is located above the upper Hubbard band (UHB) from the $\rm B$ manifold and both are nearly separated.

\jbmsec{As for the AB Hamiltonian, 
we compare separately \jbmsectwo{Hg/Ca compounds} on the one hand and Bi compounds on the other hand, as explained above.
We find three main trends:
(1)  \textit{Ab initio} $U/|t_1|$ increases when $N_{\ell}$ and $T_c^{\rm exp}$ increase, 
which suggests a positive correlation between $T_c^{\rm exp}$ and $U/|t_1|$, although this correlation remains empirical since values of $T_c^{\rm exp}$ are estimated from experiment,
and it is desirable to solve the present LEHs in future studies
to clarify the difference in SC between compounds. 
Nonetheless, this positive correlation is consistent with the previous \misec{report} for the dependence of the SC order parameter $\Delta_{\rm SC}$ on $U/|t_1|$ \misec{in simple models~\cite{Ido2017}}. 
(2) Also, $U/|t_1|$ decreases when $\delta$ increases,
which \misec{partly explains} the progressive disappearance of SC when $\delta$ exceeds $\delta_{\rm opt}$.
(3) In addition, for $N_{\ell} \geq 2$, off-site interactions between electrons at neighboring CuO$_2$ layers are comparable to that within a CuO$_2$ layer. 
We discuss contributions of these inter-CuO$_2$ layer interactions to both $\Delta_{\rm SC}$ and the stability of the SC state:
A possible scenario is that inter-CuO$_2$ layer interactions contribute to destabilize charge-ordered states which compete with the SC state.
The present quantitative estimates will allow to investigate the latter dependence, as well as detailed clarification about the severe competition with the charge ordering/phase separation when they are solved by an accurate solver.
}
\\

This paper is organized as follows:
\jbmsec{Sec.}~\ref{sec:meth} describes our method and computational details.
The first part of the LEH derivation starting from the Kohn-Sham \jbmsec{(KS) level with LDA or GGA} and improving it to the $GW$ level supplemented with LRFB correction~\cite{Hirayama2019} is outlined in Sec.~\ref{sec:prep}.
In Sec.~\ref{sec:1band}, we start from the $GW$+LRFB electronic structure to derive $\rm AB$ LEH
\jbmsec{for \jbmsectwo{Hg/Ca compounds} in Sec.~\ref{sec:1bandhgca}, and for \jbmsectwo{Bi compounds} in Sec.~\ref{sec:1bandbi}.}
In Sec.~\ref{sec:disc}, we analyze material dependence of the derived LEH parameters, 
and their effect on $\Delta_{\rm SC}$ and stability of SC state.
\jbmsec{In Appendix~\ref{app:mace}, we give a reminder of the methodology \cite{Hirayama2018, Hirayama2019} outlined in Sec.~\ref{sec:meth}, give details about intermediate steps of the derivation of the single-orbital $\rm AB$ Hamiltonian, and benchmark our results with respect to the all-electron implementation \cite{Hirayama2018,Hirayama2019}.
}
\jbmsec{
In Appendix \ref{app:doping}, we discuss in detail the effect of hole doping on the electronic structure, and decrease in $U/|t_1|$ when hole doping increases.
In Appendix \ref{app:bi}, we 
discuss in detail the crystal structure of Bi compounds, including the \misec{uncertainty} on $d^{z}_{\rm Oap}$ and $d^{z}_{\rm buck}$, 
and estimate the variation in $U/|t_1|$ with $d^{z}_{\rm Oap}$ and $d^{z}_{\rm buck}$ for Bi compounds.
}
\jbmsec{In Appendix~\ref{app:abb}, we discuss} the validity of the single-orbital $\rm AB$ Hamiltonian derived in Sec.~\ref{sec:1band}.
\jbmsec{In Appendix~\ref{app:ode}, we propose} an approximation (not used in this paper, but useful for future studies) to reduce the computational cost of the MACE for compounds with large number of bands as the cuprates with $N_{\ell} \geq 2$, without loss of accuracy.

\section{Methods and computational details}
\label{sec:meth}

\begin{figure}[!htb]
\includegraphics[scale=0.36]{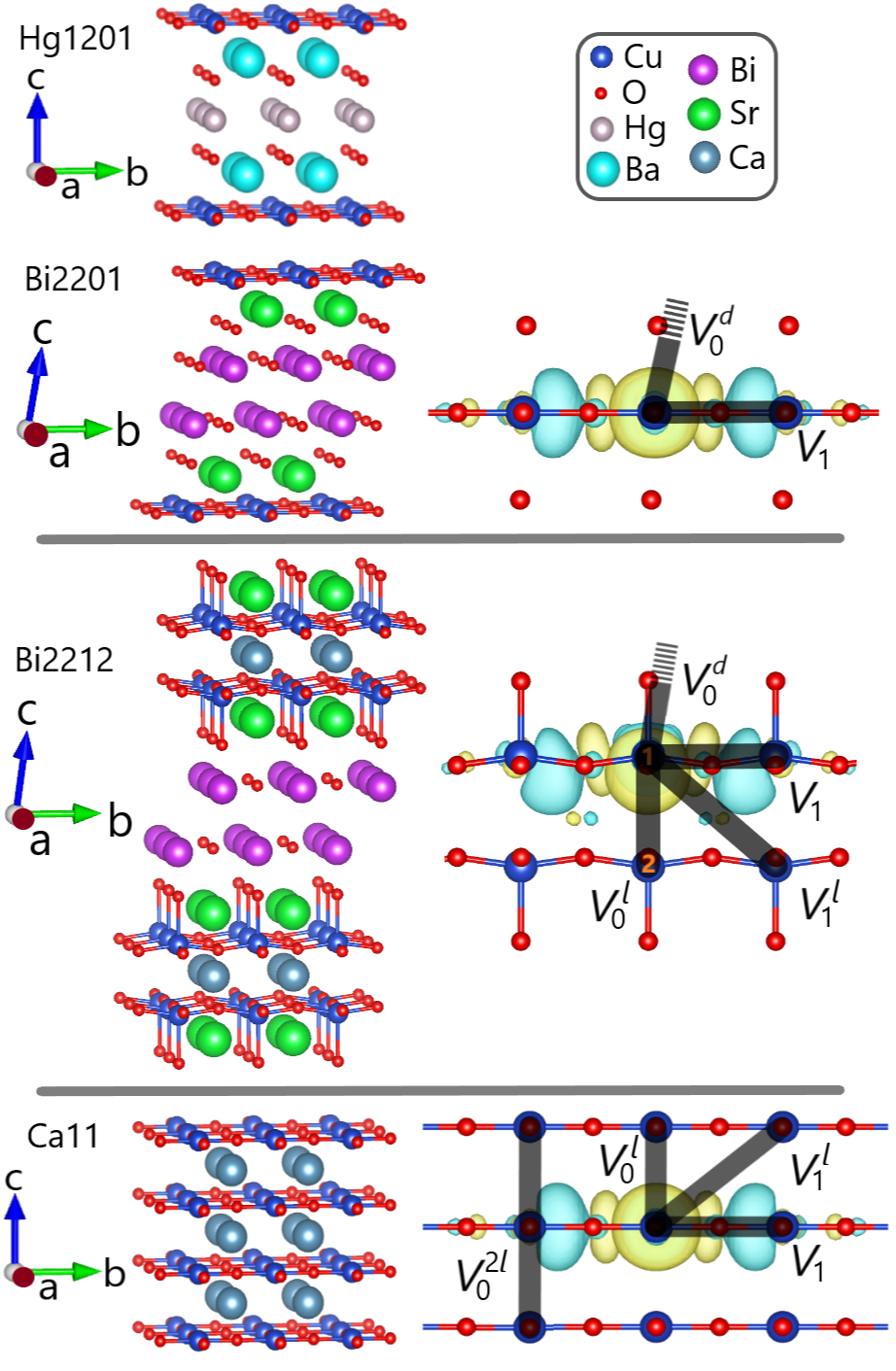}
\caption{
Left panel: 
Crystal structures of Hg1201, Bi2201, Bi2212 and Ca11, and directions of primitive vectors (${\bf a}$,${\bf b}$,${\bf c}$).
\jbmb{Colored balls used for each atomic species are defined in the upper right inset.}
Right panel:
Simplified representation of CuO$_2$ layer structures for Bi2201, Bi2212 and Ca11, with only Cu atoms, in-layer and apical O atoms.
We also show schematic representations of a few intralayer ($V_{n}$), interlayer ($V^{l}_{n}$), second interlayer ($V^{2l}_{n}$) and distant interlayer ($V^{d}_{n}$) interaction \jbm{parameter}s by thick black bonds, and isosurfaces of $\rm AB$ \jbmsec{maximally localized} Wannier orbitals within the $\rm AB$ Hamiltonian
\jbmt{}{of Hg1201 and Ca11 at hole doping per AB orbital $\delta=0.0$, and of Bi2201 and Bi2212 at $\delta=0.2$.}
(absolute value is 0.01, yellow is positive, blue is negative).
The simplified representation for Hg1201 is not shown here.
\jbmb{For Bi2212, we show only one of the two AB orbitals.}
}
\label{fig:crystal}
\end{figure}

Effective LEHs in the present paper have the form
\begin{align}
& \hat{\mathcal{H}}  = \sum_{(i,{\bf R},\sigma)}\sum_{(j,{\bf R'},\sigma')} t_{ij}^{\sigma \sigma'}(\jbmsectwo{{\bf R'-R}}) c^{\dagger}_{i\sigma{\bf R}} c^{}_{j\sigma'{\bf R'}} \nonumber \\
& +  \jbmt{}{\frac{1}{2}}\sum_{(i,{\bf R},\sigma)}\sum_{(j,{\bf R'},\sigma')} U_{ij}^{\sigma \sigma'}(\jbmsectwo{{\bf R'-R}}) c^{\dagger}_{i\sigma{\bf R}} c^{\dagger}_{j\sigma'{\bf R'}} c_{j\sigma'{\bf R'}} c_{i\sigma{\bf R}}, \nonumber 
\end{align}
where ${\bf R}$ is the coordinate of the unit cell in the space [abc] expanded in the (${\bf a},{\bf b},{\bf c}$) frame in Fig. \ref{fig:crystal}. The indices $i,j$ denote the orbitals within the unit cell, and $\sigma,\sigma'$ denote the spin indices.
By using these notations, $c^{\dagger}_{i\sigma{\bf R}}$ and $c^{}_{i\sigma{\bf R}}$ are respectively the creation and annihilation operators in the spin-orbital coordinate ($i,\sigma$) at ${\bf R}$, and $t_{ij}^{\sigma \sigma'}(\jbmsectwo{{\bf R'-R}})$ and $U_{ij}^{\sigma \sigma'}(\jbmsectwo{{\bf R'-R}})$ are respectively the hopping and direct interaction \jbm{parameter}s between spin-orbitals ($i,\sigma$) at ${\bf R}$ and ($j,\sigma'$) at ${\bf R'}$, which satisfy translational symmetry \jbmsec{so that we may restrict the calculation to $t_{ij}^{\sigma \sigma'}(\jbmsectwo{{\bf R}})$ and $U_{ij}^{\sigma \sigma'}(\jbmsectwo{{\bf R}})$.}
If ${\bf R}={\bf R}_0=[000]$ and $i=j$, we abbreviate $U_{ii}^{\sigma\neq\sigma'}({\bf R}_0)$ as the onsite interaction $U$ for the $\rm AB$ Hamiltonian, and $U_i$ for multi-orbital Hamiltonians.

In the case of the $\rm AB$ Hamiltonian, there is only one $\rm AB$ orbital per unit cell for Hg1201, Bi2201 and Ca11 (\oldyy{$i=j=1$}), but two for Bi2212 ($i,j=1,2$, as in Fig. \ref{fig:crystal}).
\oldyy{For a comparison of energy scales in the series of the cuprates,}
we separate $U_{ij}^{}({\bf R})$ into four different categories of \jbm{parameter}s\jbm{, denoted as} onsite, intralayer, interlayer (\oldyy{``$l$"}) and distant interlayer (\oldyy{``$d$"}) \jbm{parameter}s.
A few \oldyy{typical} interaction \jbm{parameter}s are represented schematically in Fig. \ref{fig:crystal}.
Intralayer, off-site interaction \jbm{parameter}s are $V_{n}=U_{11}^{}({\bf R}_n)$, where ${\bf R}_n$ gives the position of the $n^{\rm th}$ nearest neighbour orbital within the CuO$_2$ layer (we have ${\bf R}_1=[100]$, ${\bf R}_2=[110]$, ${\bf R}_3=[200]$, ${\bf R}_4=[210]$, ${\bf R}_5=[220]$ and ${\bf R}_6=[300]$).
Interlayer interaction \jbm{parameter}s are defined only for Bi2212 and Ca11, as $V^{l}_{n}=U_{12}^{}({\bf R}_n)$ for Bi2212 and $V^{l}_{n}=U_{11}^{}({\bf R}_n + {\bf c})$ for Ca11\jbmb{, where  ${\bf c}=[001]$ is defined as \jbmb{in} Fig. \ref{fig:crystal}.}
Distant interlayer \jbm{parameter}s for Hg1201, Bi2201 and Bi2212 are defined as those between different CuO$_2$ layers separated by a block layer\jbm{, that is,} 
$V^{d}_{n}=U_{11}^{}({\bf R}_n + {\bf c})$ for Hg1201 and Bi2201 and $V^{d}_{n,ij}=U_{ij}^{}({\bf R}_n + {\bf c})$ for Bi2212.
For Ca11, there are no block layers; instead of $V^{d}_{n}$, \jbmb{we define the second interlayer \jbm{parameter}s as}
$V^{2l}_{n}=U_{11}^{}({\bf R}_n + 2{\bf c})$.
The one-particle part $t_{ij}^{}({\bf R})$ is classified into intralayer ($t_n$), interlayer ($t^{l}_n$) and distant interlayer ($t^{d}_n$) or second interlayer ($t^{2l}_{n}$) hopping \jbm{parameter}s, which are defined similarly.
\jbmsec{Regarding interaction parameters other than $U_{ij}^{\sigma \sigma'}(\jbmsectwo{{\bf R'-R}})$ (including Hund, exchange and pair hopping parameters, and parameters beyond two-body interaction), they are assumed to play minor roles and ignored\jbmt{ ; however,}{. For the sake of completeness, we give the Hund interaction parameters in the Supplemental Material. However, in the AB Hamiltonian, the amplitude of the intersite Hund interaction does not exceed $\sim 0.04$ eV for all compounds, that is, $4\%$ of the direct interaction parameter.
On the other hand,
} the superexchange energy $J \sim 4|t_1|^2/U$ between neighbouring AB orbitals is not negligible, and we will discuss it as well.}

Distant interlayer \jbm{parameter}s are usually neglected in the $\rm AB$ Hamiltonian, even within the most recent MACE methodology~\cite{Hirayama2018,Hirayama2019}. 
Neglecting $t^{d}_n$ \jbm{parameter}s is justified by their small amplitude ($\lesssim 0.01$ eV for \jbmsectwo{Bi compounds}).
Although $V^{d}_{n}$ \jbm{parameter}s are small ($V^{d}_{0}/U \lesssim 8\%$ in this paper), they are not always practically negligible.
\oldyy{It is worth noting that the distant interlayer Coulomb interactions lead to the screening of the intralayer interactions due to the dielectric or metallic responses from the distant layers.}
It may be possible to include the screening effect from layers outside the 2D Hamiltonian 
by using the dimensional downfolding procedure \cite{Nakamura2010}, which we do not consider here.
In the following, we will focus on intralayer and interlayer parameters \jbmsec{; other parameters such as $V^{d}_{n}$ and $V^{2l}_{n}$ may be found in the Supplemental Material.}

We compute the effective parameters $t_{ij}^{}({\bf R})$ and $U_{ij}^{}({\bf R})$ by using our new implementation of the MACE scheme~\cite{Aryasetiawan2004,Imada2010,Hirayama2013,Hirayama2018,Hirayama2019} within the \texttt{RESPACK} code~\cite{Nakamura2020}.
\jbmsec{We quickly summarize the scheme here ; details are reminded in Appendix \ref{app:mace}.}
We first compute the electronic structure at the \jbmsec{KS} level.
Then, the medium-energy (M) space, composed of $3d$-like bands from Cu and $2p$-like bands from O, is improved at the $GW$+LRFB level, by following the methodology in Ref. \onlinecite{Hirayama2019}; other bands are left at the KS level. 
We then use the resulting $GW$+LRFB electronic structure as a starting point to derive the $\rm AB$ Hamiltonian at cRPA and c$GW$ levels.
The cRPA allows to remove the double counting in the screening in the two-particle part, whereas the c$GW$ allows to remove the exchange and correlation double counting term in the one-particle part.

\begin{figure*}[!htb]
\includegraphics[scale=1.0]{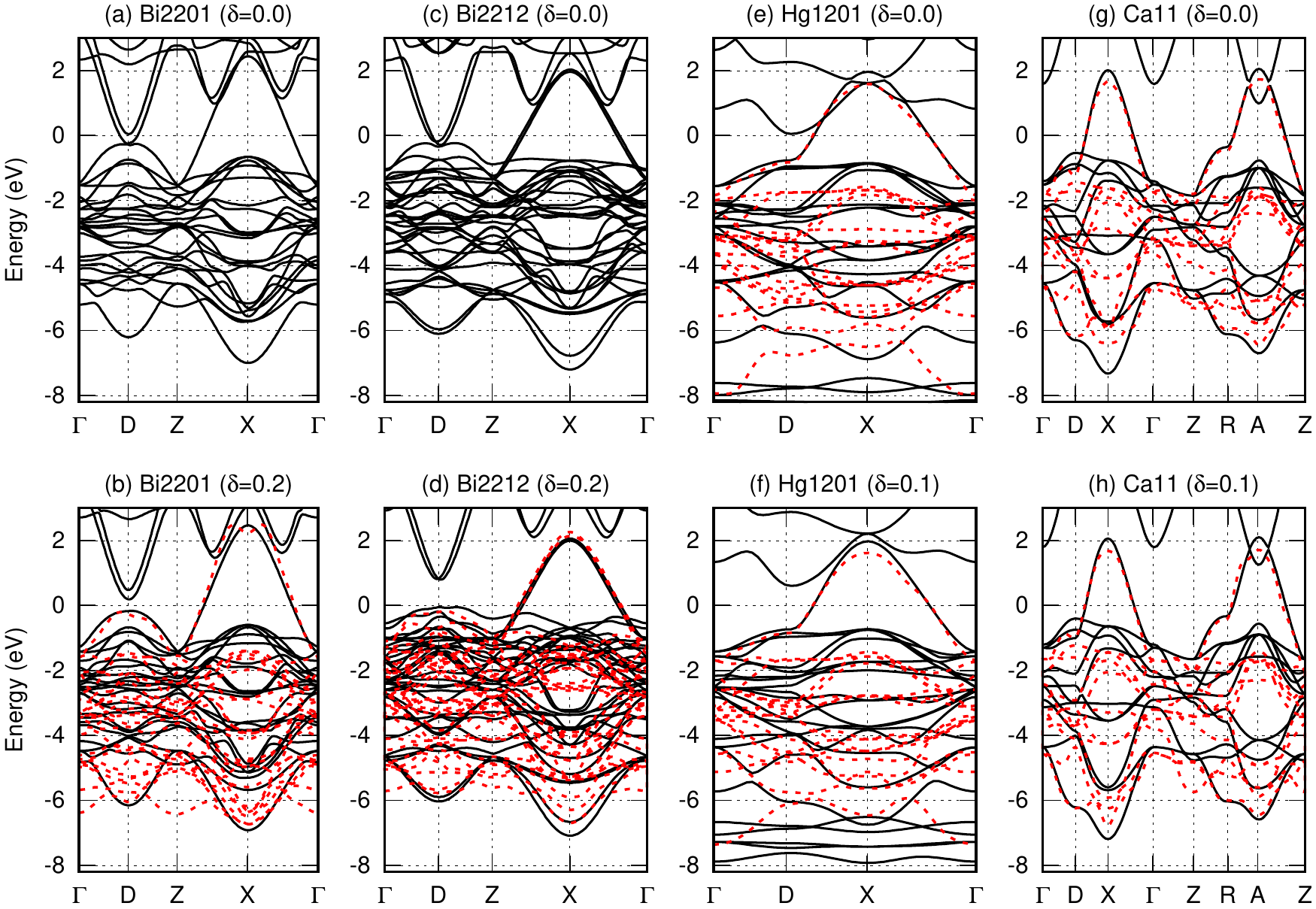}
\caption{
\jbmsec{Band structures of 
\jbmsectwo{Bi compounds} at hole doping per AB orbital \misec{$\delta=0.0$, (a), (c)} and $\delta=0.2$, \misec{(b), (d)}, and \jbmsectwo{Hg/Ca compounds} at $\delta=0.0$, (e), (g) and $\delta=0.1$ (f), (h),
at the KS level (all bands, in solid black \misec{curves}) and $GW$+LRFB level (improvement of bands within M space, in dashed red \misec{curves}).
For \jbmsectwo{Bi compounds} \misec{in (a) and (c)}, we show the intermediary KS band structures at $\delta=0.0$ \jbmt{\misec{discussed in Appendix A}}{for completeness, although we do not use them as a starting point in the derivation of the LEH.}
}
High-symmetry points are defined in Cartesian coordinates, in units of $2\pi/|{\bf a}|$, $2\pi/|{\bf b}|$ and $2\pi/|{\bf c}|$, as
$\Gamma = [0.0 \ 0.0 \ 0.0]$, D $=[0.5 \ 0.0 \ 0.0]$, Z $=[1.0 \ 0.0 \ 0.0]$ and X $=[0.5 \ 0.5 \ 0.0]$. 
For Ca11, we redefine Z $=[0.0 \ 0.0 \ 0.5]$, and we define R $=[0.5 \ 0.0 \ 0.5]$ and A $=[0.5 \ 0.5 \ 0.5]$. 
}
\label{fig:band}
\end{figure*}

Now, we give computational details. 
Structural data is taken from 
\jbmsec{Ref.~\onlinecite{Torardi1988} for Bi2201,
Ref.~\onlinecite{Torrance1988} for Bi2212,
Ref.~\onlinecite{Putilin1993} for Hg1201 and
Ref.~\onlinecite{KARPINSKI1994} for Ca11.}
DFT calculations are done with {\sc Quantum ESPRESSO} \cite{QE-2009,QE-2017} and optimized norm-conserving Vanderbilt pseudopotentials (PPs)\footnote{
We used the PPs X\_ONCV\_PBE-1.0.upf (X = Bi, Sr, Ca, Hg, Ba, Cu\jbmsec{, O, Pb, Au and K}) from the http://www.quantum-espresso.org
distribution.} \cite{Schlipf2015} 
using the GGA-PBE functional \cite{Perdew1996}.
For Ca11 and Hg1201, we regenerate the PPs by using the \texttt{ONCVPSP} code \cite{Hamann2013}, and switch the functional to the Perdew-Zunger LDA \cite{Perdew1981} to compare with the calculations using the all-electron implementation \cite{Hirayama2018,Hirayama2019}, for which the LDA was used. Nonetheless, we have checked that using either the LDA or GGA has little influence on the KS electronic structure.
\jbmsec{Hole doping is simulated as follows:
We use the implementation of the virtual crystal approximation (VCA) \cite{nordheim1931electron} in {\sc Quantum ESPRESSO} \cite{QE-2009,QE-2017}. 
As for Bi compounds, we simulate hole doping by interpolating the PPs for Bi and Pb, with proportions of $1-\delta_{\rm PP}$ for Bi and $\delta_{\rm PP}$ for Pb.
This corresponds to the experimental substitution of Bi by Pb in the SC phase \cite{Amano2004,Hobou2009}.
Similarly, we interpolate the PPs for Hg and Au in the case of Hg1201, and Ca and K in the case of Ca11, with proportions of $1-\delta_{\rm PP}$ for Hg or Ca and $\delta_{\rm PP}$ for Au or K.
Thus, $\delta_{\rm PP}$ corresponds to the hole doping per atom of dopant (Bi, Hg or Ca) in the unit cell.
This yields the total hole doping in the unit cell $\delta_{\rm tot}=2\delta_{\rm PP}$ for Bi compounds (since there are two Bi atoms in the unit cell), and $\delta_{\rm tot}=\delta_{\rm PP}$ for \jbmsectwo{Hg/Ca compounds} (since there is only one Hg or Ca atom in the unit cell).
We deduce the hole doping per AB orbital $\delta=0.5\delta_{\rm tot}$ for Bi2212 (in which there are two Cu atoms per unit cell) and $\delta=\delta_{\rm tot}$ for other compounds (in which there is only one Cu atom per unit cell).
}
The full Brillouin zone is sampled with a $12 \times 12 \times 12$ $k$-point grid at the DFT level,
reduced to $6 \times 6 \times 3$ \oldyy{($6 \times 6 \times 6$)} in $GW$ and c$GW$ calculations for Hg1201, Bi2201 and Bi2212 \oldyy{(Ca11)}.
We consider 100 bands for Ca11 (from $\sim -4.1$ Ha to $\sim + 2.7$ Ha with respect to the Fermi level), 250 bands for Bi2201 ($-4.1/+2.0$ Ha), 340 bands for Bi2212 ($-4.1/+2.1$ Ha) and 190 bands for Hg1201 ($-4.3/+1.8$ Ha).
We use a Fermi-Dirac smearing of $0.0272$ eV in the calculation of occupation numbers.
The KS exchange-correlation potential is extracted from the KS electronic structure by sampling the unit cell with a grid of size $120 \times 120 \times 450$ for Bi2201, $120 \times 120 \times 5\jbmsectwo{4}0$ for Bi2212, $150 \times 150 \times 150$ for Ca11 and $120 \times 120 \times 4\jbmsectwo{5}0$ for Hg1201. 
The plane wave cutoff energy is $100$ Ry for wavefunctions and $8$ Ry for polarization.
We compute the RPA and cRPA polarizations for $100$ real frequencies and $30$ imaginary frequencies, by considering the exponential grid in Ref. \onlinecite{Nohara2009} on both real and imaginary axes; the modulus of frequency has the maximum value $\sim 18.5$ Ha for Bi2201, $\sim 18.8$ Ha for Bi2212, $\sim 20.5$ Ha for Ca11 and $\sim 18.5$ Ha for Hg1201.
the $GW$ and c$GW$ self-energies are calculated by using the contour deformation technique \cite{Godby1988,Giantomassi2011}.

\section{Preprocessing of initial electronic structure at KS level to proceed to $GW$+LRFB level}
\label{sec:prep}

Here, we discuss the preprocessing of the starting electronic structure
\jbmsectwo{before the improvement of the M space from the KS level to the $GW$+LRFB level.}
\\

\jbmsec{
Fig. \ref{fig:band} shows the KS (GGA) band structures for \misec{Bi2201, Bi2212, Hg1201 and Ca11.}
}
\jbmsectwo{
We \misec{employ the doping concentration in the experimental SC phase close to the optimal value to derive the LEH} 
as discussed in the Introduction.}

\misec{The M space is composed of $23$ bands for Bi2201 and $34$ bands for Bi2212.}
\jbmsectwo{
\jbmt{We}{For Bi2201, we} consider $\delta_{\rm PP}=0.1$ in order to obtain \jbmt{$\delta=0.2$}{$\delta=\delta_{\rm tot}=0.2$},
in agreement with the experimental optimal value $\delta_{\rm opt} \sim 0.19$ for substituted Bi2201 \cite{ARAO2005351}.
}
\jbmsec{
\jbmt{In}{For Bi2212, we consider $\delta_{\rm PP}=0.2$ in order to obtain $\delta_{\rm tot}=0.4$ and $\delta=0.2$, 
so that $\delta=0.2$ is the same for Bi2212 and Bi2201. Indeed, in} order to compare the two compounds reliably \jbmt{}{and on equal footing}, the parameter which should be identical is the hole doping per AB orbital $\delta$, and not the total hole doping in the unit cell $\delta_{\rm tot}$.
Also, the choice of $\delta=0.2$ for Bi2212 is reasonable since $\delta_{\rm opt} \sim 0.27$ \cite{fukase1990ultrasonic,Fang1992}\jbmt{, and 
we keep the same value $\delta=0.2$ as for Bi2201 in order to compare both compounds on equal footing.}{. Thus, considering $\delta=0.2$ for both compounds is in correct agreement with the optimally doped experimental SC phase, and allows to compare both compounds on equal footing.}
}

\jbmsec{In the case of \jbmsectwo{Hg/Ca compounds}, 
the KS (LDA) band structures for $\delta=0.0$ are shown in panels (e) and (g) of Fig.~\ref{fig:band}.
The M space is composed of 17 and 11 bands, respectively. 
The crystal structures \cite{Putilin1993,KARPINSKI1994} correspond to the experimental high-$T_c^{\rm exp}$ SC phases.
\misec{We consider $\delta=0.1$ in addition to $\delta=0.0$, as discussed in the Introduction.
Resulting band structures are shown in panels (f) and (h) of Fig.~\ref{fig:band}.}
The choice of  $\delta \leq 0.1$ is justified for both compounds as we mentioned in Introduction.
}
\\

Then, we preprocess the M space at the $GW$+LRFB level, following the methodology in Ref. \onlinecite{Hirayama2019}.
\jbmsec{Details are given in Appendix \ref{app:mace}, along with a reminder of the MACE scheme ; here, we briefly outline the scheme.
The M space is preprocessed at the $GW$ level:
We calculate the frequency-dependent $GW$ self-energy, from which we construct the $GW$ electronic structure 
(which includes M bands preprocessed at the quasiparticle $GW$ level, and bands outside the M space left at the KS level).
}
We next improve the $GW$ electronic structure at the $GW$+LRFB level as follows.
We restart from the $GW$ electronic structure and derive a three-orbital LEH ($xp$ Hamiltonian) at the cRPA level and c$GW$ level with the SIC (c$GW$\jbmt{}{$-$}SIC) \cite{Hirayama2015}. 
We solve this $xp$ Hamiltonian with the \texttt{mVMC} code \cite{Misawa2014,Tahara2008,MISAWA2019447} to deduce the LRFB correction $\Delta\mu$ of the charge transfer energy $\Delta E_{xp}$ between $x$ and $p$ \jbmsec{orbitals, from the exchange splitting effect in the antiferromagnetic phase of the mother compound.
We then combine $\Delta \mu$ with the previously calculated $GW$ self-energy to obtain the $GW$+LRFB \cite{Hirayama2019} electronic structure (which includes M bands preprocessed at the quasiparticle $GW$+LRFB level, and bands outside the M space left at the KS level).
The corresponding band structures for the M space are shown in Fig.~\ref{fig:band}.
}
We note that, for Hg1201 \jbmsec{at $\delta=0.0$}, the $GW$+LRFB band structure \jbmsec{in Fig.~\ref{fig:band}(f)} is in good agreement with Fig.~7 in Ref. \onlinecite{Hirayama2019}.

\section{Single-orbital $\rm AB$ Hamiltonian}
\label{sec:1band}

We now proceed to the main process to derive the $\rm AB$ Hamiltonian for all four aforementioned compounds.
\jbmsec{
As discussed in Sec.~\ref{sec:prep}, we will compare separately 
(1) \jbmsectwo{Hg/Ca compounds} for $\delta=0.0$ and $\delta=0.1$ in Sec.~\ref{sec:1bandhgca} and
(2) \jbmsectwo{Bi compounds} for $\delta=0.2$ in Sec.~\ref{sec:1bandbi}.
Here, we discuss the differences between AB Hamiltonians and their origin ;
implications of the differences in AB Hamiltonians regarding the difference in SC between compounds will be discussed later in Sec.~\ref{sec:disc}.
}
\jbmsec{
We start from the $GW$+LRFB electronic structure, 
and compute the AB \jbmsectwo{maximally localized Wannier (MLW)} orbitals as described in Appendix~\ref{app:benchmark}.
The outer window consists in the M space, from which we exclude the 4, 7, 10 and 2 lowest bands for Hg1201, Bi2201, Bi2212 and Ca11, respectively.
Then, we compute the two-particle part at cRPA level and one-particle part at c$GW$ level,
which yields the final effective Hamiltonian at the c$GW$+LRFB(AB) level, 
as described in Appendix~\ref{app:benchmark}.
}
\\

\paragraph*{Validity of the $\jbmb{\rm AB}$ Hamiltonian ---}
Before showing results for the $\jbmb{\rm AB}$ Hamiltonian, we
\jbmsec{discuss the restriction to the single-orbital picture.}
In Appendix~\ref{app:abb}, we extend the $\jbmb{\rm AB}$ Hamiltonian to the $\jbmb{\rm AB}\jbmb{\rm B}$ Hamiltonian, which includes the $\jbmb{\rm AB}$ orbital plus two \jbmsec{bonding ($\jbmb{\rm B}$)} orbitals.
\jbmb{We analyze the competition between (i) the average onsite Coulomb repulsion on $\jbmb{\rm AB}$ and $\jbmb{\rm B}$ manifolds, and (ii) the charge transfer energy between $\jbmb{\rm AB}$ and $\jbmb{\rm B}$ manifolds.
We show that (i) is weaker than or barely equal to (ii).
As a consequence, the upper Hubbard band from the $\jbmb{\rm B}$ manifold remains well below the Fermi level, and
it is nearly separated from the lower Hubbard band from $\jbmb{\rm AB}$ orbital.
\jbmsec{This suggests the $\rm B$ manifold does not essentially contribute to low-energy physics, and it is reasonable to restrict the LEH to the $\rm AB$ orbital.}
}

\subsection{\jbmsec{Comparison of \jbmsectwo{Hg/Ca compounds}}}
\label{sec:1bandhgca}

\begin{figure}[!htb]
\includegraphics[scale=0.75]{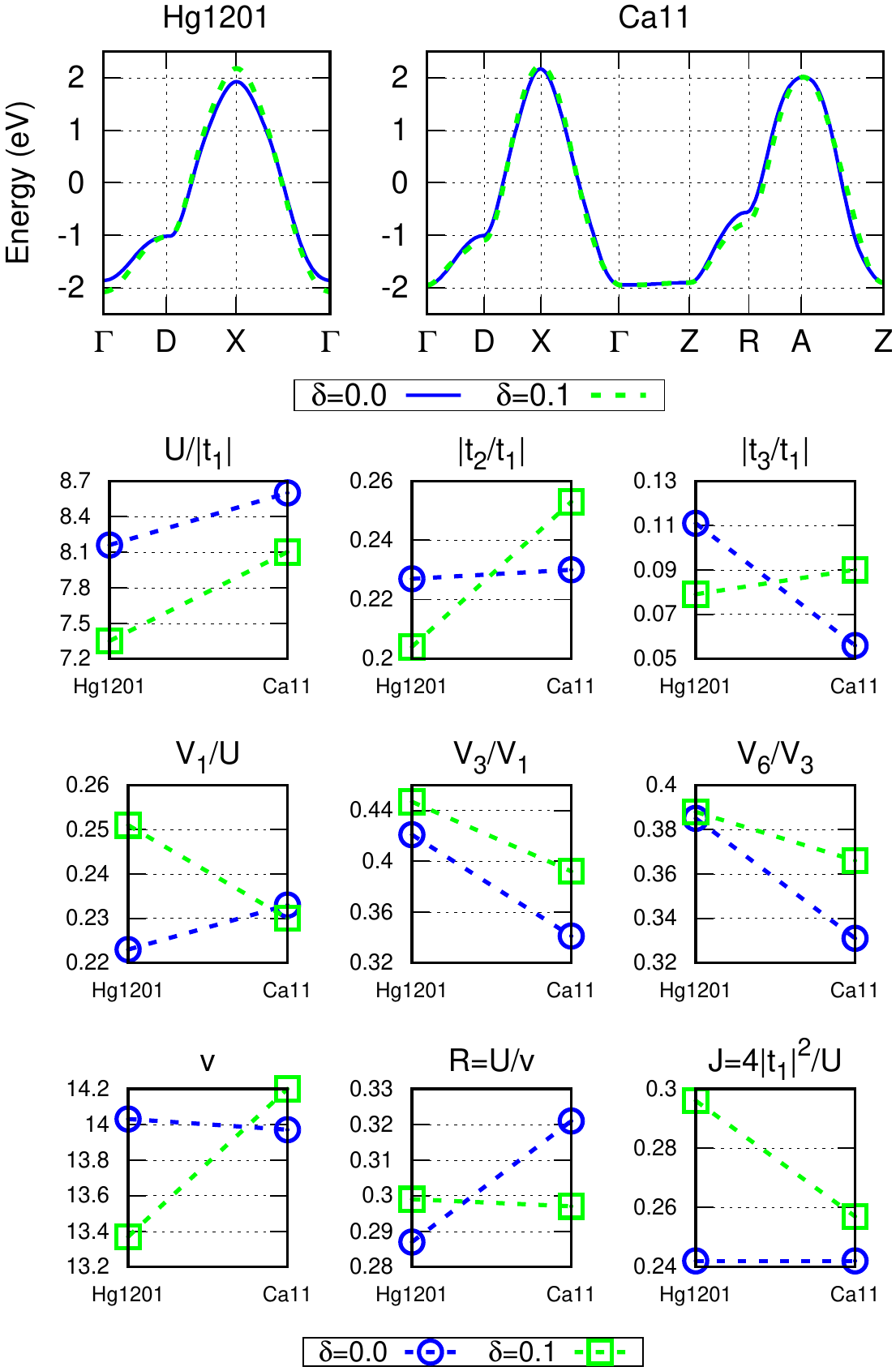}
\caption{
\jbmsec{
Upper panel : Band structure of the AB Hamiltonian at the c$GW$+LRFB(AB) level, corresponding to the one-particle part in Table \ref{tab:1nbandhgca}.
High-symmetry points are defined as in Fig. \ref{fig:band}.
Lower panel : important quantities and ratios between effective parameters in Table \ref{tab:1nbandhgca}.
}
}
\label{fig:charthgca}
\end{figure}

\jbmsec{
First, we discuss the comparison of \jbmsectwo{Hg/Ca compounds}.
Results for the AB Hamiltonian are summarized in Table \ref{tab:1nbandhgca} and Fig. \ref{fig:charthgca}. 
Table \ref{tab:1nbandhgca} shows the irreducible effective parameters ; the complete list is given in Supplemental Material.
Fig. \ref{fig:charthgca} shows band structures for the one-particle part, as well as important quantities and ratios between effective parameters
\jbmsec{In the case of Hg1201, we benchmark our result with respect to the all-electron implementation \cite{Hirayama2019} in Appendix  \ref{app:benchmark}.}
}
\\

\jbmsec{
The overall trends are summarized as follows.
(1) At $\delta=0.0$, $U/|t_1|$ increases from Hg1201 to Ca11, concomitantly with the increase in $N_{\ell}$ and $T^{\rm exp}_c$.
(2) At $\delta=0.1$, (2a) $U/|t_1|$ decreases with respect to $\delta=0.0$ for both compounds.
However, (2b) $U/|t_1|$ is still larger for Ca11 with respect to Hg1201.
(3) The decay of intralayer interactions with distance 
becomes faster from Hg1201 to Ca11.
(4) In addition, non-negligible interlayer effective parameters appear for Ca11. 
}

\begin{table}[!htb]
\centering
\resizebox{\columnwidth}{!}{
\begin{tabular}{ccccccccc}
\hline
\multicolumn{9}{c}{\jbmsec{No hole doping ($\delta=0.0$)}}\\
\hline
& \jbmsec{$U/|t^{}_{1}|$} & $U^{}_{}$ & $V^{}_{1}$ & $V^{}_{2}$ & $V^{}_{3}$ & $V^{}_{4}$ & $V^{}_{5}$ & $V^{}_{6}$ \\
Hg1201  &  \jbmsec{8.16} & 4.029 & 0.900 & 0.520 & 0.379 & 0.329 & 0.265 & 0.146 \\
Ca11  &  \jbmsec{8.60} & 4.482 & 1.044 & 0.554 & 0.356 & 0.288 & 0.201 & 0.118 \\
& & $V^{l}_{0}$ & $V^{l}_{1}$ & $V^{l}_{2}$ & $V^{l}_{3}$ & $V^{l}_{4}$ & $V^{l}_{5}$ & $V^{l}_{6}$ \\
Ca11  &  & 0.749 & 0.521 & 0.396 & 0.289 & 0.247 & 0.184 & 0.108 \\
& & $t^{l}_{0}$ & $t^{}_{1}$ & $t^{}_{2}$ & $t^{}_{3}$ & $t^{}_{4}$ & $t^{}_{5}$ & $t^{}_{6}$\\
Hg1201 &  & --- & -0.494 & 0.112 & -0.055 & 0.018 & 0.002 & -0.002 \\
Ca11  &  & -0.059 & -0.521 & 0.120 & -0.029 & 0.008 & -0.007 & -0.001 \\
\hline
\multicolumn{9}{c}{\jbmsec{Hole doping per AB orbital $\delta=0.1$}}\\
\hline
& \jbmsec{$U/|t^{}_{1}|$} & \jbmsec{$U^{}_{}$} & \jbmsec{$V^{}_{1}$} & \jbmsec{$V^{}_{2}$} & \jbmsec{$V^{}_{3}$} & \jbmsec{$V^{}_{4}$} & \jbmsec{$V^{}_{5}$} & \jbmsec{$V^{}_{6}$} \\
\jbmsec{Hg1201}  & \jbmsec{7.35} & \jbmsec{3.999} & \jbmsec{1.002} & \jbmsec{0.596}  &  \jbmsec{0.448} & \jbmsec{0.389} & \jbmsec{0.320} & \jbmsec{0.174} \\
\jbmsec{Ca11}  & \jbmsec{8.10} & \jbmsec{4.221} & \jbmsec{0.969} & \jbmsec{0.539} & \jbmsec{0.380} & \jbmsec{0.316} & \jbmsec{0.241} & \jbmsec{0.139} \\
& & \jbmsec{$V^{l}_{0}$} & \jbmsec{$V^{l}_{1}$} & \jbmsec{$V^{l}_{2}$} & \jbmsec{$V^{l}_{3}$} & \jbmsec{$V^{l}_{4}$} & \jbmsec{$V^{l}_{5}$} & \jbmsec{$V^{l}_{6}$} \\
\jbmsec{Ca11}  & & \jbmsec{0.739} & \jbmsec{0.530} & \jbmsec{0.414} & \jbmsec{0.322} & \jbmsec{0.283} & \jbmsec{0.227} & \jbmsec{0.129}  \\
& & \jbmsec{$t^{l}_{0}$} & \jbmsec{$t^{}_{1}$} & \jbmsec{$t^{}_{2}$} & \jbmsec{$t^{}_{3}$} & \jbmsec{$t^{}_{4}$} & \jbmsec{$t^{}_{5}$} & \jbmsec{$t^{}_{6}$}\\
\jbmsec{Hg1201}  & & \jbmsec{---} & \jbmsec{-0.544} & \jbmsec{0.111} & \jbmsec{-0.043} & \jbmsec{0.010} & \jbmsec{0.000} & \jbmsec{-0.004}  \\
\jbmsec{Ca11}  & & \jbmsec{-0.053} & \jbmsec{-0.521} & \jbmsec{0.132} & \jbmsec{-0.047} & \jbmsec{0.008} & \jbmsec{0.000} & \jbmsec{-0.014} \\
\hline
\end{tabular}
}
\caption{
\jbmsec{
Effective Hamiltonian parameters (in eV) for the AB Hamiltonians of \jbmsectwo{Hg/Ca compounds} at $\delta=0.0$ \jbmsec{and $\delta=0.1$}. The complete list of parameters is given in Supplemental Material.
}
\jbmsec{We also give the values of $U/|t^{}_{1}|$.}
}
\label{tab:1nbandhgca}
\end{table}

\jbmsec{
(1) First, we discuss the increase in $U/|t_1|$ from Hg1201 to Ca11 at $\delta=0.0$.
The value of $U/|t_1|$ increases by $\sim 5\%$ from Hg1201 to Ca11. 
This increase is not caused by the variation in $|t_1|$, which increases by $\sim 5\%$ from Hg1201 to Ca11.
Instead, it is caused by the variation in $U$, which increases by $\sim 11\%$ from Hg1201 to Ca11.
The value of $U$ is controlled by two parameters: The bare interaction $v$, which has a similar value of $\sim 14.0$ eV for both compounds, and the \jbmt{\misec{screening} ratio at cRPA level}{c$GW$+LRFB screening ratio} $R=U/v$, which increases by $\sim 12\%$ from Hg1201 to Ca11,
so that the ES  is reduced from Hg1201 to Ca11.
This reduction in ES is clearly responsible for the larger $U/|t_1|$ in Ca11.
Let us discuss the origin of this reduction.
The increase in $R$ from Hg1201 to Ca11 still holds within three-orbital Hamiltonians (ratios $R_x$ and  $R_p$ in Table \ref{tab:3band} of Appendix~\ref{app:mace}, and ratios $R_{\rm AB}$ and $R_{\rm B}$ in Table \ref{tab:xabpb} of Appendix~\ref{app:abb}), 
which suggests the ES channel between AB-like band and B-like bands is not responsible for the increase in $R$.
Instead, the main factor likely to be effective to the ES is caused by substitution of apical O atoms by apical Cu atoms from Hg1201 to Ca11:
The number of apical O atoms per Cu atom is $N_{\rm Oap}=2$ for Hg1201 ($N_{\ell} = 1$),
but $N_{\rm Oap}=0$ for Ca11 ($N_{\ell} = \infty$) since both apical O atoms are replaced by apical Cu atoms from neighbouring CuO$_2$ planes, as seen in Fig. \ref{fig:crystal}.
This explains the increase in $R$.
Indeed, 
\misec{the $p$ orbitals at apical O contribute to cRPA/c$GW$\jbmt{}{/c$GW$+LRFB} screening while the screening by the AB band at the CuO$_2$ plane is excluded in the estimate of $U$, resulting in poorer screening by cRPA/c$GW$\jbmt{}{/c$GW$+LRFB} for larger $N_{\ell}$. It indicates the importance of estimating the interaction by the cRPA/c$GW$\jbmt{}{/c$GW$+LRFB} screening.}
More complex factors may also enter.
For instance, the distance between neghbouring CuO$_2$ planes in Ca11 is $d^{z}_{\rm Cu/Cu} = 3.18$ \AA~\cite{KARPINSKI1994} 
whereas, in Hg1201, the distance between Cu and apical O is $d^{z}_{\rm Oap}=2.80$ \AA~\cite{Putilin1993}:
The larger value of $d^{z}_{\rm Cu/Cu}$ in Ca11 with respect to $d^{z}_{\rm Oap}$ in Hg1201 may also contribute to reduce the ES, 
in addition to the decrease in $N_{\rm Oap}$.
Conversely, for $N_\ell \geq 2$, the presence of non-correlated orbitals in Cu and in-plane O ions may contribute to increase the ES of neighbouring layers.
}
\\

\jbmsec{
(2) Now, we discuss the case of $\delta = 0.1$.
First, we see that (2a) $U/|t_1|$ is reduced with respect to $\delta=0.0$ for both compounds.
Namely, $U/|t_1|$ decreases by $\sim 12\%$ for Hg1201 and by $\sim 6\%$ for Ca11,
so that (2b) at $\delta=0.1$, $U/|t_1|$ is $\sim 10\%$ larger for Ca11 with respect to Hg1201.
However, the cause of (2a) is different for both compounds:
In the case of Hg1201, $v$ is reduced by $\sim 5\%$ and $|t_1|$ increases by $\sim 10 \%$, whereas $U$ is almost identical:
The dominant mechanism in the reduction of $U/|t_1|$ is the delocalization of AB orbital, which increases $|t_1|$.
In the case of Ca11, $v$ and $|t_1|$ remain similar (difference is less than $\sim 2\%$) but $U$ decreases by $\sim 6\%$:
The dominant mechanism is the increase in ES.
As discussed in Appendix \ref{app:doping}, the difference between dominant mechanisms in both compounds is due to 
complex effects introduced by the LRFB correction in the preprocessing of the starting electronic structure.
Nonetheless, if we consider the starting electronic structure at KS or $GW$ level (without the LRFB correction), 
the trend (2a) should remain valid and the underlying cause of the decrease in $U/|t_1|$ upon hole doping is more intuitive.
As explained in Appendix \ref{app:doping}, \jbmsectwo{the hole doping of AB orbital 
decreases Cu$3d$/O$2p$ charge transfer energies.
If we do not consider complex effects introduced by the LRFB, 
the decrease in charge transfer energies contributes to delocalize the AB orbital 
and increase the ES, reducing $U/|t_1|$.
}
}
\\

\jbmsec{
(3) Also, the decay of intralayer interactions with distance becomes faster from Hg1201 to Ca11, as seen in Fig.~\ref{fig:charthgca}:
Values of $V_3/V_1$ and $V_6/V_3$ decrease from Hg1201 to Ca11 at both $\delta=0.0$ and $\delta=0.1$.
Finally, (4) non-negligible interlayer effective parameters appear for Ca11:
$V^{l}_{0}$ has intermediate values between $V^{}_{1}$ and $V^{}_{2}$.
This is caused by the proximity of CuO$_2$ planes due to $N_\ell = \infty$.
We note that values of $V^{l}_{n}$ decay with distance (from $\sim 0.7$ eV for $n=0$ to $\sim 0.1$ eV for $n=6$), similarly to values of $V^{}_{n}$.
}

\subsection{\jbmsec{Comparison of \jbmsectwo{Bi compounds}}}
\label{sec:1bandbi}

\begin{figure}[!htb]
\includegraphics[scale=0.75]{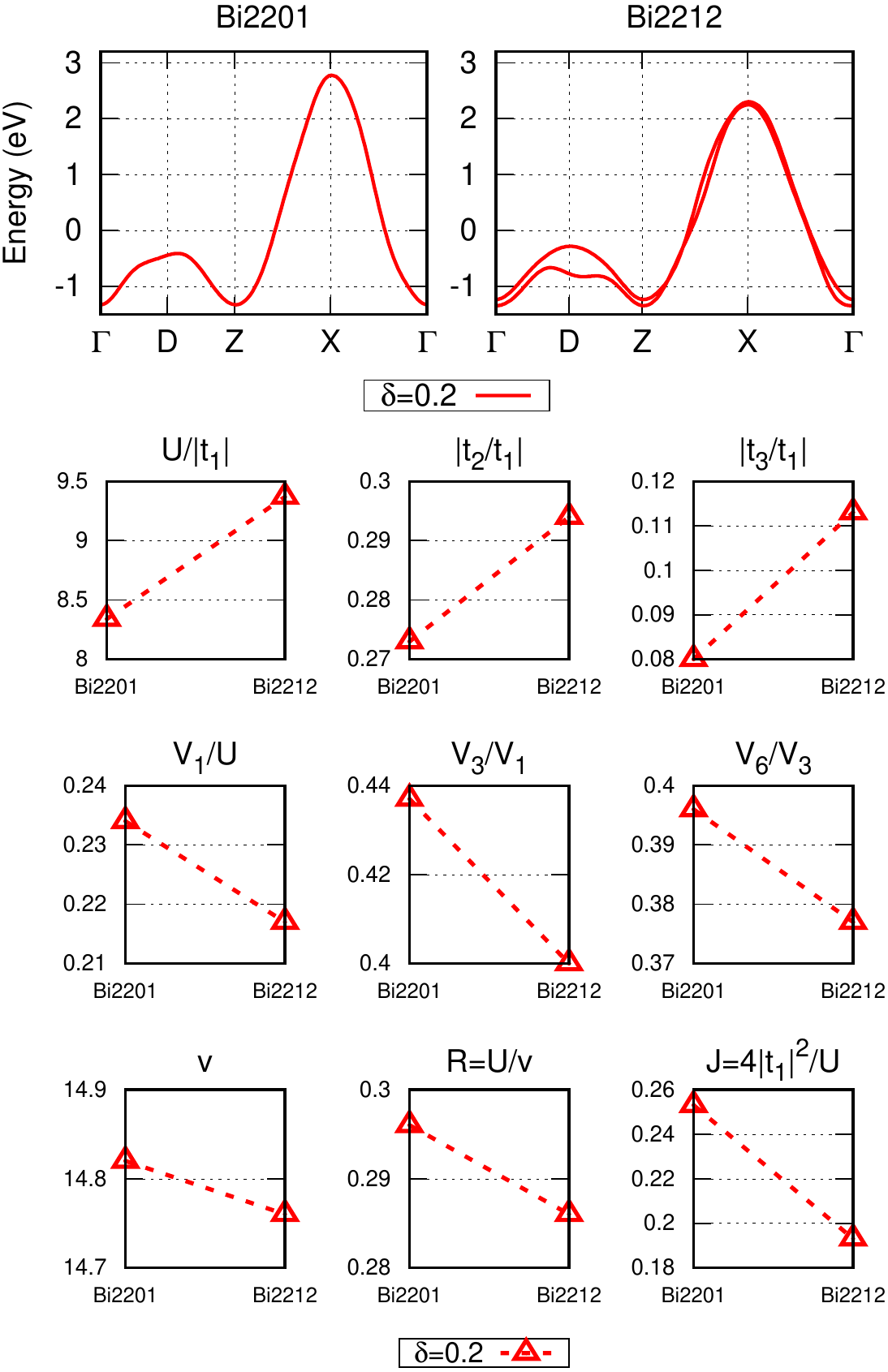}
\caption{
\jbmsec{
Upper panel : Band structure of the AB Hamiltonian at the c$GW$+LRFB(AB) level, corresponding to the one-particle part in Table \ref{tab:1nbandbi}.
High-symmetry points are defined as in Fig. \ref{fig:band}.
Lower panel : important quantities and ratios between effective parameters in Table \ref{tab:1nbandbi}.
}
}
\label{fig:chartbi}
\end{figure}

\jbmsec{Now, we discuss the comparison of \jbmsectwo{Bi compounds}. 
Results for the AB Hamiltonian are summarized in Table \ref{tab:1nbandbi} and Fig. \ref{fig:chartbi}.
Table \ref{tab:1nbandbi} shows the irreducible effective parameters ; the complete list is given in Supplemental Material.
Fig. \ref{fig:chartbi} shows band structures for the one-particle part, as well as important quantities and ratios between effective parameters.
}
\\

\jbmsec{
The overall trends at $\delta=0.2$ are summarized as follows:
(1) $U/|t_1|$ increases from Bi2201 to Bi2212, concomitantly with the increase in $N_{\ell}$ and $T^{\rm exp}_c$.
(2) The decay of interlayer interactions with distance becomes faster from Bi2201 to Bi2212.
(3) In addition, non-negligible interlayer effective parameters appear for Bi2212.
Regarding (2) and (3), the discussion is similar to that for (3) and (4) in Sec.~\ref{sec:1bandhgca} and we do not \misec{repeat} it.
In the following, we discuss (1) in detail.
}
\\

\begin{table}[!htb]
\centering
\resizebox{\columnwidth}{!}{
\begin{tabular}{ccccccccc}
\hline
\multicolumn{9}{c}{\jbmsec{Hole doping \jbmsec{per AB orbital} $\delta=0.2$}}\\
\hline
& \jbmsec{$U/|t^{}_{1}|$} & $U^{}_{}$ & $V^{}_{1}$ & $V^{}_{2}$ & $V^{}_{3}$ & $V^{}_{4}$ & $V^{}_{5}$ & $V^{}_{6}$ \\
\jbmsec{Bi2201} & \jbmsec{8.34} & \jbmsec{4.393} & \jbmsec{1.030} & \jbmsec{0.602} & \jbmsec{0.450} & \jbmsec{0.395} & \jbmsec{0.334} & \jbmsec{0.178} \\
\jbmsec{Bi2212}  & \jbmsec{9.37} & \jbmsec{4.226} & \jbmsec{0.915} & \jbmsec{0.518} & \jbmsec{0.366} & \jbmsec{0.312} & \jbmsec{0.253} & \jbmsec{0.138} \\
& & $V^{l}_{0}$ & $V^{l}_{1}$ & $V^{l}_{2}$ & $V^{l}_{3}$ & $V^{l}_{4}$ & $V^{l}_{5}$ & $V^{l}_{6}$ \\
\jbmsec{Bi2212}  & & \jbmsec{0.643} & \jbmsec{0.463} & \jbmsec{0.368} & \jbmsec{0.291} & \jbmsec{0.262} & \jbmsec{0.220} & \jbmsec{0.120} \\
& & $t^{l}_{0}$ & $t^{}_{1}$ & $t^{}_{2}$ & $t^{}_{3}$ & $t^{}_{4}$ & $t^{}_{5}$ & $t^{}_{6}$\\
\jbmsec{Bi2201} & & --- & \jbmsec{-0.527} & \jbmsec{0.144} & \jbmsec{-0.042} & \jbmsec{0.016} & \jbmsec{-0.014} & \jbmsec{-0.002} \\
\jbmsec{Bi2212}  & & \jbmsec{-0.098} & \jbmsec{-0.451} & \jbmsec{0.133} & \jbmsec{-0.051} & \jbmsec{-0.001} & \jbmsec{0.006} & \jbmsec{0.001} \\
\hline		
\end{tabular}
}
\caption{
\jbmsec{
Effective Hamiltonian parameters (in eV) for the AB Hamiltonians of \jbmsectwo{Bi compounds} at $\delta=0.2$.
The complete list of parameters is given in Supplemental Material.
\jbmsec{We also give the values of $U/|t^{}_{1}|$.}
} 
}
\label{tab:1nbandbi}
\end{table}

\jbmsec{
(1) At $\delta=0.2$, $U/|t_1|$ increases by $\sim 13 \%$ from Bi2201 to Bi2212.
This increase is not caused by the variation in $U$, which decreases by $\sim 4\%$ from Bi2201 to Bi2212.
Instead, the increase is due to the variation in $|t_1|$, which decreases by $\sim 14\%$ from Bi2201 to Bi2212.
This decrease in $|t_1|$ is caused by the buckling of in-plane Cu-O-Cu bonds.
Indeed, although the in-plane cell parameter along ${\bf a}$ is the same for Bi2201 and Bi2212 ($\sim 3.81$~\AA), in-plane O atoms are slightly distorted out of the CuO$_2$ plane in Bi2212 because of GdFeO$_3$-type tilting of CuO$_6$ octahedron~\cite{ImadaRMP}, which increases the distance between Cu and in-plane O atoms. 
This contributes to decrease the hopping amplitude $|t_1|$.
However, the localization of MLW orbitals is not affected, \misec{and} $v$ varies by less than $1\%$ between both compounds.
\misec{Despite the reduction of the characteristic energy scale \jbmt{$t_1$}{$|t_1|$} for Bi2212 \jbmt{than}{with respect to} Bi2201, 
experimental $T_c$ is enhanced for Bi2212, which may be ascribed to \jbmt{}{the} increase of $U/|t_1|$, 
with which $T_c$ may nonlinearly increase (see discussions in Sec.\ref{sec:disc} for details). 
}
}

\misec{
The minor (4\%) decrease of $U$ for Bi2212 than Bi2201 is counterintuitive in terms of the discussion in Sec.~\ref{sec:1bandhgca} because a part of the neighboring apex oxygen layer is replaced by CuO$_2$ plane with increasing $N_{\ell}$ from Bi2201 to Bi2212. 
The origin is discussed in  Appendix~\ref{app:bi}). 
}

\misec{
There exists experimental uncertainty of $d^{z}_{\rm Oap}$. For Bi2201 the uncertainty range is between 2.26 and 2.60~\AA \ \cite{Torrance1988,Torardi1988,Ito1998,Amano2004}
while in the case of Bi2212, though smaller the range ($d^{z}_{\rm Oap}\sim 2.30-2.50$ \AA~\cite{Cicco1993}) also exists.
However, we have employed nearly the lower bound 2.30 \AA \ for Bi2212 and nearly \jbmt{upperbound}{the upper bound} 2.58 \AA \ for Bi2201. 
Since $U$ monotonically increase\jbmt{}{s} with $d^{z}_{\rm Oap}$ (see Appendix \ref{app:bi} and Fig.~\ref{fig:biapp}), 
$U$ could decrease for Bi2201 and increase for Bi2212. 
This could make larger difference in $U/|t_1|$ between Bi2212 and Bi2201. 
Therefore, we can safely say that $U/|t_1|$ is larger for Bi2212 than Bi2201.
}

\section{Discussion}
\label{sec:disc}

\begin{table*}[!htb]
\begin{tabular}{cccccccccccccccc}
& & \multicolumn{1}{c}{Starting point}  & \multicolumn{1}{c}{Compound} & \jbmb{$T_{c}^{\rm exp}$ (K)}  & \jbmsec{$N_\ell$} & \multicolumn{1}{c}{\ \ \ \ } & $|U/t^{}_{1}|$ & \jbmsec{$J$ (eV)} & \multicolumn{1}{c}{\ \ \ \ } & \jbmb{$|t_1|$ (eV)} & $|t^{}_{2}/t^{}_{1}|$ & $|t^{}_{3}/t^{}_{1}|$  & \multicolumn{1}{c}{\ \ \ \ } & $|V^{}_{1}/U|$ & $|V^{}_{3}/V^{}_{1}|$ \\
	\hline
& Ref. \onlinecite{Hirayama2018} & $GW$ & \jbmb{La$_2$CuO$_4$} & $\sim 45$ & \jbmsec{$1$} & \multicolumn{1}{c}{\ \ \ \ } & 10.36 & \jbmsec{0.185} & \multicolumn{1}{c}{\ \ \ \ } & 0.48 & 0.15 & 0.21 & \multicolumn{1}{c}{\ \ \ \ } & 0.25 & 0.56 \\
& Ref. \onlinecite{Hirayama2018} & $GW$ & Hg1201     & $\sim 90$  & \jbmsec{$1$}  & \multicolumn{1}{c}{\ \ \ \ } & 9.49 & \jbmsec{0.194} & \multicolumn{1}{c}{\ \ \ \ } & 0.46 &  0.26 & 0.16 & \multicolumn{1}{c}{\ \ \ \ } & 0.22 & 0.54 \\
& Ref. \onlinecite{Hirayama2019} & $GW$+LRFB & Hg1201 & $\sim 90$ & \jbmsec{$1$}  & \multicolumn{1}{c}{\ \ \ \ } & 7.56 & \jbmsec{0.270} & \multicolumn{1}{c}{\ \ \ \ } & 0.51 &  0.25 & 0.15 & \multicolumn{1}{c}{\ \ \ \ } & 0.22 & 0.39 \\
& This work & $GW$+LRFB & Hg1201 & $\sim 90$ & \jbmsec{$1$}  & \multicolumn{1}{c}{\ \ \ \ } & 8.16 & \jbmsec{0.240} & \multicolumn{1}{c}{\ \ \ \ } & 0.49 &  0.23 & 0.11  &\multicolumn{1}{c}{\ \ \ \ } &  0.22 &  0.42 \\
\hline
\end{tabular}
\caption{
\jbmsec{
Reminder of the results for AB Hamiltonians from Refs.~\cite{Hirayama2018,Hirayama2019} for La$_2$CuO$_4$ and Hg1201, at $\delta=0.0$.
}
In Ref. \onlinecite{Hirayama2018}, the $GW$ electronic structure is used as a starting point instead of the $GW$+LRFB one ; the framework of the rest of the calculation is similar.
\jbmsec{We also remind our result for Hg1201 at $\delta=0.0$ for comparison, and show values of $J=4|t_1|^2/U$.
}
}
\label{tab:ratio}
\end{table*}

In Sec. \ref{sec:1band}, we discussed the origins of the differences in the $\rm AB$ Hamiltonians for different compounds.
Material dependence of the LEH parameters clarified in the present paper will contribute to understanding of material dependence of the SC on the microscopic level and hence to understanding of the universal mechanism of SC when they are solved by reliable quantum many-body solvers in future.
However, even before solving them, one can gain insight into the materials dependence of 
the experimentally observed SC phases from the derived $\rm AB$ Hamiltonians.

\jbmsec{Here, we discuss the insights provided by material dependence of the $\rm AB$ Hamiltonian, regarding SC.}
We first discuss two points; (1) amplitude of the SC order parameter $\Delta_{\rm SC}$,
\jbmsectwo{(2)} competition of the SC state with charge inhomogeneous states at low temperatures\jbmsectwo{, and 
(3) decrease in $U/|t_1|$ upon hole doping.}
\jbmsec{Finally, (4) we discuss 
\misec{origins and material dependence of} $|t_1|$ and associated quantities such as the \misec{magnetic superexchange} constant.
}
\\

On (1), $\Delta_{\rm SC}$ is given by the long-range value of \jbmsec{$P_{dd}$}, as $\Delta_{\rm SC}=\sqrt{P_{dd}(r\rightarrow \infty)}$.
According to Refs. \onlinecite{Ohgoe2020} and \onlinecite{Ido2017}, $\Delta_{\rm SC}$ strongly increases with $|U/t_1|$.
This simple trend is in good agreement with the correlation between higher $T_c^{\rm exp}$ and larger \jbmsec{$|U/t_1|$} in our results and the previous cRPA studies \cite{Teranishi2018,Nilsson2019}.
On the other hand, there exists exceptions such as La$_2$CuO$_4$ (La201), where \jbmsec{$|U/t_1|$ is larger than for Hg1201 despite the lower $T_c$ ; as an illustration, results from Refs.~\onlinecite{Hirayama2019,Hirayama2018} are reminded in Table \ref{tab:ratio}.}
\jbmsec{However, this is consistent with previous studies~\cite{Hirayama2018,Hirayama2019,Ido2018,Imada2021} which argue that} too large $|U/t_1|$ leads rather to the CO or charge inhomogeneous states than the SC.

\misec{Finally, we mention that a recent study \cite{Iwano2022} of the bilayer $t-t'$ Hubbard model shows that $P_{dd}(r\rightarrow \infty)$ is similar for $N_\ell=1$ and $N_\ell=2$ at least at the optimal doping. This suggests that $\Delta_{\rm SC}$ at optimal doping may not depend on inter-CuO$_2$ layer quantities, but rather on intra-CuO$_2$ layer quantities such as $U/|t_1|$. However, this should be tested by solving the AB Hamiltonians on the \textit{ab initio} basis here beyond the $t-t'$ Hubbard model.}

\jbmsec{(2)} Besides $U$, off-site intralayer interactions $V_n$ are important as well.
For Hg1201, $|V_1/U|$ and $|V_3/V_1|$ were shown \cite{Ohgoe2020} to have a significant effect on SC: 
Starting from the ``only $U$" case, $V_1$ strongly reduces the value of $P_{dd}$ (from $\sim 5 \times 10^{-2}$ to $\sim 2 \times 10^{-3}$), but $V_3$ partly compensates this reduction ($P_{dd}$ increases to $\sim 7 \times 10^{-3}$).
Despite the overall reduction of $P_{dd}$ and thus $\Delta_{\rm SC}$ 
due to $V_n$ parameters, the latter more destabilizes the competing CO state, and relatively favor the SC ground state.
In our results, 
$|V_1/U|$ \jbmsec{slightly decreases with $N_{\ell}$ and $T_c^{\rm exp}$ (or remains similar in the case of \jbmsectwo{Hg/Ca compounds} at $\delta=0.0$),} whereas $|V_3/V_1|$ decreases with $N_{\ell}$ and $T_c^{\rm exp}$.
This is counterintuitive since a lower value of $|V_3/V_1|$ is not expected to favor a SC ground state or high value of $\Delta_{\rm SC}$ \cite{Ohgoe2020}.
However, for $N_{\ell} \geq 2$, the effect of $V_3$ may be enhanced by off-site interlayer interactions.
Indeed, the intralayer effect of $V_3$ is to destabilize the CO state by causing frustration in the charge ordering.
In the case of $N_{\ell} \geq 2$, the interlayer interaction becomes comparable or larger than $V_3$.
\jbmsec{We remarked in Sec.~\ref{sec:1bandhgca} that values of $V^{l}_{n}$ decay with distance:
This suggests interlayer interaction does not act merely as a uniform background, 
and short-range interlayer interaction may play a role in the stability of CO states.
}
It is plausible that $V^{l}_{n}$ parameters enhance the frustration effect on top of $V_3$, and would contribute to further destabilization of the CO state. 
This is an intriguing issue to be examined in the future.

 \misec{Two intriguing issues are left for future studies:}
(1) Do $V^{l}_{n}$ parameters affect the value of the SC order parameter $\Delta_{\rm SC}$?
(2) Do $V^{l}_{n}$ parameters contribute to further \oldyy{destabilization of} CO states?
The $\rm AB$ Hamiltonians presented in this paper may be used as a basis to investigate these questions.

\jbmsec{(3)
Also, we have clarified in Sec.~\ref{sec:1bandhgca} and Appendix \ref{app:doping} that $U/|t_1|$ decreases when hole doping $\delta$ increases.
}
\misec{At least in the overdoped region beyond the optimal doping, the reduction of $U$ may play roles for the reduction of $T_c$ with $\delta$ \cite{Chakravarty2001}.}

\jbmsec{(4) Then, we discuss the possible role of $|t_1|$ and superexchange energy $J \sim 4|t_1|^2/U$.
As discussed in Sec.~\ref{sec:1bandbi}, in the case of Bi2212,
$|t_1| \sim 0.45$ eV is reduced compared to Bi2201 for which $|t_1| \sim 0.53$ eV (and also other compounds for which $|t_1| \geq 0.49$ eV).
This reduction in $|t_1|$ is caused by the buckling of in-plane Cu-O-Cu bonds.}
\misec{It is obvious that if all the energy scale\jbmt{}{s} of the Hamiltonian parameters are scaled by $\lambda$, $T_c$ should also be scaled by $\lambda$. Therefore, $T_c$ must be scaled by the energy unit defined by \jbmt{$t_1$}{$|t_1|$} itself in addition to the dimensionless parameter $U/|t_1|$.}
\misec{Nevertheless, since the superconducting order is expected to increase nonlinearly with $U/|t_1|$, the reduced \jbmt{$t_1$}{$|t_1|$} may lead to enhanced $T_c$, which is consistent with Bi2212 showing \jbmt{$T_c$ higher}{higher $T_c$} than Bi2201.}

\jbmsec{Second, let us quickly discuss the possible role of $J$, which is represented in Figs.~\ref{fig:charthgca} and \ref{fig:chartbi}: 
\misec{ Ref. \cite{ivashko2019strain} claimed that} (i) a larger $J$ may be favorable to SC ; reciprocally, smaller values of $J$ and hence $|t_1|$ may be destructive for SC.
However, in our results for Bi compounds, $J \sim 0.192$ eV for Bi2212 is smaller than \jbmsec{$J \sim 0.253$} eV for Bi2201 despite the much \misec{higher} $T^{\rm exp}_c$ for Bi2212 ($\sim 84$ K) compared to Bi2201 ($\sim 10$ K), so that $J$ shows a negative correlation with $T^{\rm exp}_c$ \misec{again}.
Regarding \jbmsectwo{Hg/Ca compounds}, at $\delta=0.0$, $J \sim 0.242$ eV for both compounds. 
However, at $\delta=0.1$, $J\sim 0.296$ eV for Hg1201 but $J \sim 0.256$ eV for Ca11, so that $J$ shows a negative correlation with $T^{\rm exp}_c$.
In any case, we do not find a positive correlation between $J$ and $T^{\rm exp}_c$.
\misec{This suggests that $J$ is not a primary component to control $T_c$.}
\misec{This is, however, in accordance with the trend of higher $T_c$ for larger $U/|t_1|$ mentioned already if $t_1$ is the same.}
}

\misec{Although the essence does not change, the parameters of our derived LEH could be substantially altered by the atomic coordinate\jbmt{}{s}, 
particularly sensitively by the apex oxygen position, which has experimental uncertainty in the case of Bi compounds in the literature. 
It is desired to experimentally determine the positions more precisely. 
In particular, we have shown the Hamiltonian parameters for Bi2201 in Table~\ref{tab:1nbandbi} when we assume $d^z_{\rm Oap}=2.58$ \AA, while it could be realistically $\sim2.4$~\AA, which makes difficult to compare with Ca11 and Hg 1201 quantitatively because $d^z_{\rm Oap}=2.4$~\AA~would give roughly 20\% reduction of $U/|t_1|$ as is inferred from Fig.~\ref{fig:biapp}, resulting in $U/|t_1|\sim 6.7$, much smaller than the cases of Hg1201 and Ca11, which could be the origin of lower $T_c$.}

\section*{Summary}

We have derived and compared LEHs for Bi2201, Bi2212\jbmsec{, Hg1201} and Ca11 to gain insights into the mechanism of cuprate superconductivity from the differences of the Hamiltonian parameters of the \jbmsec{compounds} that show diverse superconducting transition temperatures. 
\jbmsec{In the case of Hg1201, we have benchmarked our result with respect to Ref. \onlinecite{Hirayama2019}.}
The Hamiltonians are also derived by aiming at serving for future studies to obtain detailed physical properties by accurate quantum many-body solvers. 
\\

We have employed the following steps for the derivation:

(1) Preprocessed quasiparticle electronic structure is used as a starting point for the derivation of the LEH. 
\misec{To derive the LEH, we employ} \jbmsectwo{hole doping per AB orbital $\delta \leqslant 0.1$ for Hg/Ca compounds and $\delta=0.2$ for Bi compounds,
which is close to the experimental optimal value \misec{for superconductivity}.
}
We improved the exchange and correlation from the KS level to the $GW$+LRFB level.

(2) By using the result of (1), the single-\jbmsec{orbital} AB LEHs are derived on the cGW+LRFB level.
In addition to (2), three-orbital $\rm AB\rm B$ LEHs were obtained, which showed that $\rm B$ orbitals \jbmsectwo{might} have few effect on the low-energy physics except for the indirect \jbmt{cRPA-type}{c$GW$+LRFB} screening effect to renormalize the effective interaction within the AB subspace\jbmsec{.}
\jbmsec{\misec{The sufficiency to restrict} to the single-orbital picture appears reasonable, which \jbmsec{supports} 
the focus to the single-orbital $\rm AB$ Hamiltonian.
}
We then compared the \jbmsec{variations in} 
effective parameters 
\jbmsec{from Bi2201 to Bi2212 on the one hand, and from Hg1201 to Ca11 on the other hand, within the single-band $\rm AB$ Hamiltonian.
The main findings are the following: 
\misec{Experimental SC transition temperature $T^{\rm exp}_c$ becomes higher for larger $U/|t_1|$ concomitant\jbmt{}{ly} with larger \jbmt{number of layers in the unit cell $N_{\ell}$}{number $N_{\ell}$ of laminated CuO$_2$ planes between the two neighboring block layers}.}
In addition, \misec{$T^{\rm exp}_c$ is higher if the off-site interactions $V_n$ decays faster as a function of the distance} 
or non-negligible interlayer parameters emerge for $N_{\ell} \geq 2$.
Furthermore, 
$U$ decreases upon hole doping, \misec{which may play a role for the decrease of $T_c$ and disappearance of SC at overdoped hole concentration  
as observed experimentally}.
}
Our result on the systematic dependence of the {\it ab initio} LEH parameters 
\jbmsec{suggests} that larger \jbmsec{$U/|t_1|$} favors larger SC order parameter $\Delta_{\rm SC}$ and hence higher $T_c$, whereas 
\misec{tuning of intra- and inter-layer interactions 
could amplify $T_c$.}
The derived \textit{ab initio} LEHs provide a reliable starting point to investigate these hypotheses by solving them by a reliable quantum many-body solver.

\section*{Acknowledgements}

We thank Kazuma Nakamura for useful discussions.
This work was supported by MEXT as Program for Promoting Researches on the Supercomputer Fugaku (Basic Science for Emergence and Functionality in Quantum Matter ­Innovative Strongly-Correlated Electron Science by Integration of Fugaku and Frontier Experiments­, JPMXP1020200104) and used computational resources of supercomputer Fugaku provided by the RIKEN Center for Computational Science (Project ID\oldyy{: hp200132, hp210163, and hp220166}).
Part of the computation was done \oldyy{using the facilities of the Supercomputer Center, the Institute for Solid State Physics, the} University of Tokyo.
We also acknowledge the financial support of JSPS Kakenhi Grant Nos. 16H06345 \misec{and 22A202 (``Foundation of Machine Learning Physics")}. 
\jbmsec{Fig. \ref{fig:crystal} was drawn by using software \texttt{VESTA} \cite{Momma2011}.}

\appendix

\section{\jbmsec{Reminder of the MACE procedure and details about intermediary quantities}}
\label{app:mace}

\jbmsec{Here, as a complement to Sec. \ref{sec:meth},
we give a reminder of the MACE scheme~\cite{Aryasetiawan2004,Imada2010,Hirayama2013,Hirayama2018,Hirayama2019}, 
which we have implemented within the \texttt{RESPACK} code~\cite{Nakamura2020}.
In addition, we give details about intermediate quantities such as the three-orbital $xp$ effective Hamiltonian and the LRFB correction $\Delta \mu$.
}
\\

\jbmsec{We start from the electronic structure at the KS level, which is represented in Fig.~\ref{fig:band}.
The one-particle Green's function at KS level is 
\begin{equation}
G^{\rm \jbmt{}{KS}}(r,r',\omega) = \sum_{n,k} \frac{\psi^{{\rm KS}}_{nk}(r)\psi^{*{\rm KS}}_{nk}(r')}{\omega-\epsilon^{{\rm KS}}_{nk}+i\eta},
\label{eq:gks}
\end{equation}
\jbmt{which we denote as $G^{\rm KS}[\epsilon^{{\rm KS}}_{nk},\psi^{{\rm KS}}_{nk}]$the KS eigenvalues and eigenfunctions are $(\epsilon^{{\rm KS}}_{nk},\psi^{{\rm KS}}_{nk})$}
{and the superscript in the one-particle Green's function $G$ is that in KS eigenvalues and eigenfunctions $(\epsilon^{{\rm KS}}_{nk},\psi^{{\rm KS}}_{nk})$ in Eq.~(\ref{eq:gks}).}
$k$ is the wavevector \jbmt{}{in the full Brillouin zone}, $m,n$ are the band indices, and $\eta$ is the attenuation factor whose value is set to $0.027$ eV in our calculations.
The origin of frequency $\omega$ is the Fermi level $\epsilon_{\rm F}$, which is set to zero.
First, we preprocess the medium-energy (M) space, composed of $3d$-like bands from Cu atoms and $2p$-like bands from O atoms in the unit cell ; 
the complementary subspace, denoted as H, will be left at the KS level.
The number of bands within M space is denoted as $N_{\rm M}$, which is 5 times the number of Cu atoms in the unit cell plus 3 times the number of O atoms in the unit cell: $N_{\rm M}=11$ for Ca11, $17$ for Hg1201, $23$ for Bi2201 and $34$ for Bi2212.
}

\subsection{\jbmsec{Disentanglement of the M space and H space at the KS level}}
\label{sec:macedis}

\jbmsec{
In case the M space is entangled with other bands from H space, we first disentangle \cite{Miyake2009} the M space from these other bands.
The latter consist in the $s$-bands from Hg and Ba near X point for Hg1201 at $\delta=0.0$, the semicore bands around $\sim -7/-6$ eV \jbmt{}{near X point} for Hg1201 at $\delta = 0.1$, 
and the interstitial $s$-like band from Ca near A point for Ca11 at $\delta=0.0$ and $0.1$, as seen in Fig.~\ref{fig:band}.
}

\jbmsec{
To do so, we extract the M space by computing $N_{\rm M}$ MLW orbitals with atomic $3d$ character centered on Cu atoms and atomic $2p$ character centered on O atoms ; 
The outer window is the whole M space, plus 
one band above M space for Ca11,
two bands above M space for Hg1201 at $\delta=0.0$,
and one band below M space for Hg1201 at $\delta=0.1$.
If necessary, we use the inner window to preserve the band dispersion of \jbmt{both}{}M bands\jbmt{,
and to avoid modifications in the band dispersion of $s$-bands near Fermi level, 
so that unphysical electron pockets are not artificially \jbmt{re}{}introduced by disentanglement.}{.}
The inner window is $[-7.0 : 0.0]$ eV with respect to Fermi level for Hg1201 at $\delta=0.0$, 
$[-6.0 : +0.9]$ eV for Hg1201 at $\delta=0.1$, 
and $[-8.0 : +0.5]$ eV for Ca11 at $\delta =0.1$.
Regarding Ca11 at $\delta=0.0$, we have checked that the final AB Hamiltonian is not affected (notably, $U/|t_1|$ varies by less than $1\%$) by the choice of using the inner window, so that the latter is unnecessary.}

\jbmsec{
We minimize the spillage functional \cite{Souza2001} in the outer window to extract the M \jbmt{sub}{}space,
then minimize the spread functional \cite{Marzari1997} in the M \jbmt{sub}{}space to deduce the MLW orbitals and Wannier bands, which provide a basis spanning the M space at KS level.
Then, we recompute the KS bands which were formerly entangled with the M space ; the latter bands become orthogonal to the M space after this procedure.
We obtain the disentangled KS eigenelements $(\epsilon_{nk}^{\rm KS,dis},\psi_{nk}^{\rm KS,dis})$.
This modifies the one-particle Green's function at \jbmt{}{the} KS level $G^{\rm KS}$, allowing to separate it \jbmt{as}{into}
\begin{equation}
\jbmt{G^{\rm KS,dis} = G^{}_{\rm M}[\epsilon_{nk}^{{\rm KS,dis}},\psi_{nk}^{{\rm KS,dis}}] + G^{}_{\rm H}[\epsilon^{{\rm KS,dis}}_{nk},\psi_{nk}^{{\rm KS,dis}}],}
{G^{\rm KS,dis} = G^{{\rm KS,dis}}_{\rm M} + G^{{\rm KS,dis}}_{\rm H},}
\label{eq:gksdis}
\end{equation}
in which the subscript "M" (respectively, "H") means the summation over band index $n$ in Eq.~(\ref{eq:gks}) is restricted to the M space (respectively, H space).
In case the M bands are already disentangled from H bands (which is the case for \jbmsectwo{Bi compounds} at $\delta=0.2$, as seen in Fig.~\ref{fig:band}), we simply have
$(\epsilon_{nk}^{\rm KS,dis},\psi_{nk}^{\rm KS,dis})=(\epsilon^{{\rm KS}}_{nk},\psi^{{\rm KS}}_{nk})$ and $G^{\rm KS,dis} = G^{\rm KS}$.
}
\subsection{\jbmsec{Preprocessing of the M space at the $GW$ level}}
\label{sec:macegw}

\jbmsec{
Then, we preprocess the disentangled M space at the $GW$ level.
To do so, we compute the irreducible RPA polarization
\begin{equation}
\chi^{} = -i G^{\rm KS,dis} G^{\rm KS,dis},
\end{equation}
which is denoted as $\chi^{}[G^{\rm KS,dis}]$,
and deduce the 
RPA screened interaction
\begin{equation}
W^{} =  [I - v \chi^{}]^{-1}v,
\end{equation}
\jbmt{}{
which is denoted as $W[v,\chi]$,}
where $I$ is the identity matrix and $v$ is the bare Coulomb interaction.
Then, we compute the matrix elements of the $GW$ self-energy\footnote{\jbmsec{For this part, the computational cost may be reduced by using the approximation proposed in Appendix~\ref{app:ode}.}}
\begin{equation}
\Sigma^{GW} = i G^{\rm KS,dis} W^{},
\end{equation}
which is denoted as $\Sigma^{GW}[G^{\rm KS,dis},W^{}]$, as a function of frequency $\omega$ and within the basis of disentangled KS eigenelements.
The matrix elements are
\begin{align}
\Sigma^{GW}_{mn}(k,\omega) = \int_{\Omega} dr & \int_{\Omega} dr' \psi_{mk}^{*{\rm KS,dis}}(r) \times \nonumber \\
& \Sigma^{GW}(r,r',\omega) \psi_{nk}^{\rm KS,dis}(r'),
\end{align}
where $m,n$ are restricted to the M space, and $\Omega$ is the unit cell.
Then, we obtain the quasiparticle $GW$ self-energy, as
\begin{equation}
\tilde{\Sigma}^{GW}_{mn}(k) = \Sigma^{GW}_{mn}(k,\omega_{mn}(k)),
\end{equation}
where $\omega_{nn}(k)=\epsilon_{nk}^{\rm KS,dis}$ and $\omega_{mn}(k)=\epsilon_{\rm F}$ if $m \neq n$, as done in Ref.~\onlinecite{Hirayama2018}.
This allows to deduce the quasiparticle $GW$ one-particle part restricted to the M space, as
\begin{equation}
\jbmt{h^{GW} = h^{\rm KS} + [- V^{\rm xc} + \tilde{\Sigma}^{GW}]Z^{GW},}
{h^{GW}_{mn}(k) = h^{\rm KS}_{mn}(k) + \Big[- V^{\rm xc}_{mn}(k) + \tilde{\Sigma}^{GW}_{mn}(k)\Big]Z^{GW}_{mn}(k),}
\label{hgw}
\end{equation}
in which the KS one-particle part $h^{\rm KS}$, the KS exchange-correlation potential $V^{\rm xc}$ and the quasiparticle $GW$ self-energy $\tilde{\Sigma}^{GW}$ 
are calculated in the basis of disentangled KS eigenelements, restricted to the M space.
We also take into account the frequency dependence of \jbmt{$\omega_{nn}$}{$\Sigma^{GW}_{nn}(k,\omega)$} 
around \jbmt{$\epsilon_{nk}^{\rm KS,dis}$}{$\omega_{nn}(k)$}, by defining the perturbative renormalization factor \jbmt{}{for diagonal elements} \cite{Hirayama2013}: 
\begin{equation}
Z^{GW}_{nn}(k)=1/\Big[ 1 - \frac{\partial}{\partial \omega} \Sigma^{GW}_{nn}(k,\omega=\omega_{nn}(k)) \Big],
\end{equation}
which is denoted as \jbmt{}{$Z[\Sigma^{GW}_{nn}(k,\omega_{nn}(k))]$}.
\jbmt{for diagonal elements ; the renormalization factor is set to unity for off-diagonal elements.}{For $m \neq n$, $Z^{GW}_{mn}(k)$ is set to unity, as done in Ref.~\onlinecite{Hirayama2018}.}
Then, we diagonalize Eq.~(\ref{hgw}), which yields the $GW$ eigenelements $(\epsilon_{nk}^{GW},\psi_{nk}^{GW})$ \jbmt{}{within the M space}.
We recompute the Fermi level, then shift bands in the H space so that their position with respect to the Fermi level remains unchanged.
We obtain the preprocessed Green's function at the $GW$ level:
\begin{equation}
\jbmt{G^{GW} = G^{}_{\rm M}[\epsilon_{nk}^{GW},\psi_{nk}^{GW}] + G^{}_{\rm H}[\epsilon_{nk}^{\rm KS,dis},\psi_{nk}^{\rm KS,dis}],}
{G^{GW} = G^{GW}_{\rm M} + G^{\rm KS,dis}_{\rm H},}
\label{eq:ggw}
\end{equation}
which gives the $GW$ electronic structure, in which exchange and correlation have been improved with respect to the LDA/GGA exchange-correlation potential in the KS electronic structure.
This $GW$ electronic structure will be used as a starting point to derive the three-orbital $xp$ Hamiltonian and compute the LRFB correction, 
which will allow to further improve the starting electronic structure at the $GW$+LRFB level.
}

\begin{table*}[!htb]
\begin{tabular}{cccccccccccccccccccccc}
\hline
	& \jbmsec{$\delta$} & \jbmb{$T_c^{\rm exp}$ (K)} & $N_e$ & $n_x$ & $n_p$ & \ \ & $\Delta E_{xp}$ & $|t_{xp}|$ & $|t_{pp}|$ & \ \ & $U_{x}$ & $U_{p}$ & \jbmt{}{$V_{xp}$}  & \ \ & $v_{x}$ & $v_{p}$ & \ \ & $R_{x}$ & $R_{p}$ & $t^{\rm SIC}$ & \ \ \ \ \jbmsec{$\Delta \mu$}\\
\hline
Hg1201 (AE)	&	\jbmsec{0.0} & $\sim 90$ & 5.0 & 1.437 & 1.781 	& \ \ & 2.41                           & 1.26        & 0.75 & \ \ & 8.84 & 5.31 & \jbmt{}{1.99} & \ \ & 28.82 & 17.11 & \ \ & 0.31 & 0.31& -1.62 & \ \ \ \ \\
Ca11 (AE)		&	\jbmsec{0.0} & $\sim 110$ & 5.0 & 1.433(0) & 1.783(5) 	& \ \ & 2.62		        & 1.29        & 0.74 & \ \ & 9.33 & 6.16 & \jbmt{}{2.35} & \ \ & 28.99 & 17.84 & \ \ & 0.32 & 0.35& -1.19 & \ \ \ \ \\
\hline
Hg1201 & \jbmsec{0.0} & $\sim 90$ & 5.0 & 1.474 & 1.763 		& \ \ & 2.03                           	& 1.32        & 0.80 & \ \ & 8.51 & 5.35 & \jbmt{}{1.93} & \ \ & 25.19 & 17.03 & \ \ & 0.34 & 0.31& -1.56 & \ \ \ \ \jbmsec{1.10}\\
Ca11		& \jbmsec{0.0} & $\sim 110$ & 5.0 & 1.454 & 1.773 		& \ \ & 2.36			& 1.36	& 0.77 & \ \ & 9.72 & 6.30 & \jbmt{}{2.46} & \ \ & 25.35 & 17.58 & \ \ & 0.38 &  0.36 & -1.48 & \ \ \ \ \jbmsec{1.50}\\
\hline
\jbmsec{Hg1201} & \jbmsec{0.1} & \jbmsec{$\sim 90$} & \jbmsec{$5.0$} & \jbmsec{ 1.490} & \jbmsec{1.755} & \ \ & \jbmsec{1.76}			& \jbmsec{1.32}	& \jbmsec{0.80} & \ \ & \jbmsec{8.92} & \jbmsec{5.53} & \jbmt{}{2.07} & \ \ & \jbmsec{25.21} & \jbmsec{17.02} & \ \ & \jbmsec{ 0.35} & \jbmsec{0.32} & \jbmsec{ -1.80} & \ \ \ \ \jbmsec{1.05} \\
\jbmsec{Ca11} & \jbmsec{0.1} & \jbmsec{$\sim 110$} & \jbmsec{$5.0$} & \jbmsec{1.474} & \jbmsec{1.763} 	& \ \ & \jbmsec{2.10}			&  \jbmsec{1.35} & \jbmsec{0.82} & \ \ \ \ & \jbmsec{8.88} & \jbmsec{5.76} &\jbmt{}{2.20} & \ \ & \jbmsec{25.23} & \jbmsec{17.14} & \ \ & \jbmsec{0.35} & \jbmsec{0.34} & \jbmsec{-1.47} & \ \ \ \  \jbmsec{1.15} \\
\hline
Bi2201 & \jbmsec{0.2} & $\sim 10$ & 5.0 & \jbmsec{1.450} & \jbmsec{1.775} 			& \ \ & \jbmsec{2.54}			& \jbmsec{1.42}	& \jbmsec{0.85} & \ \ & \jbmsec{9.48} & \jbmsec{6.19} & \jbmt{}{2.27} & \ \ & \jbmsec{25.55} & \jbmsec{17.72} & \ \ & \jbmsec{0.37} & \jbmsec{0.35} & \jbmsec{-1.37} & \ \ \ \ \jbmsec{1.41}\\
\jbmsec{Bi2212}   & \jbmsec{0.2} & \jbmsec{$\sim 84$} & \jbmsec{$5.0$} & \jbmsec{1.458} & \jbmsec{1.771} & \ \ & \jbmsec{2.40} & \jbmsec{1.35} & \jbmsec{0.86} & \ \ & \jbmsec{9.00} & \jbmsec{5.85} & \jbmt{}{1.98} & \ \ & \jbmsec{25.70} & \jbmsec{17.73} & \ \ & \jbmsec{0.35} & \jbmsec{0.33} & \jbmsec{-1.38} &  \ \ \ \ \jbmsec{1.33} \\
\hline
\end{tabular}
\caption{
Three-orbital $xp$ Hamiltonian. We give the \jbmsec{hole doping \jbmsec{per Cu atom} $\delta$ which is considered in the starting $GW$ electronic structure. We remind the values of  $T_c^{\rm exp}$ (in K), and give the }total number of electrons \jbmsec{\jbmsec{per CuO$_2$ plane and unit cell and within the}} $xp$ subspace ($N_e$), $GW$ occupation numbers for $x$ and $p$ orbitals ($n_x$ and $n_p$), charge transfer energy between the $x$ and $p$ levels ($\Delta E_{xp}$), amplitudes of hoppings between neighbouring $x$ and $p$ orbitals ($|t_{xp}|$) and two $p$ orbitals within the unit cell ($|t_{pp}|$), intra-orbital effective interaction ($U_{x}$ and $U_{p}$) and bare interaction ($v_{x}$ and $v_{p}$) for $x$ and $p$ orbitals\jbmt{}{, and inter-orbital effective interaction $V_{xp}$ between $x$ and $p$ orbitals in the unit cell}. We also give the intra-orbital \jbmt{}{c$GW$} screening ratio $R_{i}=U_{i}/v_{i}$, and the SIC $t^{\rm SIC} = -U_{x}n_{x}/2 + U_{p}n_{p}/2$.
\jbmsec{Upper panel \jbm{shows the $xp$ Hamiltonian} obtained by using the all-electron (AE) implementation \cite{Hirayama2018} \jbm{for \jbmsectwo{Hg/Ca compounds}, }for comparison.
Then, we show in lower panels the $xp$ Hamiltonian obtained with our PP implementation, and values of the LRFB correction $\Delta \mu$ obtained by solving the $xp$ Hamiltonian.
We note that $N_e$ is renormalized to the undoped value $N_e=5.0$ even if $\delta \neq 0.0$, and $n_x$, $n_p$ and $t^{\rm SIC}$ are recalculated accordingly (other parameters are not changed).}
}
\label{tab:3band}
\end{table*}

\subsection{\jbmsec{Three-orbital $xp$ Hamiltonian \misec{with} LRFB correction}}
\label{app:xp}

\jbmsec{
Now, we derive the three-orbital Hamiltonian which is used in Sec. \ref{sec:prep} for the calculation of the LRFB correction.
This Hamiltonian is denoted as $xp$ in this paper ; it is equivalent to the three-orbital Hamiltonian in Ref. \onlinecite{Hirayama2018}, and is sometimes denoted as $dpp$ Hamiltonian in the literature.
}
\jbmsec{
We restart from the $GW$ electronic structure in Eq.~(\ref{eq:ggw}), 
and construct three MLW orbitals (six in the case of Bi2212), with atomic \jbmt{Cu}{}$x$ character centered on Cu atoms, and atomic \jbmt{O}{}$p$ character centered on in-plane O atoms.
We use the whole M space as the outer window.
We obtain the $xp$ subspace, from which we disentangle other bands within the M space (denoted as N). 
The disentangled eigenelements are denoted as $(\epsilon_{nk}^{GW,{\rm dis}},\psi_{nk}^{GW,{\rm dis}})$,
and the associated Green's function is
\begin{equation}
\jbmt{
G^{GW,{\rm dis}} = G^{}_{xp}[ \epsilon_{nk}^{GW,{\rm dis}},\psi_{nk}^{GW,{\rm dis}}]+ G^{}_{\rm N}[ \epsilon_{nk}^{GW,{\rm dis}},\psi_{nk}^{GW,{\rm dis}}] + G^{}_{\rm H}[\epsilon_{nk}^{\rm KS, dis},\psi_{nk}^{\rm KS, dis}].}
{
G^{GW,{\rm dis}} = G^{GW,{\rm dis}}_{xp} + G^{GW,{\rm dis}}_{\rm N} + G^{\rm KS, dis}_{\rm H}.}
\end{equation}
Then, we compute \jbmt{the RPA polarization  $\chi^{\rm RPA}[G^{GW,{\rm dis}}]$}{$\chi^{\rm RPA}=\chi[G^{GW,{\rm dis}}]$} 
and \jbmt{}{the} cRPA polarization 
\jbmt{$\chi^{\rm cRPA} = \chi^{}[G^{GW,{\rm dis}}] - \chi_{xp}[G^{}_{xp}]$,} 
{$\chi^{\rm cRPA} = \chi^{\rm RPA} - \chi_{xp}$}
where \jbmt{$\chi_{xp}[G^{}_{xp}]$}{$\chi_{xp}=\chi[G^{GW,{\rm dis}}_{xp}]$} is the double counting screening term consisting in screening channels which are internal to the $xp$ subspace. 
From $\chi^{\rm RPA}$ and $\chi^{\rm cRPA}$, we deduce respectively the RPA screened interaction 
\jbmt{$W^{\rm RPA}[\chi^{\rm RPA}]$}{$W^{\rm RPA}[v,\chi^{\rm RPA}]$} and cRPA effective interaction \jbmt{$W^{\rm cRPA}[\chi^{\rm cRPA}]$}{$W^{\rm cRPA}[v,\chi^{\rm cRPA}]$}.
We evaluate $W^{\rm cRPA}(\omega=0)$ in the basis of MLW orbitals $w_{i{\bf R}}$, where $i$ is the Wannier band index and ${\bf R}$ is the position of the primitive cell which contains the MLW orbital, to deduce \jbmt{}{the two-particle part of the $xp$ effective Hamiltonian, as}
\begin{widetext}
\begin{equation}
U_{ij}^{}(\jbmsectwo{{\bf R'-R}}) = \int_{\Omega} dr  \int_{\Omega} dr ' w_{i{\bf R}}^{*}(r) w_{j{\bf R'}}^{*}(r') W^{\rm cRPA}(r,r',\omega=0) w_{i{\bf R}}(r) w_{j{\bf R'}}(r'),
\label{eq:h2p}
\end{equation}
\end{widetext}
\jbmt{}{and we may restrict the calculation to $U_{ij}({\bf R})$ by using the translational invariance.} 
Then, we compute the $GW$ self-energy \jbmt{$\Sigma^{GW}[G^{GW,{\rm dis}},W^{\rm RPA}]$}{$\Sigma^{}[G^{GW,{\rm dis}},W^{\rm RPA}]$}
and remove the double counting exchange and correlation term as follows. 
The latter term is the self-energy $\Sigma_{xp}$ which is internal to the $xp$ subspace.
At the $GW$ level, this term is 
\begin{equation}
\Sigma_{xp} = G^{GW,{\rm dis}}_{xp} W_{xp}
\end{equation}
where $G^{GW,{\rm dis}}_{xp}$ is the intra-$xp$ subspace Green's function, and 
\jbmt{}{
$W_{xp} = W[W^{\rm cRPA}(\omega=0),\chi_{xp}]$ is the intra-$xp$ subspace RPA screened interaction,
in which $W^{\rm cRPA}(\omega=0)$ plays the role of the intra-$xp$ subspace bare interaction.
}
We deduce the constrained self-energy as $\Delta\Sigma = \Sigma^{} - \Sigma_{xp}$.
We compute the matrix elements of $h^{\rm KS}$, $V^{\rm xc}$ and $\Delta\Sigma$ in the basis of $(\epsilon_{nk}^{GW,{\rm dis}},\psi_{nk}^{GW,{\rm dis}})$
restricted to the $xp$ subspace, and evaluate both diagonal and off-diagonal elements at $\omega_{mn}(k)=\epsilon_{\rm F}$ 
as done in Ref.~\onlinecite{Hirayama2018} to deduce \jbmt{}{the quasiparticle constrained self-energy} $\tilde{\Delta \Sigma}$.
We obtain the c$GW-$SIC one-particle part in the basis of  $(\epsilon_{nk}^{GW,{\rm dis}},\psi_{nk}^{GW,{\rm dis}})$ \jbmt{}{restricted to the $xp$ subspace}, as
\begin{equation}
\jbmt{h^{{\rm c}GW-{\rm SIC}} = [h^{\rm KS} - V^{\rm xc} + \tilde{\Delta \Sigma} - t^{\rm SIC}]Z^{{\rm c}GW},}
{h^{{\rm c}GW-{\rm SIC}}_{mn}(k) = \big[h^{\rm KS} - V^{\rm xc} + \tilde{\Delta \Sigma} - t^{\rm SIC}\big]_{mn}(k) Z^{{\rm c}GW}_{mn}(k),}
\label{eq:hcgw}
\end{equation}
in which we apply the perturbative renormalization factor \jbmt{}{$Z^{{\rm c}GW}=Z[\Delta \Sigma^{}_{nn}(k,\epsilon_{\rm F})]$} to diagonal elements, and the SIC term \cite{Hirayama2015} is computed in the basis of Wannier orbitals as
\begin{equation}
t^{\rm SIC}_{i} = - U_{ii}^{}({\bf 0}) n_{i}/2
\end{equation}
where $n_{i}$ is the occupation number of the MLW orbital $i$, then $t^{\rm SIC}$ is rotated to the basis of $(\epsilon_{nk}^{GW,{\rm dis}},\psi_{nk}^{GW,{\rm dis}})$.
Finally, we \jbmt{evaluate}{rotate} $h^{{\rm c}GW-{\rm SIC}} $ \jbmt{in}{to} the basis of MLW orbitals to deduce the one-particle part of the effective Hamiltonian, as
\begin{equation}
t_{ij}(\jbmsectwo{{\bf R'-R}}) = \int_{\Omega} dr w_{i {\bf R}}(r) h^{{\rm c}GW-{\rm SIC}}(r) w_{j {\bf R'}}(r),
\label{eq:h1p}
\end{equation}
\jbmt{}{and we may restrict the calculation to $t_{ij}({\bf R})$ by using the translational invariance.} 
}
\\

\jbmsectwo{Table \ref{tab:3band} shows the most important effective parameters for the $xp$ Hamiltonian. 
Here, we (1) compare to $xp$ Hamiltonians obtained by using the all-electron \jbmt{}{(AE)} implementation \cite{Hirayama2018,Hirayama2019} \jbmt{}{at $\delta=0.0$},
and (2) discuss a technical subtlety in the case of $\delta \neq 0.0$.}
\\

(1) \jbmsec{First, we focus on \jbmsectwo{Hg/Ca compounds} at $\delta=0.0$, and benchmark our result with respect to the AE implementation.}
For Hg1201, our results compare to those in Ref. \onlinecite{Hirayama2018}. 
Also, for Ca11, we derived the $xp$ Hamiltonian by using the AE implementation, for comparison.

(1.i) The onsite bare interaction is different for the $x$ orbital. Typically, $v_{x} \sim 29$ eV in the AE result and $\sim 25$ eV in our result.
This is due to the PPs used in our calculations; we have checked that the difference in $v_{x}$ may be corrected by regenerating the PPs with reduced cutoff radii, but at the expense of increasing the cutoff energy for wavefunctions from $100$ Ry to at least $200$ Ry.
Nonetheless, effective interaction amplitudes are in good agreement; the difference is less than $5\%$.
Also, the soft PPs allow computational tractability for
compounds with a larger number of atoms in the unit cell, such as Bi2212.

(1.ii) The one-particle part is in good agreement, except the charge transfer energy $\Delta E_{xp}$ which is a bit underestimated in our new implementation\jbm{: T}he difference is $\sim 10\%$ for Ca11 and $\sim 19\%$ for Hg1201.
For Ca11, the difference in $\Delta E_{xp}$ does not exceed $10\%$, which remains acceptable. 
For Hg1201, although the difference is a bit larger, the final $\rm AB$ Hamiltonian is well reproduced, so that we \jbmt{regard}{}deem it acceptable as well.
This discrepancy is partly due to the difference in $GW$ occupation numbers, which are more covalent in the PP result with respect to the AE result. These occupation numbers are used to calculate the SIC, as  $t^{\rm SIC}_{i} = \jbmb{-}U_{i} n_{i}/2$ \cite{Hirayama2015}, so that, before applying the renormalization factor \cite{Hirayama2013}, $\Delta E_{xp}$ is modified by $t^{\rm SIC} = \jbmb{t^{\rm SIC}_{x} - t^{\rm SIC}_{p}}$ (which is negative in practice).
For $U_x=8.51$ eV and $U_p=5.35$ eV (our values for Hg1201), a modification $n_x \rightarrow n_x -\delta$ and $n_p \rightarrow n_p + \delta/2$ with $\delta$ as small as $0.04$ electrons (corresponding to the difference between our result and the AE result) leads to an increase in $\Delta E_{xp}$ as large as $0.12$ eV, which partly explains the larger $\Delta E_{xp}$ for the AE result.
\\

\jbmsec{(2) Then, we discuss the case of $\delta \neq 0.0$.
In that case, the $xp$ Hamiltonian should, in principle, consider a total number of electrons \jbmsec{per CuO$_2$ plane and unit cell and} \jbmt{and in}{within the} $xp$ subspace $N_e=5.0-\delta$ in order to be consistent with the starting electronic structure.
Nonetheless, we renormalize $N_e=5.0-\delta$ to the undoped value $N_e=5.0$.
Here, we explain why.
}

\jbmsec{For} Hg1201 \cite{Hirayama2019} and Ca11 \jbmsec{at $\delta=0.0$}, the ground state of the $xp$ Hamiltonian \jbmsec{has} antiferromagnetic order, whereas the $GW$ electronic structure is paramagnetic. 
As stated in Ref. \onlinecite{Hirayama2019}, the difference in the character of the ground state introduces the correction from the exchange splitting (which is not taken into account at the $GW$ level).
In order to obtain a comparable correction for \jbmsec{$\delta \neq 0.0$}, 
we must obtain a ground state with similar character (that is, antiferromagnetic).
However, \jbmsec{considering $N_e = 5.0 - \delta$ poses} a practical problem: The antiferromagnetic state is not the ground state anymore at \jbmsec{e.g. $N_e=4.8$}, so that we cannot obtain a correction \jbmt{for \jbmsectwo{Bi compounds} which is comparable to that for \jbmsectwo{Hg/Ca compounds}}{which is comparable to that at $\delta=0.0$}.
In order to avoid this, we consider the following refinement: 
\jbmsec{we renormalize $N_e = 5.0 - \delta$} to the undoped value \jbmsec{$N_e = 5.0$}, 
and recalculate the $GW$ occupation numbers and SIC \cite{Hirayama2015} \jbmsec{accordingly.}
\jbmsec{The only parameter which is modified with respect to the $xp$ Hamiltonian at $N_e = 5.0-\delta$ is the charge transfer energy $\Delta E_{xp}$.}
\jbmt{In addition, we neglect interlayer parameters for Bi2212 and Ca11 in the $xp$ Hamiltonian, and consider only parameters within a given CuO$_2$ plane.}{}

\subsection{\jbmsec{LRFB correction to improve the $GW$ electronic structure}}

\jbmsec{Then, we compute} the LRFB correction \cite{Hirayama2019} of the $GW$ electronic structure.
We take the three-orbital $xp$ Hamiltonian in Table \ref{tab:3band}.
\jbmt{}{We neglect interlayer parameters for Bi2212 and Ca11 in the $xp$ Hamiltonian, and consider only parameters within a given CuO$_2$ plane.}
We solve this $xp$ Hamiltonian with the \texttt{mVMC} code \cite{Misawa2014,Tahara2008,MISAWA2019447} as described in Ref. \onlinecite{Hirayama2019}, to deduce the LRFB correction $\Delta\mu$ of the charge transfer energy $\Delta E_{xp}$ between the $x$ and $p$ levels.
We obtain the \jbmsec{value of $\Delta \mu$ which allows}
to reproduce the $GW$ occupations\jbmsec{ ; the value is given in Table \ref{tab:3band}.}
This correction is used to improve the $GW$ electronic structure at the $GW$+LRFB level.
\jbmsec{
We start from $\Delta \mu$ in the basis of MLW orbitals, which is \jbmt{}{$\Delta \mu_i=0$ for $i=x$ and $\Delta \mu_i=\Delta \mu$ for $i=p$}.
We rotate $\Delta \mu$ to the basis of $(\epsilon_{nk}^{GW,{\rm dis}},\psi_{nk}^{GW,{\rm dis}})$, in which we modify Eq.~(\ref{eq:hcgw}) as
\begin{equation}
\jbmt{h^{cGW} = [h^{\rm KS} - V^{\rm xc} + \tilde{\Delta \Sigma}]Z^{cGW} + \Delta \mu.}
{h^{{\rm c}GW-{\rm SIC + LRFB}}_{mn}(k) = h^{{\rm c}GW-{\rm SIC}}_{mn}(k) + \Delta \mu_{mn}(k).}
\end{equation}
\jbmt{so that we include the effect of}{First, we take into account the effect of} $Z^{{\rm c}GW}$ in $\Delta \mu$ by considering the one-particle quantity
\begin{equation}
\jbmt{\tilde{\Delta \mu} = \Delta \mu/Z^{\rm cGW}}
{\tilde{\Delta \mu}_{mn}(k) = \Delta \mu_{mn}(k)/Z^{{\rm c}GW}_{mn}(k).}
\end{equation}
\jbmt{in the basis of \jbmt{eigenelements of the $xp$ subspace.}{$(\epsilon_{nk}^{GW,{\rm dis}},\psi_{nk}^{GW,{\rm dis}})$}.}{}
Then, we rotate $\tilde{\Delta \mu}$ to the basis of \jbmt{}{$(\epsilon_{nk}^{\rm KS,dis},\psi_{nk}^{\rm KS,dis})$}, and modify Eq.~(\ref{hgw}) \jbmt{}{in the basis of $(\epsilon_{nk}^{\rm KS,dis},\psi_{nk}^{\rm KS,dis})$,} as
\jbmt{}{\begin{align}
 h^{GW{\rm +LRFB}}_{mn}(k) = h^{GW}_{mn}(k) + \tilde{\Delta} \mu_{mn}(k)Z^{GW}_{mn}(k).
\label{eq:hgwlrfb}
\end{align}
}
Then, Eq.~(\ref{eq:hgwlrfb}) is diagonalized as in Appendix \ref{sec:macegw} to obtain the $GW$+LRFB electronic structure,
whose eigenelements are denoted as $(\epsilon_{nk}^{GW{\rm +LRFB}},\psi_{nk}^{GW{\rm+LRFB}})$.
We recompute the Fermi level, then shift bands in the H space so that their position with respect to the Fermi level remains unchanged.
We obtain the preprocessed Green's function at the $GW$+LRFB level:
\begin{equation}
G^{GW{\rm+LRFB}} = G^{GW{\rm+LRFB}}_{\rm M} + G^{\rm KS,dis}_{\rm H}.
\label{eq:ggw}
\end{equation}
}

%

\subsection{Single-orbital AB Hamiltonian and benchmark for Hg1201}
\label{app:benchmark}


\jbmsec{We then use the resulting $GW$+LRFB electronic structure as a starting point to derive the $\rm AB$ Hamiltonian at cRPA and c$GW$+LRFB levels; the cRPA allows to remove the double counting \jbmt{}{term} in the screening in the two-particle part, whereas the c$GW$ allows to remove the exchange and correlation double counting term in the one-particle part.
}
We restart from the $GW$+LRFB electronic structure, and construct one MLW orbital (two in the case of Bi2212), 
with atomic Cu $x$ character centered on Cu atoms.
The outer window is the M space from which we exclude the lowest bands (given in Sec.~\ref{sec:1band}), in order to avoid catching bonding character.
We minimize the spillage functional \cite{Souza2001} to extract the AB subspace, then minimize the spread functional \cite{Marzari1997} in the AB subspace to deduce the \jbmt{}{AB} MLW orbitals.
\jbmt{This yields the AB subspace, from which we disentangle other bands within the M space.}
{Then, we disentangle other bands within the outer window from the AB subspace.}
These other bands, together with \jbmt{}{the} unmodified lowest bands which were excluded from the outer window,
are denoted as $\tilde{N}$.
This yields the disentangled eigenelements  $(\epsilon_{nk}^{GW{\rm +LRFB(AB)}},\psi_{nk}^{GW{\rm+LRFB(AB)}})$, and the associated Green's function:
\begin{widetext}
\begin{equation}
G^{GW{\rm +LRFB(AB)}}_{} = G^{GW{\rm +LRFB(AB)}}_{\rm AB} + G^{GW{\rm +LRFB(AB)}}_{\rm \tilde{N}} + G^{\rm KS,dis}_{\rm H},
\end{equation}
\end{widetext}
which \jbmt{is denoted as}{represents} the \oldyy{``}$GW$+LRFB($\rm AB$)" electronic structure.
As an illustration, for Bi2201, we show the comparison between $GW$+LRFB and $GW$+LRFB($\rm AB$) band structures (restricted to M space) in \oldyy{the panel (a) of} Fig. \ref{fig:abb} \oldyy{of} Appendix~\ref{app:abb}.
Then, we 
\jbmt{start from the $GW$+LRFB(AB) electronic structure, and compute the two-particle part at the cRPA level and corresponding one-particle part at the c$GW$ level, denoted as c$GW$+LRFB(AB).}{compute $\chi^{\rm RPA} = \chi[G^{GW{\rm +LRFB(AB)}}_{}]$ and $\chi_{\rm AB} = \chi[G^{GW{\rm +LRFB(AB)}}_{\rm AB}]$, 
from which we deduce $\chi^{\rm cRPA} = \chi^{\rm RPA} - \chi_{\rm AB}$. 
We obtain $W^{\rm RPA}=W[v,\chi^{\rm RPA}]$ and $W^{\rm cRPA}=W[v,\chi^{\rm cRPA}]$, from which we deduce the two-particle part by using Eq.~(\ref{eq:h2p}).
Then, we compute $W_{\rm AB}=W[W^{\rm cRPA}(\omega=0),\chi^{\rm AB}]$ and the constrained self-energy $\Delta \Sigma = \Sigma[G^{GW{\rm +LRFB(AB)}}_{},W^{\rm RPA}] - \Sigma[G^{GW{\rm +LRFB(AB)}}_{\rm AB},W_{\rm AB}]$.
Finally, we deduce the c$GW$ one-particle part by using Eq.~(\ref{eq:hcgw}) and Eq.~(\ref{eq:h1p}).}
We do not include the SIC term \jbmt{}{in Eq.~(\ref{eq:hcgw})}, since it is only useful for multi-orbital Hamiltonians with nondegenerate energy levels.
Results are presented in Sec.~\ref{sec:1bandhgca} and Sec.~\ref{sec:1bandbi}.
\jbmsec{
Here, we discuss the benchmark of the AB Hamiltonian for Hg1201 with respect to the AE result \cite{Hirayama2019} at zero hole doping.
}

\begin{table}[!htb]
\centering
\begin{tabular}{cccccccc}
\hline
& $U^{}_{}$ & $V^{}_{1}$ & $V^{}_{2}$ & $V^{}_{3}$ & $V^{}_{4}$ & $V^{}_{5}$ & $V^{}_{6}$ \\
AE  & 3.846 & 0.834 & 0.460 & 0.318 & 0.271 & 0.209 & 0.233 \\
PP  & 4.029 & 0.900 & 0.520 & 0.379 & 0.329 & 0.265 & 0.146 \\
& $t^{l}_{0}$ & $t^{}_{1}$ & $t^{}_{2}$ & $t^{}_{3}$ & $t^{}_{4}$ & $t^{}_{5}$ & $t^{}_{6}$\\
AE & --- & -0.509 & 0.127 & -0.077 & 0.018 & 0.004 & -0.004 \\
PP & --- & -0.494 & 0.112 & -0.055 & 0.018 & 0.002 & -0.002 \\
\hline
\end{tabular}
\caption{
\jbmsec{
Effective Hamiltonian parameters for the AB Hamiltonian of Hg1201 at $\delta=0.0$, taken from Table \ref{tab:1nbandhgca}.
The complete list of parameters is given in Supplemental Material.
Our result is denoted as PP.
We also show the all-electron (AE) result from Ref. \onlinecite{Hirayama2019}, for comparison.
} 
}
\label{tab:1nband1}
\end{table}

\jbmsec{Results are summarized in Table \ref{tab:1nband1}.}
The two-particle part is close for both calculations: The difference in $U$ does not exceed $\sim 0.18$ eV (less than $5\%$), and the difference in $V_{n}$ does not exceed $\sim 0.08$ eV.
As for the one-particle part, the difference in hoppings does not exceed $\sim 0.02$ eV, although the  value of $|t_3|$ is underestimated.
The difference in values of $|U/t_1|$ is only $\sim 8\%$, and values of $|t_2/t_1|$, $V_1/U$ and $V_3/V_1$ are well reproduced.
Thus, the overall agreement is deemed acceptable.
\jbmt{}{
Nonetheless, we quickly discuss the possible origin of the small difference between our result and the all-electron result in Ref. \onlinecite{Hirayama2019}, for the AB Hamiltonian in Table \ref{tab:1nband1} but also the $xp$ Hamiltonian presented earlier in Table \ref{tab:3band}.
In addition to possible differences introduced by the pseudopotential approach used in our calculations,
we mention that, in the all-electron calculation, a slightly different methodology was used:
The constrained $GW$ self-energy is calculated by using a two-iteration scheme.
First, the renormalization of the low-energy subspace by bands outside the M space is calculated.
Second, the renormalization of the low-energy subspace by other bands inside the M space is calculated.
In our calculations, we do not consider this two-iteration procedure, which is complex and computationally expensive.
Instead, we calculate the renormalization of the low-energy subspace by all other bands directly.
}

\section{\jbmsec{Effect of hole doping on the quasiparticle electronic structure and final AB Hamiltonian}} 
\label{app:doping}

\jbmsec{
In Sec.~\ref{sec:1bandhgca}, we showed that $U/|t_1|$ decreases when hole doping increases, for both \jbmsectwo{Hg/Ca compounds}.
Here, we discuss the underlying causes of \misec{this effect}.
}

\begin{figure}
\includegraphics[scale=0.5]{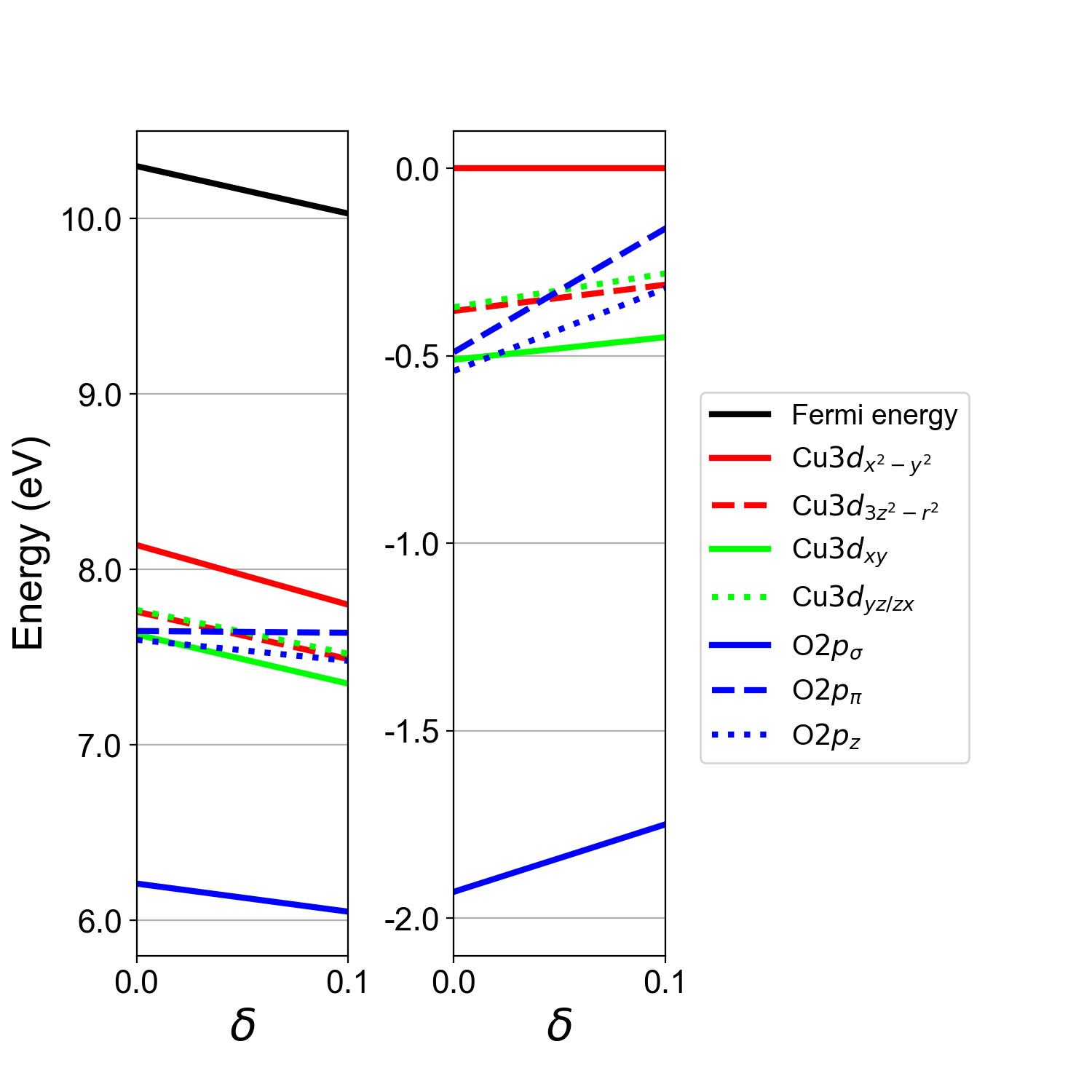}
\caption{
\jbmsec{Onsite energies of atomic-like MLW orbitals spanning the M space of Ca11 at KS level, for $\delta=0.0$ and $\delta=0.1$.
The left panel shows absolute onsite energies and the KS Fermi energy.
The right panel shows onsite energies with respect to the onsite energy for Cu$3d_{x^2-y^2}$ (that is, the charge transfer energies between Cu$3d_{x^2-y^2}$ and other orbitals \misec{if we reverse the sign to plus): For instance, the solid blue curve gives $-\Delta E_{xp}$, where $\Delta E_{xp}$ is the charge transfer energy  between Cu$3d_{x^2-y^2}$ and O$2p_{\sigma}$ orbitals.}
}
\jbmt{}{We note that the Cu$3d_{x^2-y^2}$ orbital has an onsite energy at $\sim 2.0-2.5$ eV below the Fermi level, 
because the Cu$3d_{x^2-y^2}$ orbital is atomic-like, and thus, has character in both the AB band (near the Fermi level) and the bonding bands (several eV below the Fermi level). This is illustrated by the partial density of states for the Cu$3d_{x^2-y^2}$ ("$x$") orbital of Bi2201 in the panel (c) of Fig.~\ref{fig:abb}.}
}
\label{fig:ca11-onsitee-doping}
\end{figure}

\jbmsec{
First, we discuss the starting electronic structure at KS level, as a function of hole doping.
When hole doping increases, we observe the following general trends in the KS electronic structure:
(1) The absolute (i.e. not renormalized with respect to the Fermi energy) onsite energies of Cu$3d$ orbitals are reduced ; 
(2) the value of the KS Fermi energy is reduced ; 
(3) the absolute onsite energies of O$2p$ orbitals are also reduced, but the reduction is weaker than that of Fermi energy and Cu$3d$ orbitals.
This is illustrated in the case of Ca11 in Fig.~\ref{fig:ca11-onsitee-doping} \jbmsectwo{(left panel)}: 
From $\delta=0.0$ to $\delta=0.1$, the onsite energies of Cu$3d$ orbitals and Fermi energy decrease concomitantly by \jbmt{$\sim 0.6$}{$\sim 0.35$} eV,
whereas the decrease in onsite energies of O$2p$ orbitals does not exceed \jbmt{$\sim 0.2$}{$\sim 0.15$} eV.
Also, we have checked that these trends are still valid in the case of Hg1201. 
}

\jbmsec{
Let us discuss \misec{underlying physics} of (1), (2) and (3) \misec{and possible mechanism of them}.}
\misec{It was shown that the doped hole primarily goes into the O2$p$ orbital~\cite{Hirayama2019}. 
Then the reduction of the Coulomb energy for the Cu3$d_{x^2-y^2}$ electrons \jbmt{are}{is} $-N_{\rm O}V_{dp}\delta/2$ for the hole concentration $\delta$ and the coordination number of O around Cu\jbmt{,}{} $N_{\rm O}=4$\jbmt{}{,} which leads to larger reduction than the reduction of the Coulomb energy for the O2$p_\sigma$ electron\jbmt{}{s} ($\sim U_p(2^2/2-(2-\delta/2)^2)\sim -2U_p\delta$), where $ V_{dp}$ and $U_p$ are estimated to be $\sim 2$ and $\sim 6$ eV, respectively\jbmt{}{, according to the values given in Table~\ref{tab:3band}}. Presumable partial hole doping to Cu3$d$ orbital makes the reduction of  Cu3$d_{x^2-y^2}$ orbital even larger. This is a very rough classical estimate but may capture the qualitative physics.} 

\jbmsec{
Now, we explain why (1) contributes to decrease the value of $U/|t_1|$.
As a consequence of (1), the charge transfer energies between O$2p$ and Cu$3d$ orbitals are reduced \jbmt{}{as seen in the right panel of Fig.~\ref{fig:ca11-onsitee-doping}}, which causes two distinct mechanisms.
(i) The \jbmt{}{O$2p$ $\rightarrow$ Cu$3d$} cRPA screening is stronger and the value of $R=U/v$ is smaller.
(ii) The bare interaction $v$ is reduced.
This is due to the delocalization of the AB MLW orbital, which is caused by the decrease in $\Delta E_{xp}$.
This mechanism also contributes to decrease $U$ (which partly depends on $v$) and increase $|t_1|$ (which is stronger when the AB orbital is delocalized).
The trends (i) and (ii) \misec{lead\jbmt{s}{} to} the decrease in $U/|t_1|$ when hole doping increases, and these trends are preserved at $GW$ level.
As an illustration, in the case of (ii), we show the value $\Delta E_{xp}^{GW}$ of $\Delta E_{xp}$ at $GW$ level in Table \ref{tab:deltae}, which decreases upon hole doping for both \jbmsectwo{Hg/Ca compounds}.}
\jbmsec{
However, at $GW$+LRFB level, complex effects \misec{arise from} the LRFB correction, as shown in  in Table \ref{tab:deltae}.
Indeed, we show the value $\Delta E_{xp}^{GW{\rm+LRFB}}$ of $\Delta E_{xp}$ at $GW$+LRFB level:
In the case of Ca11, (iii) $\Delta E_{xp}^{GW{\rm+LRFB}}$ is larger for $\delta=0.1$ than for $\delta=0.0$, which breaks the trend (ii).
This is caused by the $\sim 30\%$ larger value of $\Delta \mu$ at $\delta=0.0$ compared to $\delta=0.1$:
Indeed, a rough estimation of $\Delta E_{xp}^{GW{\rm+LRFB}}$ (which neglects the renormalization factor) is given by $\Delta E_{xp}^{GW}-\Delta \mu$,
and if $\Delta \mu$ is larger, then $\Delta E_{xp}^{GW{\rm+LRFB}}$ will be smaller.
The value of $\Delta \mu$ is obtained by solving the three-orbital $xp$ Hamiltonian, in which two parameters mainly control the output value of $\Delta \mu$:
(a) The first one is the value $\Delta E_{xp}^{{\rm c}GW-{\rm SIC}}$ of $\Delta E_{xp}$ at c$GW-$SIC level.
A larger value of $\Delta E_{xp}^{{\rm c}GW-{\rm SIC}}$ hinders fluctuations between $x$ and $p$ orbitals, so that a larger value of $\Delta \mu$ will be necessary to reproduce the occupation numbers at $GW$ level.
(b) The second one is $U_x-U_p$, which gives a rough estimation of the interaction energy cost to move an electron from a $p$ orbital to a $x$ orbital.
Similarly, a larger value of $U_x-U_p$ hinders fluctuations, which increases the output value of $\Delta \mu$.
And, we note that both $\Delta E_{xp}^{{\rm c}GW-{\rm SIC}}$ and $U_x-U_p$ have the largest values in the case of Ca11 at $\delta=0.0$, which explains the larger value of $\Delta \mu$ in that case.
}

\jbmsec{
In the previous paragraph, we have clarified the origin of (iii) the difference in $\Delta \mu$ for Ca11 at $\delta=0.0$, which alters the intuitive trends (i) and (ii) at $GW$+LRFB level compared to KS and $GW$ levels.
Interestingly, (iii) does not alter the fact that (iv) $U/|t_1|$ decreases when hole doping increases.
Furthermore, if we consider a starting electronic structure at the simpler KS or $GW$ level instead of the $GW$+LRFB level, (iv) is expected to remain valid due to (i) and (ii) and the absence of (iii).
Thus, (iv) is a robust trend since it remains valid for any level of sophistication of the starting electronic structure.
}

\begin{table}[!htb]
\begin{tabular}{lcccccc}
\hline
Compound & \ \ \ & \multicolumn{2}{c}{Hg1201} & \ \ \ & \multicolumn{2}{c}{Ca11}\\ 
\jbmsec{$\delta$}                                                           & \ \ \ & \jbmsec{0.0} & \jbmsec{0.1} & \ \ \ & \jbmsec{0.0} & \jbmsec{0.1}\\
\hline
\jbmsec{$\Delta E_{xp}^{GW}$}                                 & \ \ \ & \jbmsec{$1.89$} & \jbmsec{$1.69$} & \ \ \ & \jbmsec{$2.02$} & \jbmsec{$1.88$}\\
\jbmsec{$\Delta E_{xp}^{{\rm c}GW-{\rm SIC}}$} & \ \ \ & \jbmsec{$2.03$} & \jbmsec{$1.76$} & \ \ \ & \jbmsec{$2.36$} & \jbmsec{$2.10$}\\
\jbmsec{$U_x-U_p$}                                                      & \ \ \ & \jbmsec{$3.16$} & \jbmsec{$3.39$} & \ \ \ & \jbmsec{$3.42$} & \jbmsec{$3.12$}\\
\jbmsec{$\Delta \mu$}                                                 & \ \ \ & \jbmsec{$1.10$} & \jbmsec{$1.05$} & \ \ \ & \jbmsec{$1.50$} & \jbmsec{$1.15$}\\
\jbmsec{$\Delta E_{xp}^{GW{\rm+LRFB}}$}           & \ \ \ & \jbmsec{$1.15$} & \jbmsec{$0.88$} & \ \ \ & \jbmsec{$0.77$} & \jbmsec{$1.02$}\\
\jbmsec{$v$}                                                                   & \ \ \ & \jbmsec{$14.03$} & \jbmsec{$13.37$} & \ \ \ & \jbmsec{$13.97$} & \jbmsec{$14.20$}\\
\hline
\end{tabular}
\caption{
\jbmsec{For \jbmsectwo{Hg/Ca compounds} at $\delta=0.0$ and $\delta=0.1$, value $\Delta E_{xp}^{GW}$ of the charge transfer energy $\Delta E_{xp}$ between atomic $x$ and $p$ orbitals at $GW$ level,
value $\Delta E_{xp}^{{\rm c}GW-{\rm SIC}}$ of $\Delta E_{xp}$ in the $xp$ Hamiltonian at the c$GW\jbmt{}{-}$SIC level,
difference between $U_x$ and $U_p$ in the $xp$ Hamiltonian,
LRFB correction $\Delta \mu$,
value $\Delta E_{xp}^{GW{\rm+LRFB}}$ of $\Delta E_{xp}$ at $GW$+LRFB level,
and onsite bare interaction $v$ for the AB orbital constructed from the $GW$+LRFB electronic structure.}
}
\label{tab:deltae}
\end{table}

\section{\jbmsec{Experimental \misec{uncertainty} on atomic coordinates for Bi compounds}}
\label{app:bi}


\jbmsec{
Here, we discuss the experimental \misec{uncertainty} on atomic coordinates for Bi compounds,
and the subsequent \misec{uncertainty} on effective parameters in the AB Hamiltonian and especially $U/|t_1|$.
Indeed, we established in Sec.~\ref{sec:1bandbi} that $U/|t_1|$ is $\sim 13\%$ larger for Bi2212 with respect to Bi2201, 
but this result \misec{has some uncertainty ascribed to the experimental uncertainty of} the atomic coordinates.
We discuss the following results:
(1) The relative displacements of atoms along $x$ and $y$ directions due to the structural distortion do not affect $U/|t_1|$.
(2) If we take into account the \misec{uncertainty} range on $d^{z}_{\rm Oap}$ (as well as $d^{z}_{\rm buck}$ for Bi2212), $U/|t_1|$ is still at least $\sim 10\%$ larger for Bi2212 compared to Bi2201.
}
\\

\begin{table}
\begin{tabular}{llll}
\hline
      & \jbmsec{${\bf x}$} & \jbmsec{${\bf y}$} & \jbmsec{${\bf z}$}\\
\jbmsec{Cu} &  \jbmsec{0.0}  &  \jbmsec{0.0} & \jbmsec{0.0}\\
\jbmsec{O(in-CuO$_2$ plane)} &   \jbmsec{0.5} &  \jbmsec{0.0} & \jbmsec{0.0}\\
\jbmsec{O(in-CuO$_2$ plane)} &   \jbmsec{0.0} &  \jbmsec{0.5} & \jbmsec{0.0}\\
\jbmsec{Sr} &  \jbmsec{$0.5+d^{xy}_{\rm Sr}$} &  \jbmsec{$0.5+d^{xy}_{\rm Sr}$} & \jbmsec{$+d^{z}_{\rm Sr}$}\\
\jbmsec{Sr} &  \jbmsec{ $0.5-d^{xy}_{\rm Sr}$} & \jbmsec{$0.5-d^{xy}_{\rm Sr}$} & \jbmsec{$-d^{z}_{\rm Sr}$}\\
\jbmsec{O(apical)} &  \jbmsec{$+d^{xy}_{\rm Oap}$} &  \jbmsec{$+d^{xy}_{\rm Oap}$} & \jbmsec{$+d^{z}_{\rm Oap}$}\\
\jbmsec{O(apical)} &  \jbmsec{$-d^{xy}_{\rm Oap}$} & \jbmsec{$-d^{xy}_{\rm Oap}$} & \jbmsec{$-d^{z}_{\rm Oap}$}\\
\jbmsec{Bi} &  \jbmsec{$-d^{xy}_{\rm Bi}$} & \jbmsec{$-d^{xy}_{\rm Bi}$} & \jbmsec{$+d^{z}_{\rm Bi}$}\\
\jbmsec{Bi} &  \jbmsec{ $+d^{xy}_{\rm Bi}$} &  \jbmsec{$+d^{xy}_{\rm Bi}$} & \jbmsec{$-d^{z}_{\rm Bi}$}\\
\jbmsec{O(BiO layer)} &   \jbmsec{$0.5-d^{xy}_{\rm OBi}$} &  \jbmsec{$0.5-d^{xy}_{\rm OBi}$} & \jbmsec{$+d^{z}_{\rm OBi}$}\\
\jbmsec{O(BiO layer)} &   \jbmsec{$0.5+d^{xy}_{\rm OBi}$} &  \jbmsec{$0.5+d^{xy}_{\rm OBi}$} & \jbmsec{$-d^{z}_{\rm OBi}$}\\
\hline
\vspace{0.02cm}
\end{tabular}
\begin{tabular}{cccc}
\hline
\jbmsec{$d^{xy}_{\rm Sr}$} & \jbmsec{$0.008$~\AA}    & \jbmsec{$d^{z}_{\rm Sr}$} & \jbmsec{$1.748$~\AA}\\
\jbmsec{$d^{xy}_{\rm Oap}$} & \jbmsec{$0.091$~\AA} & \jbmsec{$d^{z}_{\rm Oap}$} & \jbmsec{$2.585$~\AA}\\
\jbmsec{$d^{xy}_{\rm Bi}$} & \jbmsec{$0.098$~\AA}     & \jbmsec{$d^{z}_{\rm Bi}$} & \jbmsec{$4.531$~\AA}\\
\jbmsec{$d^{xy}_{\rm OBi}$} & \jbmsec{$0.319$~\AA}  & \jbmsec{$d^{z}_{\rm OBi}$} & \jbmsec{$4.580$~\AA}\\
\hline
\end{tabular}
\caption{\jbmsec{
Atomic coordinates of atoms in the primitive cell of Bi2201, taken from Ref.~\onlinecite{Torardi1988}.
Upper panel shows Cartesian atomic coordinates ; 
for $i={\rm Sr},{\rm Oap},{\rm Bi},{\rm OBi}$, 
$d^{xy}_{i}$ is the average displacement of the corresponding atom along $x$ and $y$ directions (which is zero without distortion) with respect to ideal coordinates,
and $d^{z}_{i}$ is the distance between the atom and CuO$_2$ plane along $z$ direction.
For in-CuO$_2$ plane O atoms, Ref.~\onlinecite{Torardi1988} also reports a small displacement along $z$, which does not exceed $\sim 0.1$~\AA~and is neglected here to preserve the spatial inversion symmetry.
Lower panel shows values of $d^{xy}_{i}$ and $d^{z}_{i}$ in~\AA.
The lattice vectors in Cartesian coordinates are
${\bf a} = a{\bf x}$,
${\bf b} = a{\bf y}$,
and ${\bf c} = a/2({\bf x}+{\bf y}) + c^{z}{\bf z}$,
with $a=3.796$~\AA~ and $c^{z}=3.243\times a$\cite{Torardi1988}.
}
}
\label{tab:bi01atomic}
\end{table}

\begin{table}
\begin{tabular}{llll}
\hline
      & \jbmsec{${\bf x}$} & \jbmsec{${\bf y}$} & \jbmsec{${\bf z}$}\\
\jbmsec{Ca} &  \jbmsec{0.5} &  \jbmsec{0.5} & \jbmsec{0.0}\\
\jbmsec{Cu} &  \jbmsec{0.0} &  \jbmsec{0.0} & \jbmsec{$+d^{z}_{\rm Cu}$}\\
\jbmsec{O(in-CuO$_2$ plane)} &   \jbmsec{0.5} & \jbmsec{0.0} & \jbmsec{$+(d^{z}_{\rm Cu}-d^{z}_{\rm buck})$}\\
\jbmsec{O(in-CuO$_2$ plane)} &   \jbmsec{0.0} & \jbmsec{0.5} & \jbmsec{$+(d^{z}_{\rm Cu}-d^{z}_{\rm buck})$}\\
\jbmsec{Sr} &   \jbmsec{$0.5$} &  \jbmsec{$0.5$} & \jbmsec{$+(d^{z}_{\rm Cu}+d^{z}_{\rm Sr})$}\\
\jbmsec{O(apical)} &   \jbmsec{0.0} &  \jbmsec{0.0} & \jbmsec{$+(d^{z}_{\rm Cu}+d^{z}_{\rm Oap})$}\\
\jbmsec{Bi} &  \jbmsec{0.0} & \jbmsec{0.0} & \jbmsec{$+(d^{z}_{\rm Cu}+d^{z}_{\rm Bi})$}\\
\jbmsec{O(BiO layer)} &   \jbmsec{$0.5$} &  \jbmsec{$0.5$} & \jbmsec{$+(d^{z}_{\rm Cu}+d^{z}_{\rm OBi})$}\\
\hline
\vspace{0.02cm}
\end{tabular}
\begin{tabular}{cccccc}
\hline
\jbmsec{$d^{z}_{\rm Cu}$} & \jbmsec{$1.686$~\AA}     &  \jbmsec{$d^{z}_{\rm Oap}$} & \jbmsec{$2.300$~\AA} & \jbmsec{$d^{z}_{\rm Bi}$} & \jbmsec{$4.415$~\AA} \\
\jbmsec{$d^{z}_{\rm buck}$} & \jbmsec{$0.276$~\AA} & \jbmsec{$d^{z}_{\rm Sr}$} & \jbmsec{$1.318$~\AA} & \jbmsec{$d^{z}_{\rm OBi}$} & \jbmsec{$4.415$~\AA}\\
\hline
\end{tabular}
\caption{\jbmsec{Atomic coordinates of atoms in the primitive cell of Bi2212, taken from Ref.~\onlinecite{Torrance1988}.
Upper panel shows Cartesian atomic coordinates for \jbmt{}{Ca atom and} half of the \jbmt{}{Cu/O/Sr/Bi} atoms in the primitive cell ; 
for the other half, the coordinate along ${\bf z}$ is the opposite.
Lower panel shows values of $d^{z}_{i}$ in~\AA.
The lattice vectors in Cartesian coordinates are
${\bf a} = a{\bf x}$,
${\bf b} = a{\bf y}$,
and ${\bf c} = a/2({\bf x}+{\bf y}) + c^{z}{\bf z}$,
with $a=3.812$~\AA~ and $c^{z}=4.021\times a$\cite{Torrance1988}.
}}
\label{tab:bi12atomic}
\end{table}

\paragraph{\jbmsec{Experimental SC phase: Structural distortion and \misec{uncertainty} on atomic positions}}
\jbmsec{
First, in the case of Bi compounds, a structural distortion occurs in the SC phase \cite{Torardi1988,Ito1998,Beskrovnyi1990,Cicco1993,Levin1994}, 
which tilts BiO and SrO layers along $x$ and $y$ directions, represented respectively by ${\bf a}$ and ${\bf b}$ in Fig.~\ref{fig:crystal}. 
This is due a mismatch between BiO block layers and CuO$_2$ layers \cite{Petricek1990,Shamray2009}.
In particular, displacements of atoms along $x$ and $y$ directions create an incommensurate modulation,
which lowers the symmetry of the crystal and requires the use of a supercell.
This multiplies the number of atoms $N_{\rm at}$ in the unit cell with respect to the high-symmetry primitive cell.
As a consequence, the computational cost, which scales as at least $N_{\rm at}^2$, becomes untractable even for the PP implementation.
However, it is possible to restrict to the primitive cell, by averaging displacements along $x$ and $y$ in the primitive cell.
This allows to keep a reasonable computational cost within the PP implementation.
We use this approximation in the present paper ; the atomic coordinates for Bi2201 are shown in Table \ref{tab:bi01atomic}.
In the following, we discuss two points:
(b.) The effect of displacements along $x$ and $y$ on the AB Hamiltonian is minor,
but (c. and d.) the effect of the \misec{uncertainty} on atomic positions along $z$ is more important.
}
\\

\begin{figure*}[!htb]
\includegraphics[scale=0.52]{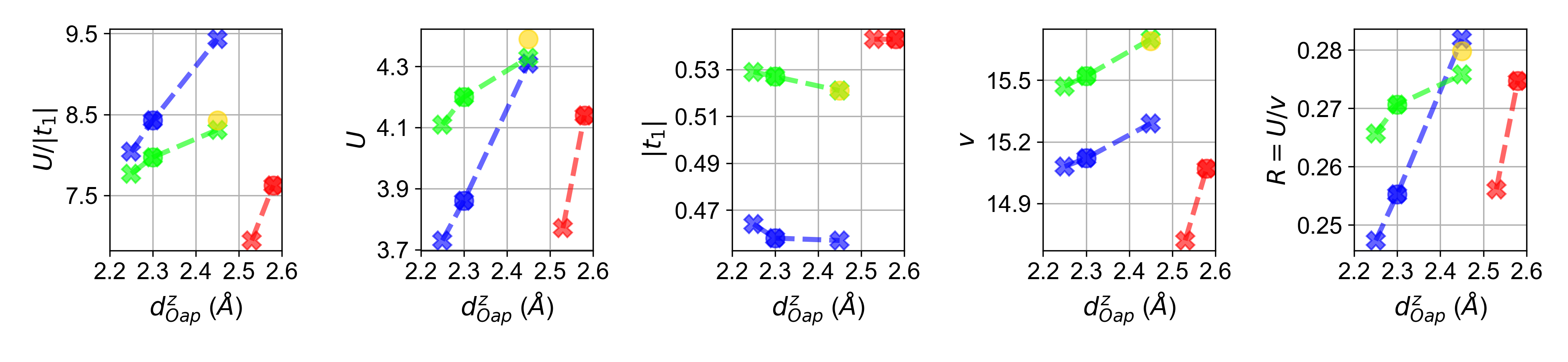}
\caption{
\jbmsec{
Values of $U/|t_1|$, $U$, $|t_1|$, onsite bare interaction $v$ and ES ratio $R=U/v$ as a function of $d^{z}_{\rm Oap}$ 
for Bi2201 (red), Bi2212 (blue), and Bi2212 with modified coordinates (green and yellow), at the GGA+cRPA level.
Red and blue dot markers represent calculations with experimental crystal structures 
in Table \ref{tab:bi01atomic} \cite{Torardi1988} for Bi2201 (red)
and in Table \ref{tab:bi12atomic} \cite{Torrance1988} for Bi2212 in which $d^{z}_{\rm buck}=0.27$ \AA~(blue).
Green and yellow dot markers consider the same crystal structure as blue for Bi2212,
but $d^{z}_{\rm buck}$ is artificially set to zero for the green marker,
and we consider $d^{z}_{\rm buck}=0.07$ \AA~and $d^{z}_{\rm Oap}=2.45$ \AA~as in Ref.~\onlinecite{Beskrovnyi1990} for the yellow marker.
Cross markers represent calculations with the same crystal structures as dot markers of the same color, 
except $d^{z}_{\rm Oap}$ which is artificially modified.
Dotted lines show extrapolations between markers of the same color. 
}
}
\label{fig:biapp}
\end{figure*}

\paragraph{\jbmsec{Effect of displacements $d^{xy}_{i}$ along $x$ and $y$ directions:}}
\jbmsec{
First, for Bi2201, we estimate the effect of displacements  $d^{xy}_{i}$ along $x$ and $y$. 
We consider the crystal structure in Table~\ref{tab:bi01atomic}
(i) without modification,
then (ii) by setting artificially $d^{xy}_{i}$ to zero,
in order to compare the AB Hamiltonian with (i) and without (ii) structural distortion.
For simplicity, we derive the AB Hamiltonian at the GGA+cRPA level:
The M space is not preprocessed from the KS level to the $GW$+LRFB level.
We start from the electronic structure at the GGA level and construct the AB MLW orbital ; 
the outer window is the M space minus the 7 lowest bands.
The one-particle part is left at the GGA level, 
and we compute the two-particle part at the cRPA level.
Results are shown in Table~\ref{tab:bidist}.
We obtain $U/|t_1|=7.62$ for (i) and $U/|t_1|=7.39$ for (ii), so that $U/|t_1|$ varies by only $3\%$.
In addition, we obtain the same value of $|t_1|=0.543$ eV for (i) and (ii).
Thus, the effect of $d^{xy}_{i}$ on the AB Hamiltonian is minor.
We note that, in (i), $d^{xy}_{\rm OBi} =0.32$~\AA~is relatively large but does not affect significantly the AB Hamiltonian.
This is because BiO block layers are well separated from CuO$_2$ planes, 
so that the distortion in BiO layer merely has a corrective effect on \jbmt{}{the} Bi$6p$ bands \jbmt{}{(located right above the AB bands in the panels (b) and (d) of Fig.~\ref{fig:band})} and the cRPA screening.
Similarly, other displacements $d^{xy}_{i} \lesssim 0.1$~\AA \ have a minor effect.
}
\begin{table}[!htb]
\centering
\begin{tabular}{cccccccc}
\hline
& \jbmsec{$U^{}_{}$} & \jbmsec{$V^{}_{1}$} & \jbmsec{$V^{}_{2}$} & \jbmsec{$V^{}_{3}$} & \jbmsec{$V^{}_{4}$} & \jbmsec{$V^{}_{5}$} & \jbmsec{$V^{}_{6}$} \\
\jbmsec{(i)} & \jbmsec{4.138} & \jbmsec{0.825} & \jbmsec{0.434} & \jbmsec{0.296} & \jbmsec{0.245} & \jbmsec{0.192} & \jbmsec{0.105} \\
\jbmsec{(ii)} & \jbmsec{4.012} & \jbmsec{0.771} & \jbmsec{0.390} & \jbmsec{0.260} & \jbmsec{0.213} & \jbmsec{0.166} & \jbmsec{0.090} \\
& \jbmsec{$t^{l}_{0}$} & \jbmsec{$t^{}_{1}$} & \jbmsec{$t^{}_{2}$} & \jbmsec{$t^{}_{3}$} & \jbmsec{$t^{}_{4}$} & \jbmsec{$t^{}_{5}$} & \jbmsec{$t^{}_{6}$}\\
\jbmsec{(i)} & --- & \jbmsec{-0.543} & \jbmsec{0.093} & \jbmsec{-0.073} & \jbmsec{0.004} & \jbmsec{0.001} & \jbmsec{-0.013} \\
\jbmsec{(ii}) & --- & \jbmsec{-0.543} & \jbmsec{0.100} & \jbmsec{-0.078} & \jbmsec{0.004} & \jbmsec{-0.001} & \jbmsec{-0.013} \\
\hline		
\end{tabular}
\caption{\jbmsec{Effective AB Hamiltonian parameters for Bi2201, both with (i) and without (ii) structural distortion along $x$ and $y$ directions.
The only difference between both calculations are the values of $d^{xy}_{i}$ in atomic coordinates from Table \ref{tab:bi01atomic}, which are unmodified in the case of (i) and set to zero in the case of (ii).
}}
\label{tab:bidist}
\end{table}

\paragraph{\jbmsec{\misec{Uncertainty} on distance $d^{z}_{\rm Oap}$ between Cu and apical O:}}
\label{par:dcuoap}
\jbmsec{
Then, we estimate the effect of the \misec{uncertainty} on $d^{z}_{\rm Oap}$.
Calculations in Sec.~\ref{sec:1bandbi} consider crystal structures 
in Tables~\ref{tab:bi01atomic} and \ref{tab:bi12atomic}, in which 
$d^{z}_{\rm Oap}=2.58$~\AA~for Bi2201 \cite{Torardi1988},
and $d^{z}_{\rm Oap}=2.30$~\AA~for Bi2212 \cite{Torrance1988}.
However, other experimental studies on the SC phase of \jbmsectwo{Bi compounds}
report different values of $d^{z}_{\rm Oap}$,
from $\sim 2.26$~\AA~to $\sim 2.60$~\AA~for Bi2201 \cite{Torrance1988,Torardi1988,Ito1998,Schlogl1993}
and from \misec{$\sim 2.25$~\AA~to $\sim 2.50$~\AA} for Bi2212 \cite{Beskrovnyi1990,Cicco1993}.
In order to estimate the effect of the \misec{uncertainty} on $d^{z}_{\rm Oap}$, 
we perform additional GGA+cRPA calculations by 
modifying artificially the value of $d^{z}_{\rm Oap}$ in Tables~\ref{tab:bi01atomic} and \ref{tab:bi12atomic} ; 
other parameters are not modified.
In the case of Bi2201, we consider $d^{z}_{\rm Oap}=2.53$~\AA ; 
In the case of Bi2212, 
we study the range $d^{z}_{\rm Oap}=2.25$~\AA~and $d^{z}_{\rm Oap}=2.45$~\AA. 
Results are shown in Fig.~\ref{fig:biapp}.
We observe a general trend: 
$U/|t_1|$, $U$, $v$ and $R$ increase with $d^{z}_{\rm Oap}$. 
In the case of $R$, a linear extrapolation of the red dotted curve down to $d^{z}_{\rm Oap} \sim 2.45$~\AA~would suggest that \textit{at equal values of $d^{z}_{\rm Oap}$}, $R$ increases (and thus, the ES decreases) when $N_{\rm Oap}$ decreases.
Possible causes are the following: 
(i) The value of $d^{z}_{\rm Oap}$ controls the position of Bi$6p$ bands with respect to the Fermi level.
A larger value of $d^{z}_{\rm Oap}$ will push the apical O atoms closer to the BiO layer, so that the negative Madelung potential from apical O anions felt by Bi$6p$ electrons will be stronger.
As a consequence, Bi$6p$ bands are destabilized and pushed farther from Fermi level, 
which may contribute to decrease the ES.
Conversely, reducing $d^{z}_{\rm Oap}$ stabilizes Bi$6p$ bands.
In addition, (ii) if $d^{z}_{\rm Oap}$ increases, the negative Madelung potential from apical O anions felt by electrons within CuO$_2$ planes will be weaker.
This stabilizes M bands, which increases the energy difference between M bands and empty bands, so that the ES decreases.
As a result, $U/|t_1|$ increases.
We note that, in the case of Bi2201, the variation in $U/|t_1|$ with $d^{z}_{\rm Oap}$ is around twice larger than that for Bi2212.
This might be related to the fact that $N_{\rm Oap}=2$ in Bi2212 whereas $N_{\rm Oap}=1$ in Bi2201.
}
\\

\paragraph{\jbmsec{\misec{Uncertainty} on amplitude $d^{z}_{\rm buck}$ of the buckling of in-plane Cu-O-Cu bonds \jbmt{}{in Bi2212}:}}
\jbmsec{
As for the amplitude $d^{z}_{\rm buck}$ of the buckling of  in-plane Cu-O-Cu bonds \jbmt{}{in Bi2212},
we consider $d^{z}_{\rm buck}=0.27$~\AA~\cite{Torrance1988} in Table \ref{tab:bi12atomic}.
However, other experimental studies \cite{Beskrovnyi1990,Levin1994} report smaller values of $d^{z}_{\rm buck}$.
For instance, in the case of Ref.~\onlinecite{Beskrovnyi1990}, we have $d^{z}_{\rm buck}=0.07$~\AA.
Thus, we estimate the effect of the \misec{uncertainty} on $d^{z}_{\rm buck}$ for Bi2212.
We start from the structure  in Table \ref{tab:bi12atomic}.
We set $d^{z}_{\rm buck}$ to zero, and perform GGA+cRPA calculations at the unmodified value $d^{z}_{\rm Oap}=2.30$~\AA~
but also at $d^{z}_{\rm Oap}=2.25$~\AA~and $d^{z}_{\rm Oap}=2.45$~\AA~;
other parameters are not modified.
Results are shown in Fig.~\ref{fig:biapp}.
We observe the following trend:
Removing the buckling decreases $U/|t_1|$. 
This is mainly due to the increase in $|t_1|$: Hoppings along $x$ and $y$ are not cut off by the buckling anymore.
We also note that $v$ increases, which contributes to increase $U$. 
Indeed, the buckling causes a delocalization along the $z$ direction, as seen in Fig.~\ref{fig:crystal}.
This delocalization along $z$ is suppressed when the buckling is removed.
}

\jbmsec{
Finally, we refine the comparison with Ref.~\onlinecite{Beskrovnyi1990} by considering the same values of $d^{z}_{\rm Oap}=2.45$~\AA~and $d^{z}_{\rm buck}=0.07$~\AA.
Result is shown in Fig.~\ref{fig:biapp}.
In that case, the value of $U/|t_1| \sim 8.43$ is identical to that from
$d^{z}_{\rm Oap}=2.30$~\AA~and $d^{z}_{\rm buck}=0.27$~\AA~\cite{Torrance1988}.
That is, the decrease in $d^{z}_{\rm buck}$ (which decreases $U/|t_1|$) is compensated by the increase in $d^{z}_{\rm Oap}$ (which increases $U/|t_1|$).
}
\\

\paragraph{\jbmsec{Summary and comparison of \jbmsectwo{Bi compounds}:}}
\jbmsec{
Without taking into account the  \misec{uncertainty} on $d^{z}_{\rm Oap}$ and $d^{z}_{\rm buck}$,
$U/|t_1|$ is $\gtrsim 10\%$ larger for Bi2212 with respect to Bi2201.
If we take into account the \misec{uncertainty} on $d^{z}_{\rm Oap}$ and $d^{z}_{\rm buck}$,
a reasonable choice is to consider the calculation for Bi2212 with 
$d^{z}_{\rm Oap}=2.45$~\AA~
and $d^{z}_{\rm buck}=0.07$~\AA~\cite{Beskrovnyi1990}.
In that case, $U/|t_1|$ does not change for Bi2212,
so that $U/|t_1|$ is still $\gtrsim 10\%$ larger for Bi2212.
In addition, $U/|t_1|$ in Bi2201 may be overestimated if we consider $d^{z}_{\rm Oap}=2.58$~\AA~\cite{Torardi1988},
so that $U/|t_1|$ is \textit{at least} $\sim 10\%$ larger for Bi2212.
}

\section{Three-orbital Hamiltonian \misec{in} comparison with $\rm AB$ Hamiltonian}
\label{app:abb}

Here, prior to the derivation of the single-orbital $\rm AB$ Hamiltonian in Sec. \ref{sec:1band}, we \jbmsec{discuss the restriction of}
the LEH to the single-orbital picture.
To do so, we extend the $\rm AB$ Hamiltonian to a three-orbital Hamiltonian, which includes $\rm B$ orbitals in addition.
We detail the case of Bi2201, then give the final three-orbital Hamiltonian for all compounds.
\\

\paragraph{Nonsuitability of the $xp$ Hamiltonian to \jbmsec{discuss the} restriction to the $\rm AB$ Hamiltonian ---}

\begin{figure*}[!htb]
\includegraphics[scale=0.75]{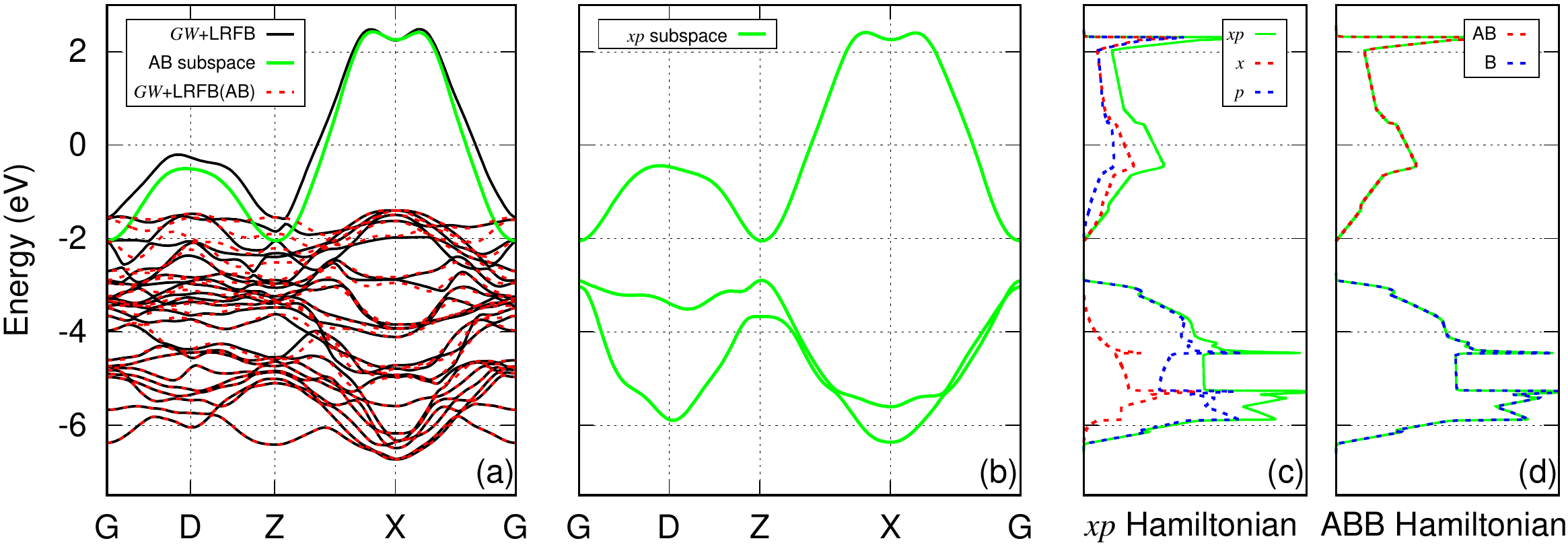}
\caption{
Band structure of Bi2201, restricted to M space, for several Hamiltonians.
\oldyy{The p}anel (a) \jbm{shows,} for the $\rm AB$ Hamiltonian, \jbm{the} band structure of the $\rm AB$ subspace (in green\jbmsec{)}
obtained from the $GW$+LRFB band structure (in black; equal to the dashed \jbmsec{red} band structure in Fig. \ref{fig:band}, panel (b)) after minimization of spillage functional.
We also represent the $GW$+LRFB($\rm AB$) band structure (in red), which includes the $\rm AB$ subspace and the 22 other bands within M space, modified by disentanglement. 
Lowest 7 bands are excluded from the outer window in the $\rm AB$ Hamiltonian, and thus, are not modified.
\oldyy{The p}anel (b) \jbm{shows the} band structure of the three-orbital ($xp$) subspace obtained from the $GW$+LRFB electronic structure, after minimization of the spillage functional. 
We stress that this subspace is rigorously the same within the $xp$ and $\rm AB\rm B$ Hamiltonians.
\oldyy{The p}anels (c) \oldyy{and} (d) \jbm{show the} total density of states for the $xp$ subspace, and partial densities of states \jbmb{for orbitals} within the $xp$ and $\rm AB\rm B$ Hamiltonians.
}
\label{fig:abb}
\end{figure*}

In Appendix~\ref{app:mace}, we derived the $xp$ Hamiltonian by starting from the $GW$ electronic structure.
Here, we quickly discuss the $xp$ Hamiltonian from the $GW$+LRFB electronic structure (same starting point as the $\rm AB$ Hamiltonian).
For Bi2201, we show the occupation numbers of \jbmsec{MLW} orbitals and intra-orbital bare interaction in \oldyy{the row \jbmb{(b)} of} Table \ref{tab:wanabb}.
The band dispersion and partial densities of states are shown in
\oldyy{the panels (b) and (c) of} Fig. \ref{fig:abb}.
These results reveal that the $xp$ Hamiltonian is not comparable to the $\rm AB$ Hamiltonian\footnote{
Furthermore, the modification of $n_x$ and $n_p$ between
\oldyy{the rows }(a) and \jbmb{(b)} of Table \ref{tab:wanabb} prevents us from performing the SIC in the derivation of the LEH.
Indeed, the self-interaction is contained within the Hartree potential at the Kohn-Sham level, so that the SIC must consider $n_x$ and $n_z$ at the Kohn-Sham level \cite{Hirayama2015}.
Values of $n_x$ and $n_z$ at the $GW$ level are close to those at the Kohn-Sham level (the difference is typically $\sim 1\%$), so that we may use them, as done in Appendix~\ref{app:mace}.
On the other hand, $n_x$ and $n_z$ are modified in \oldyy{the row} \jbmb{(b)}, which makes the SIC wrong.
}.
In fact, the $x$ orbital in the $xp$ Hamiltonian has different character from the $\rm AB$ orbital in the $\rm AB$ Hamiltonian.
In the $xp$ Hamiltonian, both $x$ and $p$ orbitals have partial density of states in both AB and B bands, due to the strong $x/p$ hybridization.
The signature of this mixing is that $p$ orbitals are not full ($n_p \sim 1.7$ instead of $2.0$).
In particular, the $x$ orbital ($n_p \sim 1.4$)
\oldyy{does not have the} purely AB character, so that it is not comparable to the $\rm AB$ orbital.

\begin{table}
\resizebox{\columnwidth}{!}{
\begin{tabular}{cccccccc}
\hline
& Starting point	& Hamiltonian & $N_e$ & $n_{x}$ & $n_{p}$  & $v_{x}$ & $v_{p}$ \\
(a) & $GW$ 		& $xp$ 				 		                        & 4.80  & \jbmsec{1.450} & \jbmsec{1.775} & \jbmsec{25.55} & \jbmsec{17.72} \\
\jbmb{(b)} & $GW$+LRFB	 &$xp$ 				 		& 4.80  & \jbmsec{1.505} & \jbmsec{1.7475}  & \jbmsec{25.55} & \jbmsec{17.72} \\
& 	&  &  \jbmb{$N_e$} & \jbmb{$n_{\rm AB}$} & \jbmb{$n_{\rm B}$}  & \jbmb{$v_{\rm AB}$} & \jbmb{$v_{\rm B}$} \\
\jbmb{(c)} & $GW$+LRFB	 &$\rm AB$ 					& 0.80  & 0.80 & ---  	& \jbmsec{14.82} & ---  \\
(d) & $GW$+LRFB 	&$\rm AB\rm B$ 	 		            & 4.80  & 0.80 & 2.00  & \jbmsec{13.41} & \jbmsec{14.21} \\
(e) & $GW$+LRFB($\rm AB$) &$\rm AB\rm B$ 	& 4.80  & 0.80 & 2.00	 & \jbmsec{14.77} & \jbmsec{14.76} \\
\hline
\end{tabular}
}
\caption{Characteristics of MLW orbitals for Bi2201, within $\jbmb{xp}$, $\rm AB$ and $\rm AB\rm B$ Hamiltonians. 
We give the total number of electrons in correlated subspace, MLW orbital occupation numbers $n_{i}$ and intra-orbital bare interaction $v_{i}$.
We also give the starting point (electronic structure before construction of MLW orbitals and disentanglement).
In the row (a), we remind the result for $xp$ Hamiltonian from $GW$ electronic structure (Table \ref{tab:3band}).
In the row (b), we show the result for $xp$ Hamiltonian from $GW$+LRFB electronic structure.
In the row (c), we remind the result for $\rm AB$ Hamiltonian from $GW$+LRFB electronic structure.
In the row (d), we show the result for $\rm AB\rm B$ Hamiltonian from $GW$+LRFB electronic structure.
In the row (e), we show the result for $\rm AB\rm B$ Hamiltonian from $GW$+LRFB($\rm AB$) electronic structure (see \oldyy{the panel (a) of} Fig. \ref{fig:abb}).
}
\label{tab:wanabb}
\end{table}

Let us discuss the nature of the $xp$ Hamiltonian in more details.
The $xp$ Hamiltonian is useful for the LRFB correction and improvement of the starting electronic structure beyond the quasiparticle $GW$ approximation, for a more accurate derivation of the LEH.
In the $xp$ Hamiltonian, the $p$ orbitals have some character within the $\rm AB$-like band at the Fermi level (as revealed by the partial density of states and occupation number); this partial AB character is responsible for the role of $p$ orbitals in the low-energy physics of the $xp$ Hamiltonian.
Still, it does not say anything about the importance of bands other than the $\rm AB$-like band (namely, the two $\rm B$-like bands) for the low-energy physics.
As a result, it does not say anything about the necessity to include more than one band (and thus, one orbital) in the LEH.
\\

\begin{table*}[!htb]
\begin{tabular}{ccccccccccccccccccccc}
\hline
	& \ \ \ \ \ & & \multicolumn{3}{c}{$\rm AB$ Hamiltonian} & \ \ \ \ \ &  \multicolumn{13}{c}{$\rm AB\rm B$ Hamiltonian}  \\
	& \ \ \ \ \ & \jbmsec{$\delta$} & $N_e=n_{\rm AB}$ & $v_{\rm AB}$ & $U_{\rm AB}$ & \ \ \ \ \ & $N_e$ & $n_{\rm AB}$ & $n_{\rm B}$ & $v_{\rm AB}$ & $v_{\rm B}$ & $U_{\rm AB}$ & $U_{\rm B}$ & $R_{\rm AB}$ & $R_{\rm B}$ & $\Delta E_{\rm ABB}$  & $\Delta E^{\rm min}_{\rm ABB}$ & $\mathcal{L}_{\rm B}$ & $\epsilon^{\rm maxH}_{\rm B}$ \\ 
\hline
Hg1201	& & \jbmsec{0.0} &  1.0 & 14.03  & 4.03  			& \ \ \ \ \  				& 5.0  					& 1.0  & 2.0 & 13.97 & 14.06 & 4.42 & 4.46 & 0.32 & 0.32 & 7.47 & 4.5 & 0.99 & -4.1 \\ 
Ca11		& & \jbmsec{0.0} &  1.0 & 13.97  & 4.48  			& \ \ \ \ \  				& 5.0  					& 1.0  & 2.0 & 13.83 & 14.26 & 5.01 & 5.15 & 0.36 & 0.36 & 8.00 & 4.8 & 1.06 & -4.3 \\ 
\hline
\jbmsec{Hg1201}	& & \jbmsec{0.1} & \jbmsec{0.9} & \jbmsec{13.37} & \jbmsec{4.00} 	& \ \ \ \ \  & \jbmsec{4.9}  	& \jbmsec{0.9}  & \jbmsec{2.0} & \jbmsec{13.30} & \jbmsec{13.76} & \jbmsec{4.38} & \jbmsec{4.52} & \jbmsec{0.33} & \jbmsec{0.33} & \jbmsec{7.71} & \jbmsec{4.7} & \jbmsec{0.95} & \jbmsec{-4.2} \\
\jbmsec{Ca11}	& & \jbmsec{0.1} & \jbmsec{0.9} & \jbmsec{14.20}  & \jbmsec{4.22}  	& \ \ \ \ \  & \jbmsec{4.9}  	& \jbmsec{0.9}  & \jbmsec{2.0}     & \jbmsec{14.10} & \jbmsec{14.20} & \jbmsec{4.70} & \jbmsec{4.80} & \jbmsec{0.33} & \jbmsec{0.34} & \jbmsec{7.96} & \jbmsec{4.6} & \jbmsec{1.03} & \jbmsec{-4.3} \\ 
\hline
\jbmsec{Bi2201}	& & \jbmsec{0.2} & \jbmsec{0.8} & \jbmsec{14.82}  & \jbmsec{4.39}  	& \ \ \ \ \  & \jbmsec{4.8}  & \jbmsec{0.8}  & \jbmsec{2.0}     & \jbmsec{14.77} & \jbmsec{14.76} & \jbmsec{5.20} & \jbmsec{5.22} & \jbmsec{0.35} & \jbmsec{0.35} &  \jbmsec{8.82} & \jbmsec{5.6} & \jbmsec{0.93} & \jbmsec{-4.6} \\
\jbmsec{Bi2212}	& & \jbmsec{0.2} & \jbmsec{0.8} & \jbmsec{14.76}  & \jbmsec{4.23}  	& \ \ \ \ \  & \jbmsec{4.8}  & \jbmsec{0.8}  & \jbmsec{2.0}     & \jbmsec{14.87} & \jbmsec{14.69} & \jbmsec{5.02} & \jbmsec{4.99} & \jbmsec{0.34} & \jbmsec{0.34} &  \jbmsec{8.68} & \jbmsec{5.3} & \jbmsec{0.94} & \jbmsec{-4.6} \\ 
\hline
\end{tabular}
\caption{
Three-orbital $\rm ABB$ Hamiltonian\jbm{. We give the} 
total number $N_e$ of electrons \jbmsec{\jbmsec{per Cu atom and two in-plane O atoms}} within $\rm AB\rm B$ subspace, $GW$+LRFB occupation numbers $n_{i}$, intra-orbital effective interactions $U_{i}$ and bare interactions $v_{i}$, \jbmt{}{c$GW$+LRFB} screening ratio $R_i = U_i / v_i$ for $i=\rm AB,\rm B$, and charge transfer energy $\Delta E_{\rm ABB}$. We also give estimations of the minimal effective value of the charge transfer energy $\Delta E^{\rm min}_{\rm ABB}$, the ratio $\mathcal{L}_{\rm B}=(U_{\rm AB}+U_{\rm B})/2\Delta E^{\rm min}_{\rm ABB}$, and the maximal energy of the upper Hubbard band for $\rm B$ manifold with respect to the Fermi level $\epsilon^{\rm maxH}_{\rm B} = \epsilon^{\rm max}_{\rm B} + U_{\rm B}/2$ .
}
\label{tab:xabpb}
\end{table*}

\paragraph{AB/B transformation\jbm{: F}rom $xp$ Hamiltonian to \jbmb{ABB} Hamiltonian ---}

In order to compare the $\rm AB$ and three-orbital Hamiltonians, we set the condition as follows\jbm{:} 
\textit{\jbm{T}he $\rm AB$ orbital must be included in the three-orbital Hamiltonian}.
This can be achieved by considering the gauge
\oldyy{degrees of freedom}
in the construction of \jbmsec{MLW} orbitals.
Once the spillage functional has been minimized \cite{Souza2001} to extract the $xp$ subspace (band dispersion in \oldyy{the panel (b) of} Fig. \ref{fig:abb}), 
we still have to choose the unitary transformation $\mathcal{U}$ within this subspace, which yields the \jbmsec{MLW} orbitals.

In the $xp$ Hamiltonian, we chose $\mathcal{U}$ which minimizes the spread functional on the $xp$ subspace.
Now, we impose the following transformation, denoted as AB/B~\cite{Hirayama2022silverarxiv}\footnote{\jbm{The AB/B transformation has been first proposed  by Hirayama \textit{et al}. \cite{Hirayama2022silverarxiv}.
}}.
\jbm{S}tarting from the $xp$ subspace, we consider the AB band (which spans the $\rm AB$ subspace) and the two other bands ($\rm B$ subspace) separately.
We project again the initial guess for the $x$ orbital on the $\rm AB$ subspace, and the initial guesses for $p$ orbitals on the $\rm B$ subspace.
Then, we minimize the spread functional separately within each subspace; we obtain $\mathcal{U}$ on each subspace, yielding one MLW orbital \jbmb{in AB subspace} and two MLW orbitals \jbmb{in B subspace}.
This Hamiltonian is denoted as $\rm AB\rm B$.
For Bi2201, partial densities of states within the $\rm AB\rm B$ Hamiltonian are shown in \oldyy{the panel (d) of} Fig. \ref{fig:abb}; occupation numbers and intra-orbital bare interaction are shown in \oldyy{the row (d) of} Table \ref{tab:wanabb}.
\jbmb{
Although the total density of states is the same as in the $xp$ Hamiltonian (since the correlated subspace obtained after spillage minimization is the same), we successfully isolate the AB orbital.
This is revealed by the occupation number ($n_{\rm AB} = 0.8$ is the same as in the $\rm AB$ Hamiltonian) and the partial density of states (the $\jbmb{{\rm AB}}$ orbital is entirely contained within the $\rm AB$-like band).
Similarly, we isolate the two B orbitals, which are entirely contained within the two B-like bands.
}

Still, this $\rm AB\rm B$ Hamiltonian (\oldyy{the row (d) of} Table \ref{tab:wanabb}) is not completely suitable for comparison with the $\rm AB$ Hamiltonian (\oldyy{the row \jbmb{(c)} of} Table \ref{tab:wanabb})\jbmb{.}
Indeed, \jbmb{the onsite bare interaction for the AB orbital is} $v_{\rm AB} \sim \jbmsec{13.4}$ eV\jbmb{, which} is underestimated with respect to the $\rm AB$ Hamiltonian ($v_{\rm AB} \sim \jbmsec{14.8}$ eV), so that the $\rm AB$ MLW orbitals in both Hamiltonians are not completely equivalent.
This comes from the difference in the $\rm AB$ subspace, encoded within the band dispersion of the $\rm AB$ band shown in \oldyy{the panels (a) and (b) of Fig. \ref{fig:abb}, where the band dispersions} are similar, but not identical.
In order to solve the discrepancy in $v_{\rm AB}$, we propose the following scheme.
Instead of starting from the $GW$+LRFB electronic structure, we start from the $GW$+LRFB($\rm AB$) electronic structure in \oldyy{the panel (a) of} Fig. \ref{fig:abb}\jbmb{. Then, we set the initial guesses,} minimize the spillage functional, extract the $xp$ subspace, and \jbmb{apply the AB/B transformation to} calculate MLW orbitals for the $\rm AB\rm B$ Hamiltonian as described previously.
Results are given in \oldyy{the row (e) of} Table \ref{tab:wanabb}.
Now, $v_{\rm AB}$ is very close to the value for the  $\rm AB$ Hamiltonian.
This improvement comes from the fact that the $\rm AB$ subspace within the $\rm AB$ Hamiltonian is already disentangled from other bands within the $GW$+LRFB($\rm AB$) electronic structure: The spillage minimization procedure is able to pinpoint and mimick this subspace.
Now, the $\rm AB$ MLW orbitals in both $\rm AB$ and $\rm AB\rm B$ Hamiltonians are completely equivalent.

\paragraph{Results for the $\rm AB\rm B$ Hamiltonian ---}
Now, we derive the $\rm AB\rm B$ Hamiltonian for all compounds, by considering the following procedure. 
We start from the $GW$+LRFB($\rm AB$) electronic structure, calculate the $\rm AB$ and $\rm B$ MLW orbitals as described previously, and disentangle the rest of the M space from the $\rm AB\rm B$ subspace. These disentangled bands, together with the $\rm AB\rm B$ subspace and other bands outside the M space (left at the KS level), constitute the $GW$+LRFB($\rm AB\rm B$) electronic structure.
We start from the latter, and calculate the two-particle part at the cRPA level and one-particle part at the c$GW$\jbmt{}{$-$}SIC level.
Importantly, the occupation numbers $n_{\rm AB}$ and $n_{\rm B}$ of MLW orbitals are strictly the same as at the KS level, so that it is possible to perform the SIC.

\begin{figure}[!htb]
\includegraphics[scale=0.35]{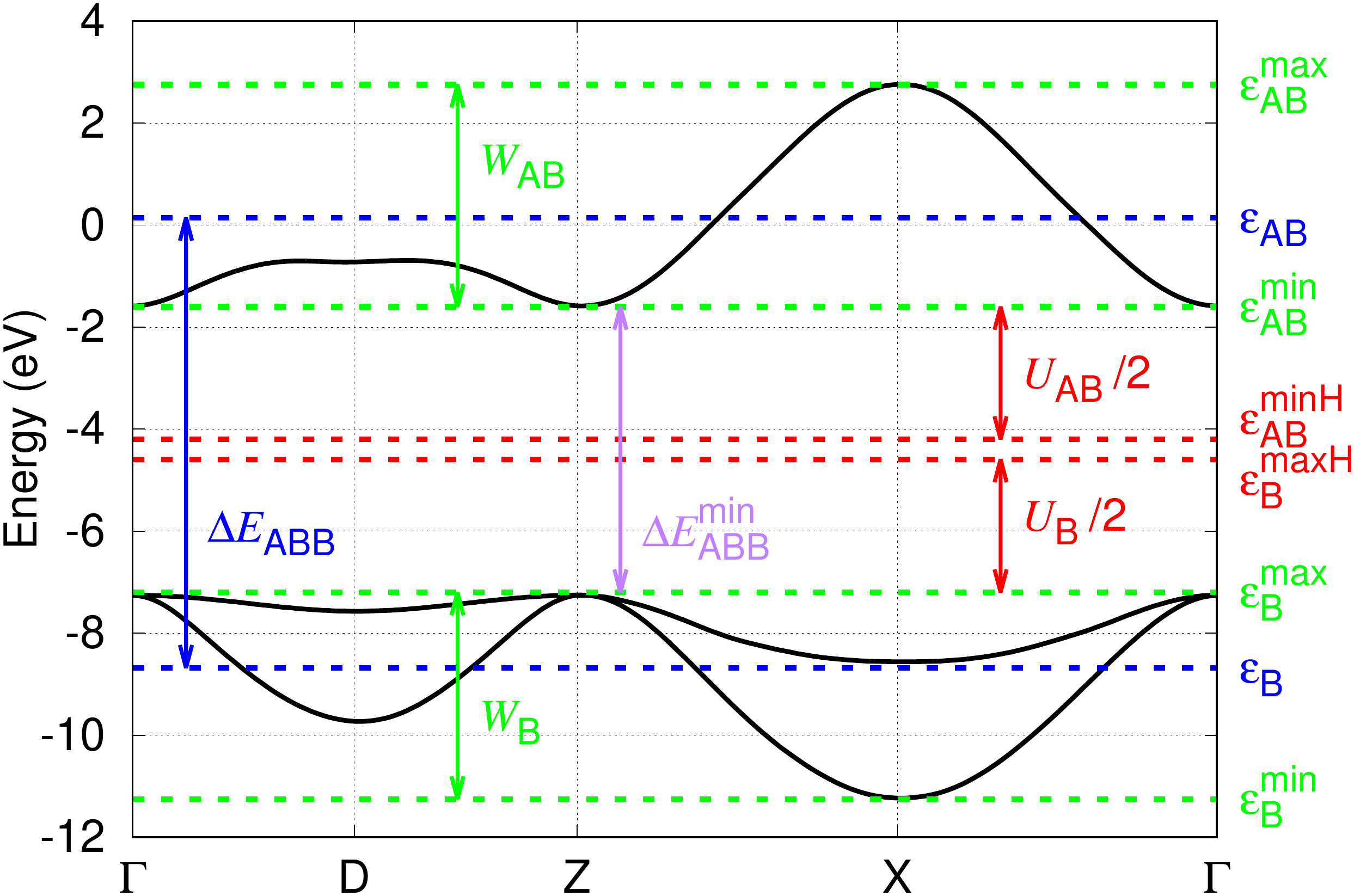}
\caption{
\jbmb{Band structure corresponding to the one-particle part of the $\rm AB\rm B$ LEH for Bi2201 at the c$GW$\jbmt{}{$-$}SIC level (in Table \ref{tab:xabpb}).
We also represent the onsite energy $\epsilon_i$, the minimum and maximum values $\epsilon^{\rm min}_{i}$ and $\epsilon^{\rm max}_{i}$ of the band energy, the partial bandwidth 
$W_i = \epsilon^{\rm max}_{i} - \epsilon^{\rm min}_{i}$ for each manifold, and the charge transfer energy $\Delta E^{}_{\rm ABB}=\epsilon^{}_{\rm AB}-\epsilon^{}_{\rm B}$.
In addition, we show the estimations of the minimal effective value $\Delta E^{\rm min}_{\rm ABB}=\epsilon^{\rm min}_{\rm AB}-\epsilon^{\rm max}_{\rm B}$ of the charge transfer energy, 
the maximum energy of the upper Hubbard band for $\rm B$ manifold $\epsilon^{\rm maxH}_{\rm B} = \epsilon^{\rm max}_{\rm B} + U_{\rm B}/2$, and 
the minimum energy of the lower Hubbard band for $\rm AB$ manifold $\epsilon^{\rm minH}_{\rm AB} = \epsilon^{\rm min}_{\rm AB} - U_{\rm AB}/2$.
}
}
\label{fig:abbhubbard}
\end{figure}

Results are shown in Table \ref{tab:xabpb}; we reproduce values of $v_{\jbmb{\rm AB}}$ and $U_{\jbmb{\rm AB}}$ from the $\jbmb{\rm AB}$ Hamiltonian \jbmsec{in Tables \ref{tab:1nbandhgca} and \ref{tab:1nbandbi} for comparison.}
First, $U_{\jbmb{\rm AB}}$ is larger than for the $\jbmb{\rm AB}$ Hamiltonian. This does not come from the $\jbmb{\rm AB}$ MLW orbital, which is equivalent in both Hamiltonians since $v_{\jbmb{\rm AB}}$ is nearly identical.
The difference comes from the \jbmt{cRPA screening}{c$GW$+LRFB screening}, which excludes the channel between $\jbmb{\rm AB}$ and $\jbmb{\rm B}$ orbitals within the $\jbmb{\rm AB}\jbmb{\rm B}$ Hamiltonian, but includes it within the $\jbmb{\rm AB}$ Hamiltonian. This screening channel contributes to \oldyy{the reduction of} $U_{\jbmb{\rm AB}}$ in the $\jbmb{\rm AB}$ Hamiltonian, with respect to the $\jbmb{\rm AB}\jbmb{\rm B}$ Hamiltonian.

Second, the charge transfer energy $\Delta E_{\rm ABB}$ between the AB and B orbitals is very large ($\Delta E_{\rm ABB} \sim 7.5-\jbmsec{9.0}$ eV) compared to the $xp$ Hamiltonian ($\Delta E_{xp} \sim \jbmsec{1.8-2.6}$ eV in \jbmsec{Appendix~\ref{app:mace}, Table~\ref{tab:3band}}).
This is due to the different character of MLW orbitals, combined to the SIC.
Indeed, the update of the onsite energies $\epsilon_i$ due to the SIC is $t^{\rm SIC}_{i} = -U_{i} n_{i}/2$ \cite{Hirayama2015}. 
For the $xp$ Hamiltonian, we have $t^{\rm SIC}_{x} = -\jbmsec{6.87}$ eV and $t^{\rm SIC}_{p} = -\jbmsec{5.49}$ eV, so that the improvement of $\Delta E_{xp}$ due to the SIC (before applying the renormalization factor \cite{Hirayama2013}) is $t^{\rm SIC}  = t^{\rm SIC}_{x} -t^{\rm SIC}_{p} \sim -\jbmsec{1.37}$ eV, as seen in Table \ref{tab:3band}.
On the other hand, for the $\rm AB\rm B$ Hamiltonian, we have $t^{\rm SIC}_{\rm AB} = -\jbmsec{2.08}$ eV and $t^{\rm SIC}_{\rm B} = -\jbmsec{5.22}$ eV. As a result, $t^{\rm SIC} = t^{\rm SIC}_{\rm AB} - t^{\rm SIC}_{\rm B} \sim +\jbmsec{3.14}$ eV, which 
explains the large value of $\Delta E_{\rm ABB}$ within the $\rm AB\rm B$ LEH compared to $\Delta E_{xp}$ within the $xp$ LEH.

For the $\rm ABB$ Hamiltonian, the large $\Delta E_{\rm ABB}$ suggests that $\rm B$ orbitals are far below the Fermi level and play no role in low-energy physics except for the \jbmt{cRPA-type}{c$GW$+LRFB} screening from the B orbitals, \jbmsec{which suggests it is reasonable to exclude B orbitals from the LEH.}
Still, we must take into account the strong electronic correlation encoded within $U_{\rm AB}$ and $U_{\rm B}$ in order to verify this.
Here, we propose the following criterion: 
\textit{It \jbmsec{may be} safe to exclude the $\rm B$ manifold from the LEH if the upper Hubbard band from the $\rm B$ manifold} 
(UHB$_{\rm B}$) 
\textit{remains well below the Fermi level, and does not overlap strongly with the lower Hubbard band from the $\rm AB$ manifold} 
(LHB$_{\rm AB}$).
That is, the charge transfer energy must be larger than or similar to the intra-orbital \misec{ECR}, and the insulating phase of the mother compound should retain the purely AB-like Mott insulating character, without B $\rightarrow$ AB charge transfer insulating behaviour.

We discuss in detail the latter point for the case of Bi2201, before discussing other compounds.
We represent \jbmb{in} Fig. \ref{fig:abbhubbard} the band structure of Bi2201 at the c$GW$\jbmt{}{$-$}SIC level, along with the onsite energy $\epsilon_{i}$ for each manifold, the charge transfer energy $\Delta E_{\rm ABB} = \epsilon_{\rm AB} - \epsilon_{\rm B}  \sim \jbmsec{8.8}$ eV, and other quantities defined below.
In order to quantify the competition between intra-orbital \misec{ECR} and charge transfer energy, a first possibility is to 
estimate the energy of the LHB$_{\rm AB}$ as $\epsilon_{\rm AB} - U_{\rm AB}/2$, and the energy of the UHB$_{\rm B}$ as $\epsilon_{\rm B} + U_{\rm B}/2$.
Equivalently, we define the dimensionless ratio
\begin{equation}
\mathcal{L}_{\rm B}^{(0)} = \frac{U_{\rm AB} + U_{\rm B}}{2 \Delta E_{\rm ABB}},
\label{eq:lp0}
\end{equation}
so that $\mathcal{L}_{\rm B}^{(0)} <1$ if $\epsilon_{\rm B} + U_{\rm B}/2 < \epsilon_{\rm AB} - U_{\rm AB}/2$.
For Bi2201, we obtain $\epsilon_{\rm AB} - U_{\rm AB}/2 \sim -\jbmsec{2.6}$ eV and $\epsilon_{\rm B} + U_{\rm B}/2 \sim -\jbmsec{6.0}$ eV, so that $\mathcal{L}_{\rm B}^{(0)} \sim \jbmsec{0.59}$ is well below $1$.

However, the ratio $\mathcal{L}_{\rm B}^{(0)}$ in Eq. (\ref{eq:lp0}) is not appropriate enough as the criterion.
For instance, it ignores the finite bandwidths $W_{\rm AB} \sim 4.2$ eV and $W_{\rm B} \sim 4.1$ eV of the ${\rm AB}$ and ${\rm B}$ manifolds at the c$GW$\jbmt{}{$-$}SIC level (as seen in Fig. \ref{fig:abbhubbard}), which are comparable to $U_{\rm AB} \sim \jbmsec{5.2}$ eV, $U_{\rm B} \sim \jbmsec{5.2}$ eV and $\Delta E_{\rm ABB} \sim \jbmsec{8.8}$ eV. 
These finite bandwidths may be retained at least partially by the LHB$_{\rm AB}$ and UHB$_{\rm B}$ when the LEH is solved, and should be taken into account in Eq. (\ref{eq:lp0}).
Thus, we modify Eq. (\ref{eq:lp0}) as follows.
We include the effect of $W_{\rm AB}=\epsilon^{\rm max}_{\rm AB}-\epsilon^{\rm min}_{\rm AB}$ and $W_{\rm B}=\epsilon^{\rm max}_{\rm B}-\epsilon^{\rm min}_{\rm B}$, where 
$\epsilon^{\rm min}_{i}$ (respectively, $\epsilon^{\rm max}_{i}$) is the minimum (respectively, maximum) value of the band energy for each manifold, as defined \jbmb{in} Fig. \ref{fig:abbhubbard}.
To do so, we replace $\Delta E_{\rm ABB} \sim \jbmsec{8.8}$ eV by $\Delta E^{\rm min}_{\rm ABB}=\epsilon^{\rm min}_{\rm AB}-\epsilon^{\rm max}_{\rm B} \sim \jbmsec{5.6}$ eV.
Equivalently, we assume that the minimum energy of the LHB$_{\rm AB}$ is
\begin{equation}
\epsilon^{\rm minH}_{\rm AB} = \epsilon^{\rm min}_{\rm AB} - U_{\rm AB}/2
\end{equation}
which is $\sim -\jbmsec{4.1}$ eV for Bi2201, and the maximum energy of the UHB$_{\rm B}$ is 
\begin{equation}
\epsilon^{\rm maxH}_{\rm B}  = \epsilon^{\rm max}_{\rm B} + U_{\rm B}/2
\end{equation}
which is $\sim -\jbmsec{4.5}$ eV for Bi2201.
We redefine Eq. (\ref{eq:lp0}) as
\begin{equation}
\mathcal{L}_{\rm B}^{} = \frac{U_{\rm AB} + U_{\rm B}}{2 \Delta E^{\rm min}_{\rm ABB}}\oldyy{,}
\label{eq:lp}
\end{equation}
and the criterion is satisfied if $\mathcal{L}_{\rm B}^{} \lesssim 1$ and $\epsilon^{\rm maxH}_{\rm B}$ is well below the Fermi energy.
Values of $\mathcal{L}_{\rm B}^{}$ and $\epsilon^{\rm maxH}_{\rm B}$ are shown in Table \ref{tab:xabpb}.
We see that $\epsilon^{\rm maxH}_{\rm B} \lesssim - 4.0$ eV for all compounds, which is well below the Fermi energy.
In addition, $\mathcal{L}_{\rm B}^{} \lesssim 1$ for all compounds.
Thus, LHB$_{\rm AB}$ and UHB$_{\rm B}$ 
are nearly separated and are not strongly entangled.
We note the fact that $\mathcal{L}_{\rm B}^{}$ as defined in Eq.~(\ref{eq:lp}) is an upper bound, and the true value of $\mathcal{L}_{\rm B}^{}$ in the practical resolution of the LEH may be lower because of the band narrowing of each UHB and LHB in comparison to $W_{\rm B}$ or $W_{\rm AB}$ due to the correlation effect as was observed in e.g. Ref. \onlinecite{Charlebois2020} for the two-dimensional Hubbard model.
Therefore, $\Delta E^{\rm min}_{\rm ABB}$ may be larger.
In any case, the criterion is satisfied for all compounds\jbmsec{.}
\jbmsec{
This suggests it may be appropriate to exclude the $\rm B$ manifold from the LEH and restrict the present study to the $\rm AB$ LEH.
Of course, this can only be confirmed by solving both AB and ABB LEHs with an accurate low-energy solver, and comparing the ground state and value of SC order parameter for both LEHs.
This issue is left for future studies, for which the AB and ABB LEHs provided in the Supplemental Material may be used as a base.
}
\\

\jbmsec{Finally, for the sake of completeness, we mention that the Hund exchange energy is non-negligible in the ABB LEH\jbmt{, although we ignore it in Eq.~\ref{eq:heff}.}{.}
\jbmt{For instance, in the case of Ca11 at $\delta=0.0$, the Hund exchange energy between fully occupied B orbitals and half-filled AB orbital is $0.79$ eV.}{For all compounds, the Hund exchange between AB and B orbitals in the unit cell is $\sim 0.70-0.80$ eV, that is, $\sim 25\%$ of the direct interaction between AB and B orbitals in the unit cell $\sim 2.3-2.8$ eV.}
However, since the B band is completely filled, the Hund's rule coupling and the exchange coupling should not play a role.
Nonetheless, the Hund exchange energy is given in the Supplemental Material for each ABB Hamiltonian.
}

\section{Reduction of the computational cost of the $GW$ correction of M space}
\label{app:ode}

Here, we propose an approximation to reduce the computational cost of the $GW$ self-energy, without loss of accuracy.
This approximation has not been used for calculations in this paper, but may be useful for future studies.

One of the most difficult parts of the calculation is the $GW$ correction of the M space\jbm{: T}he computational cost of the $GW$ self-energy for the $N_{\rm M}$ bands in the M space scales as $N_{\rm M} (N_{\rm M}+1)/2$, which becomes challenging for cuprates with large $N_{\rm M}$ (as an example, $N_{\rm M}=11$ for Ca11 and $17$ for Hg1201, but $23$ for Bi2201 and $34$ for Bi2212).
For each wavevector in the irreducible Brillouin zone, \oldyy{we} have to compute $N_{\rm M}$ diagonal elements and $N_{\rm M} (N_{\rm M}-1)/2$ off-diagonal elements (ODEs) in the upper triangle; ODEs in the lower triangle may be deduced by Hermitianity. 
The large computational cost comes from ODEs.

However, in practice, ODEs have a sparse structure, especially for highly symmetric systems.
Thus, it is desirable to anticipate which ODEs are negligible, and restrict the calculation of the $GW$ self-energy to \oldyy{finite and} important ODEs.
To do so, the idea is to use as a guideline the ODEs of the Kohn-Sham exchange-correlation potential $V^{\rm xc}$\jbm{.} 
\jbm{F}irst, matrix elements of $V^{\rm xc}$ are much cheaper to calculate than those of the $GW$ self-energy.
Second, we remark in practice that negligible (respectively, non-negligible) ODEs for the $GW$ self-energy are also negligible (respectively, non-negligible) for $V^{\rm xc}$\jbm{:} 
\jbm{A}lthough $GW$ improves exchange and correlation beyond the Kohn-Sham level, the correction is mainly quantitative and does not change the overall structure of the matrix elements.

Thus, the procedure is the following\jbm{.} (1) \jbm{C}ompute all $N_{\rm M} (N_{\rm M}+1)/2$ matrix elements for $V^{\rm xc}$; (2) determine ODEs whose amplitude is negligible, e.g. by using a cutoff energy $\epsilon$; (3) assume these ODEs are negligible in the $GW$ self-energy as well; (4) calculate the $GW$ self-energy only for diagonal elements and non-negligible ODEs.

We benchmarked this procedure in the case of Ca11\jbm{;} the cutoff energy $\epsilon=0.05$ eV allows to reproduce the $GW$ band structure with excellent accuracy, while reducing the number of ODEs and computational cost by $\sim 40\%$.
Thus, the procedure can be useful for future studies of cuprates with large $N_{\rm M}$, especially those for which $N_{\ell} \geq 2$.
The finite cutoff value $\epsilon$ may be determined for another cuprate with small $N_{\rm M}$ (e.g. Ca11).
Then, this cutoff value may be considered for the cuprate with large $N_{\rm M}$.
Of course, the procedure may also be used for systems other than cuprates.


\newpage
\bibliographystyle{unsrt}

\end{document}